\documentclass[a4paper,11pt]{article}

\usepackage{amsmath}
\usepackage{amssymb}
\usepackage{latexsym}
\usepackage{graphicx}
\usepackage{enumerate}
\usepackage{colortbl}
\usepackage{amsthm} 
\usepackage{bm}
\usepackage{mathtools}
\usepackage{caption}
\usepackage{subcaption}
\usepackage{natbib}
\usepackage{url}
\mathtoolsset{showonlyrefs=true}
\usepackage[table]{xcolor}

\usepackage[utf8]{inputenc}
\usepackage[T1]{fontenc}

\usepackage{hyperref}

\newtheorem{theorem}{Theorem}

\newtheorem{definition}{Definition}

\newtheorem{corollary}{Corollary}

\newcommand{\T}{\top}
\newcommand{\para}[1]{\vspace{0.75\baselineskip}\noindent\textbf{#1}\ }

\graphicspath{
	{./art/}
}

\begin{document}
\title{\textbf{\Large An interpretable family of projected normal distributions and a related copula model for Bayesian analysis of hypertoroidal data}}
\author{\scshape{Shogo\ Kato\,$^{a}$, \quad Gianluca\ Mastrantonio\,$^{b}$,} \vspace{0.1cm}\\
\scshape{ \quad and \quad Masayuki\ Ishikawa\,$^{c}$} \vspace{0.5cm}\\
\textit{\(^{a}\) Institute of Statistical Mathematics, Japan} \vspace{0.1cm}\\
\textit{\(^{b}\) Department of Mathematics, Polytechnic of Turin, Italy} \vspace{0.1cm}\\
\textit{\(^{c}\) Mitsui Sumitomo Insurance Co., Ltd., Japan} \vspace{0.5cm}\\ 
}
\date{\today}
\maketitle

\begin{abstract}
This paper introduces two families of probability distributions for Bayesian analysis of hypertoroidal data.
The first family consists of symmetric distributions derived from the projection of multivariate normal distributions under specific parameter constraints.
This family is closed under marginalization and hence any marginal distribution belongs to a lower-dimensional case of the same family.
In particular the univariate marginal of the family is the unimodal case of the projected normal distribution on the circle.
The second family is a flexible extension of the copula case of the first family, which can accommodate any univariate marginal distributions.
Unlike existing models derived via projection, both families have the common advantage that their parameters possess a clear and intuitive interpretation.
The use of latent variables simplifies Bayesian estimation using Markov chain Monte Carlo algorithms.
The usefulness of the proposed families is demonstrated through the analysis of a meteorological dataset.
\end{abstract}

\noindent%

\section{Introduction} \label{sec:introduction}

Data recorded as angles frequently appear in various academic fields.
Such examples include wave directions in environmental science, directions of animal movement in biology, dihedral angles in protein backbone structures in bioinformatics, and times of crimes  in political science.
In practice, angular observations are often jointly observed with other angular observations, forming what is referred to as hypertoroidal data.
For example, conformational angles in protein structure consist of two dihedral angles and one torsion angle of the side chain,
 and wave directions are typically observed  at multiple locations in the sea.
It is well-known that ordinary statistical methods for linear data cannot be directly applied to such hypertoroidal data and  new statistical methods are needed for their analysis.
See, for example, \citet{Mardia1999a}, \citet{Ley2017a} and \citet{Pewsey2021} for the general review of statistics of hypertoroidal data.

A major problem in analyzing hypertoroidal data is the lack of tractable and interpretable probability distributions that can serve as a foundation for parametric statistical analysis.
This contrasts with the analysis of hyperspherical data, for which a variety of such distributions are available; see, e.g., \citet{Scealy2019}, \citet{Kato2020} and \citet{Pewsey2021} for recent reviews.
However, if the analysis is restricted to 2- or 3-dimensional toroidal cases, there are a few distributions that are both tractable and easily interpreted.
The well-known families of distributions for the 2-dimensional case include the bivariate von Mises distributions \citep{Mardia1975b, Singh2002, Mardia2007, Kent2008, Mardia2010} and copula-based models \citep{Wehrly1980, Shieh2005, Shieh2011, Jones2015, Kato2015a}.
A 3-dimensional extension of \citet{Wehrly1980} using the wrapped Cauchy-type copula density was recently proposed by \cite{Kato2024}.
Several distributions on the general-dimensional torus have been proposed in the literature.
\cite{Mardia2008} and \cite{Mardia2014} considered a multivariate extension of the von Mises distribution, and \cite{Navarro2017} discussed its generalization.
The inverse stereographic normal distribution has been proposed by \cite{Selvitella2019}.
A tree-structured model on the general-dimensional torus was proposed by \cite{Leguey2016} in machine learning.
Models on the hypertorus using other approaches include the models of \cite{Fernandez-Duran2014} and \cite{Kim2016a}.
However, existing distributions for general-dimensional toroidal data have certain limitations, such as unclear parameter interpretations, multimodality, and/or a limited range of attainable dependence strength between variables.
In addition, a common issue is that parameter estimation via the maximum likelihood approach is often difficult due to a complicated form of the probability density function.

Recently, Bayesian approaches have been increasingly adopted for modeling directional data, particularly in relation to spatial and spatio-temporal directional data analysis.
Many of these successful approaches are based on projections or wrappings of multivariate normal distributions; see, e.g.,
\cite{Jona-Lasinio2012},
\cite{Wang2013a,Wang2014},
\cite{Wang2015}, \cite{Mastrantonio2016a}, and \cite{Hernandez-Stumpfhauser2017}.
Although these models generally do not have closed-form expressions for their probability density functions, they provide elegant solutions for Bayesian estimation by introducing latent variables.
However, these models are not designed for hypertoroidal data, which are particularly challenging to handle using frequentist approaches, except for a special case of the model proposed by \cite{Mastrantonio2018}, which nonetheless suffers from issues such as multimodality, unclear parameter interpretations, and a limited range of dependence between variables.

In this paper we propose two families of probability distributions for the Bayesian analysis of hypertoroidal data.
The first family, the (hyper-)toroidal projected normal (TPN) distribution, is derived by projecting multivariate normal distributions under specific parameter constraints.
The TPN model inherits various tractable properties of the multivariate normal distribution.
For example, unlike most of existing projected distributions for directional data, the TPN distribution is symmetric around the mean direction.
The family of TPN distributions is closed under marginalization and, in particular, the univariate marginal of the TPN distribution corresponds to the symmetric and unimodal case of the projected normal (PN) distribution \citep[see, e.g.,][Equation (3.5.49)]{Mardia1999a}.
The second family proposed in this paper is an extension of the copula case of the TPN (CTPN) family.
The CTPN is a flexible family of distributions with arbitrary univariate marginal distributions.
The bivariate case of the CTPN belongs to the well-known family of \cite{Wehrly1980} and therefore has a simple expression for its density.
The modes of the joint CTPN density can be expressed in a simple form under certain parameter conditions.
The CTPN can describe a wide range of dependence structures between variables, including independence as well as perfect positive and negative dependence.
For both the TPN and CTPN families, the parameters have clear and intuitive interpretations.
For the TPN family, each marginal mean direction and concentration is governed by a specific parameter, and an additional parameter controls the correlation between each pair of variables.
For the CTPN family, each marginal distribution is specified freely, with its mean direction and other characteristics determined by the parameters of the marginal density, while the dependence among circular variables is governed by the copula density.
As with existing projected distributions, the proposed families generally do not have closed-form expressions for their densities.
However, the introduction of latent variables greatly simplifies Bayesian estimation for both families.
We define a new prior distribution for the covariance matrix of the underlying normal distribution, needed for the constrains that must be imposed to obtain interpretable models, and we develop  Markov chain Monte Carlo (MCMC) algorithms for Bayesian estimation of the proposed families.
Through a simulation study, it is seen that the MCMC algorithm successfully recover the values of the parameters used to simulate the data for both the TPN and the copula case.  
We apply the proposed distributions to real meteorological data consisting of time series of four wind-direction and two wave-direction measurements.


The code used to produce the results of the real data example in \S \ref{sec:example} and the simulation examples in the Supplementary Material is available at \url{https://github.com/GianlucaMastrantonio/toroidal_projected_normal}.

\section{A Family of Projected Normal Distributions on the Hypertorus} \label{sec:TPN_distribution}

\subsection{Idea behind the derivation of the proposed family} \label{sec:derivation}
Before introducing the proposed family and its properties, we first explain the idea behind its derivation using the univariate and bivariate settings.
In the univariate setting, a circular variable $\Theta$ is said to follow a projected normal (PN) distribution, denoted by $\mathrm{PN}((\eta_c, \eta_s)^\T, \bm{\Psi})$, if a bivariate random vector $(X_c, X_s)^\T$ follows a bivariate normal distribution ${\cal N}_2((\eta_c, \eta_s)^\T, \bm{\Psi})$, and $\Theta$ is defined through the transformation:
\begin{equation}
	\Theta = \text{atan}^* \left(\frac{X_s}{X_c}\right),
\end{equation}
where $\text{atan}^*(y/x)$ denotes the modified arctangent function that returns an angle in $[-\pi,\pi)$ for $(x,y) \neq (0,0)$, whose sine and cosine are $y/\{ x^2+y^2 \}^{1/2}$ and $x/\{ x^2+y^2 \}^{1/2}$, respectively \citep[see][Equation (2.2.4)]{Mardia1999a}.
It is known that when $\bm{\Psi} = \bm{I}_2$, the PN is symmetric around the mean direction
$ \mu = \text{atan}^* ( \eta_s / \eta_c ), $
and the concentration parameter 
$
	\kappa = \{ \eta_c^2 + \eta_s^2 \}^{1/2}
$
is a function of the mean vector $(\eta_c, \eta_s)^\T$ \citep[][Equation (3.5.49)]{Mardia1999a}.
The starting point of our proposal, which to the best of our knowledge has never been pointed out before, is that, under the assumption $\bm{\Psi} = \bm{I}_2$, we can construct a PN-distributed variable by directly using $(\mu, \kappa)$ as parameters. This is possible if we assume that $(X_c, X_s)^\T \sim {\cal N}_2((\kappa, 0)^\T, \bm{I}_2)$, with
$ \Theta = \mu + \text{atan}^* ( X_s / X_c ). $

We can now move to the bivariate setting. If we derive two circular variables $\Theta_1  $ and $\Theta_2$ from the linear variables $\{ X_{c,1},X_{c,2},X_{s,1},X_{s,2} \}$ via
\begin{equation}
	\Theta_1 = \mu_1 + \text{atan}^* \left(\frac{X_{s,1}}{X_{c,1}}\right), \quad 
    \Theta_2 = \mu_2 + \text{atan}^* \left(\frac{X_{s,2}}{X_{c,2}}\right), 
\end{equation}
a simple way to introduce dependence between $\Theta_1$ and $\Theta_2$ while keeping the marginal distribution invariant (see Theorem \ref{thm:marginals} below for a formal proof) is to let $(X_{c,1},X_{c,2})$ and $ (X_{s,1},X_{s,2})$ be independent and assume  
\begin{equation}
	\begin{pmatrix}
	    X_{c,1} \\ X_{c,2}
	\end{pmatrix}
    \sim {\cal N}_2 (\bm{\kappa},\bm{\Sigma}_c), \quad
    \begin{pmatrix}
	    X_{s,1} \\ X_{s,2}
	\end{pmatrix}
    \sim {\cal N}_2 (\bm{0},\bm{\Sigma}_s),
\end{equation}
where $\bm{\kappa} \in (\mathbb{R}^+)^2$, and both $\bm{\Sigma}_c$ and $\bm{\Sigma}_s$ are correlation matrices.
Depending on the values of $\rho_{c} = [\bm{\Sigma}_c]_{1,2}$ and $\rho_{s} = [\bm{\Sigma}_s]_{1,2}$, different types of dependence may be observed, where $[\bm{\Sigma}]_{j,k}$ denotes the $(j,k)$ element of $\bm{\Sigma}$.
For interpretability and parameter parsimony, we aim to introduce a single parameter $\rho$ such that, for positive values, the two circular variables tend to be positively related, whereas for negative values, they tend to be negatively related.

From  \citet{Wang2013a}, we know that $E \{ \sin (\Theta_1 - \mu_1 ) \} = E \{ \sin (\Theta_2 - \mu_2 ) \} = 0$.  
It is evident that $X_{s,j}$ and   
$\sin (\Theta_j - \mu_j )$
 have the same sign and, moreover, the larger the value of $X_{s,j}$, the larger the value of $\sin (\Theta_j - \mu_j )$.
Hence, a positive $\rho_s$ induces a positive correlation between the two sine transformations and hence between $\Theta_1-\mu_1$ and $\Theta_2-\mu_2$, whereas a negative $\rho_s$ has the opposite effect.
To induce the appropriate correlation between the two circular variables, we also need to define the correlation between the two $X_{c,j}$'s which induces a dependence between the cosines.
In this case, regardless of whether the relationship between $\Theta_1 - \mu_1 $ and $\Theta_2 - \mu_2 $ is positive or negative, it is natural to assume that $\cos (\Theta_j - \mu_j)$ increases with $X_{c,j}$ for $j=1,2$, due to the identity $\cos (\Theta_j-\mu_j) = \cos \{ - (\Theta_j-\mu_j)\}$.
 This suggests that, for the sign of $\rho_c$ to reflect the relationship between $\Theta_1$ and $\Theta_2$, the parameter $\rho_c$ should always be positive. 
For this reason, we set $[\bm{\Sigma}_s]_{1,2} = \rho$ and $[\bm{\Sigma}_c]_{1,2} = |\rho|$, which is the basic idea of our proposal. A generalization to arbitrary dimensions is proposed below.


\subsection{Definition and basic properties} \label{sec:definition}
  
\begin{definition} \label{def:model}
	Let $\bm{X}_c=(X_{c,1},\ldots,X_{c,d})^\T$ and $\bm{X}_s=(X_{s,1},\ldots,X_{s,d})^\T$ be two independent $d$-dimensional random vectors distributed as ${\cal N}_d (\bm{\kappa}, \bm{\Sigma}_c)$ and ${\cal N}_d(\bm{0}, \bm{\Sigma})$, respectively, where $\bm{\kappa} = (\kappa_1, \ldots, \kappa_d)^\T$, $\kappa_1, \ldots, \kappa_d \geq 0$, and $\bm{\Sigma}_c$ and $\bm{\Sigma}$ are correlation matrices with the $(j,k)$ elements $[\bm{\Sigma}]_{j,k} = \rho_{jk}$ and $[\bm{\Sigma}_c]_{j,k} = |\rho_{jk}|$ $(1\leq j,k \leq d)$.
    Define the circular random variables
    \begin{equation}
	\Theta_1 = \mu_1 + \text{atan}^* \left(\frac{X_{s,1}}{X_{c,1}}\right), \quad \ldots, \quad
    \Theta_d = \mu_d + \text{atan}^* \left(\frac{X_{s,d}}{X_{c,d}}\right), 
\end{equation}
%
with $\Theta_j \in [-\pi, \pi)$ and $\mu_j \in [-\pi,\pi)$.
    Then the random vector $(\Theta_1, \ldots, \Theta_d)$ consists of $d$-dimensional circular variables taking values on the hypertorus $[-\pi,\pi)^d$, and we refer to their joint distribution as the (hyper-)toroidal projected normal (TPN) distribution.
\end{definition}

As is clear from the definition, random variates from the TPN distributions can be efficiently generated using those from multivariate normal distributions.

For convenience, we write $(\Theta_1,\ldots,\Theta_d) \sim \mathrm{TPN}_d (\bm{\mu},\bm{\kappa},\bm{\Sigma})$, with $\bm{\mu}=(\mu_1,\ldots,\mu_d)^\T$,  if the random vector $(\Theta_1,\ldots,\Theta_d)$ follows the distribution in Definition \ref{def:model}.
For any $\bm{\Sigma}$, $\bm{\Sigma}_c$ is always positive definite when $d=2$, but this is not generally the case when $d \geq 3$.
Nevertheless, our numerical simulations suggest that when the positive definite matrices $\bm{\Sigma}$ with $d\geq 3$ are generated from Inverse Wishart distributions, the corresponding $\bm{\Sigma}_c$ is often positive definite as well.
In Bayesian estimation, appropriate prior distributions must be specified for both $\bm{\Sigma}$ and $\bm{\Sigma}_c$,  see \S \ref{sec:MCMC}.

Basic properties of the TPN family are provided below. 
Here and throughout this section, assume that $\bm{\Theta} = (\Theta_1,\ldots,\Theta_d)^\T \sim \mathrm{TPN}_d (\bm{\mu},\bm{\kappa},\bm{\Sigma})$.
\begin{theorem} \label{thm:symmetry}
	The distribution of $\bm{\Theta}$ is symmetric about $\bm{\mu}$, namely, $ \bm{\Theta} - \bm{\mu} \stackrel{d}{=} \bm{\mu} - \bm{\Theta} $.
\end{theorem}
The symmetry of our model leads to clear interpretation of the parameters, which include not only $\bm{\mu}$ itself but also $\bm{\kappa}$ and $\bm{\Sigma}$; see Theorem \ref{thm:univariate_marginal} and Figs.~\ref{fig:univariate_density} and \ref{fig:bivariate_density} below.


\begin{theorem} \label{thm:marginals}
	Let $\bm{\Theta}_1 = (\Theta_1,\ldots,\Theta_{d_1})$ and $\bm{\Theta}_2 = (\Theta_{d_1+1},\ldots,\Theta_d)$.
	Then
	$$
	\bm{\Theta}_1 \sim \mathrm{TPN}_{d_1}(\bm{\mu}_1, \bm{\kappa}_1, \bm{\Sigma}_{11}) \quad \text{and} \quad \bm{\Theta}_2 \sim \mathrm{TPN}_{d_2}(\bm{\mu}_2, \bm{\kappa}_2, \bm{\Sigma}_{22}),
	$$
	where $\bm{\mu}_1=(\mu_1,\ldots,\mu_{d_1})^\T$, $\bm{\mu}_2=(\mu_{d_1+1},\ldots,\mu_d)^\T$,  $\bm{\kappa}_1 = (\kappa_1, \ldots, \kappa_{d_1})^\T$, $\bm{\kappa}_2 = (\kappa_{d_1+1}, \ldots, $ $\kappa_d)^\T$, $d_1+d_2=d$, and $\bm{\Sigma}_{11}$ and $\bm{\Sigma}_{22}$ are $d_1 \times d_1$ and $d_2 \times d_2$ sub-matrices of $\bm{\Sigma}$, respectively, defined by
	$$
	\bm{\Sigma} =
	\begin{pmatrix}
		\bm{\Sigma}_{11} & \bm{\Sigma}_{12} \\
		\bm{\Sigma}_{12}^\T & \bm{\Sigma}_{22}
	\end{pmatrix},
	$$
	with $\bm{\Sigma}_{12}$ being a $d_1 \times d_2$ matrix.
\end{theorem}
Any permutation of the random vector $(\Theta_1,\ldots,\Theta_d)$ also has the TPN distribution due to the symmetric form of the distribution and is hence closed under marginalization.

\begin{theorem} \label{thm:univariate_marginal}
	The marginal distribution of $\Theta_j$ $(j=1,\ldots,d)$ is a symmetric and unimodal case of the PN distribution with density
	\begin{equation} \label{eq:univariate_density}
	f(\theta_j) = \frac{1}{\surd (2\pi)} \phi (\kappa_j) + \kappa_j \cos (\theta_j-\mu_j) \phi \{ \kappa_j \sin (\theta_j-\mu_j) \} \Phi \{ \kappa_j \cos (\theta_j-\mu_j) \},
	\end{equation}
	where $\phi$ and $\Phi$ denote the probability density function and cumulative distribution function of the standard normal distribution, respectively.
\end{theorem}
The proof for Theorem \ref{thm:univariate_marginal} is straightforward and follows directly from \citet[\S 3.5.6]{Mardia1999a}: it is therefore omitted.

From Theorems \ref{thm:symmetry}--\ref{thm:univariate_marginal}, it follows that our model can be considered a toroidal extension of the symmetric and unimodal PN distribution on the circle.
Theorem \ref{thm:marginals} also gives the distribution of any possible combination of two circular variables, which are marginally TPN distributed.
Then, given the discussion in \S \ref{sec:derivation}, we can clearly see that $\mu_j$ is the mean direction of $\Theta_j$, the concentration of $\Theta_j$ depends only on $\kappa_j$, while the magnitude and sign of $\rho_{jk}$ measure the dependence between $\Theta_j$ and $\Theta_k$.


\begin{figure}[t]
\begin{center}
	\includegraphics[width=7cm,height=5.25cm,clip]{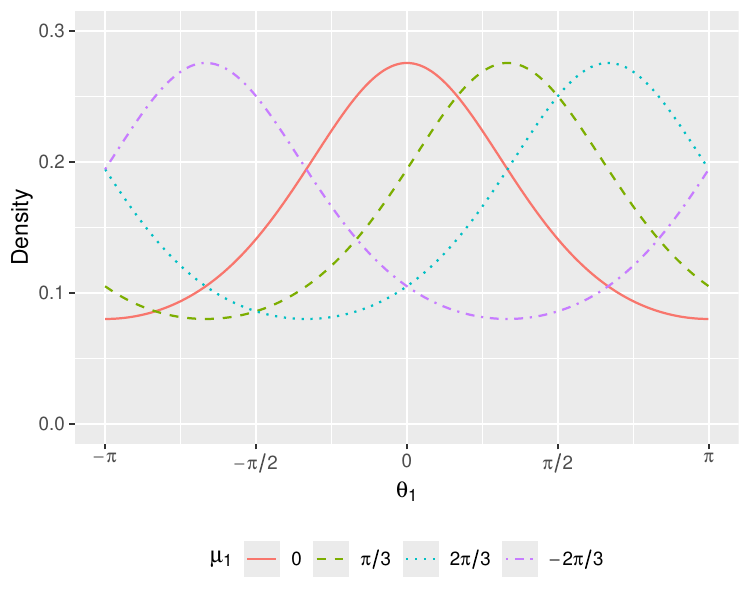} \hspace{0.5cm} 
	\includegraphics[width=7cm,height=5.25cm,clip]{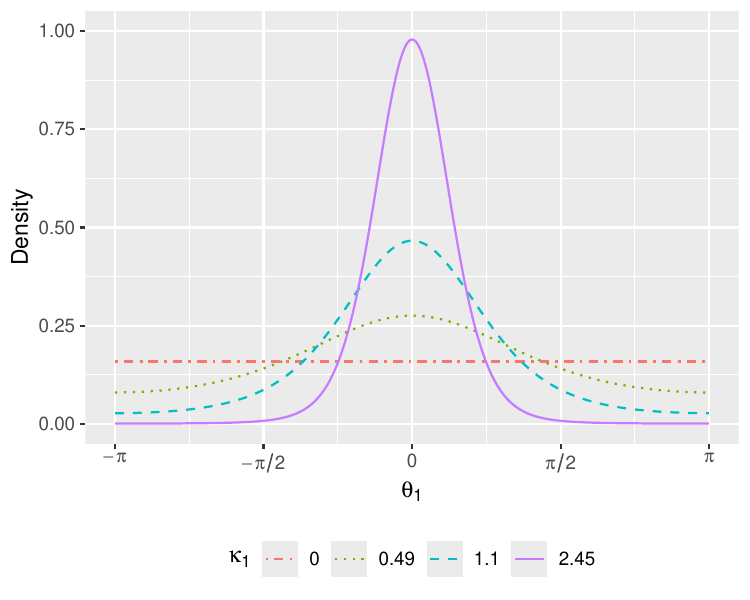}  \\ 

	{\footnotesize \hspace{0.5cm} (a) \hspace{7.1cm} (b) }
	\caption{Plots of the univariate marginal density \eqref{eq:univariate_density} of $\Theta_1$ for: (a) $\kappa_1=0.49$ with four values of $\mu_1$; and (b) $\mu_1=0$ with four values of $\kappa_1$.}
\label{fig:univariate_density}
\end{center}
\end{figure}

Figure~\ref{fig:univariate_density} shows the univariate marginal density \eqref{eq:univariate_density} of $\Theta_1$ for selected values of the parameter.
The four values of $\kappa_1$ used in Fig.~\ref{fig:univariate_density}, namely, $(0, 0.49, 1.1, 2.45)$, approximately correspond to mean resultant lengths  $|E(e^{{ i} \Theta_1})|$ of $(0, 0.3, 0.6, 0.9)$, respectively.
{The mean resultant length is a measure of concentration for circular random variable, ranging from zero for uniformity to one for a point mass
\citep[see, e.g.,][\S 3.4.2]{Mardia1999a}.}
As seen in Fig.~\ref{fig:univariate_density}(a), the parameter $\mu_1$ represents the mode of the univariate density \eqref{eq:univariate_density}.
The density \eqref{eq:univariate_density} reaches its maximum value at $\theta_1 = \mu_1$ and its minimum value at $\theta_1 = \mu_1 + \pi$.
Also the density \eqref{eq:univariate_density} is symmetric about $\theta_1 = \mu_1$.
Figure~\ref{fig:univariate_density}(b) shows how parameter $\kappa_1$ controls the concentration of the univariate marginal distribution.
If $\kappa_1 = 0$, then the marginal distribution of $\Theta_1$ is uniform on the circle.
The greater the value of $\kappa_1$, the greater the concentration of the marginal density of $\Theta_1$ around $\theta_1 = \mu_1$.
As $\kappa_1 \to \infty$, the marginal distribution of $\Theta_1$ converges to a point distribution with a singularity at $\theta_1 = \mu_1$.

\begin{figure}
	\centering {\footnotesize
		\hspace{2cm} $\kappa_1=0$ 
            \hspace{2.3cm} $\kappa_1=0.49$ 
            \hspace{2.3cm} $\kappa_1=1.1$ 
            \hspace{2.1cm} $\kappa_1=2.45$ 
            \vspace{0.10cm} 
            \\
            
		\raisebox{1.6cm}{$\rho_{12}=0$} \hspace{0.4cm}
	\includegraphics[width=3.3cm,height=3.3cm]{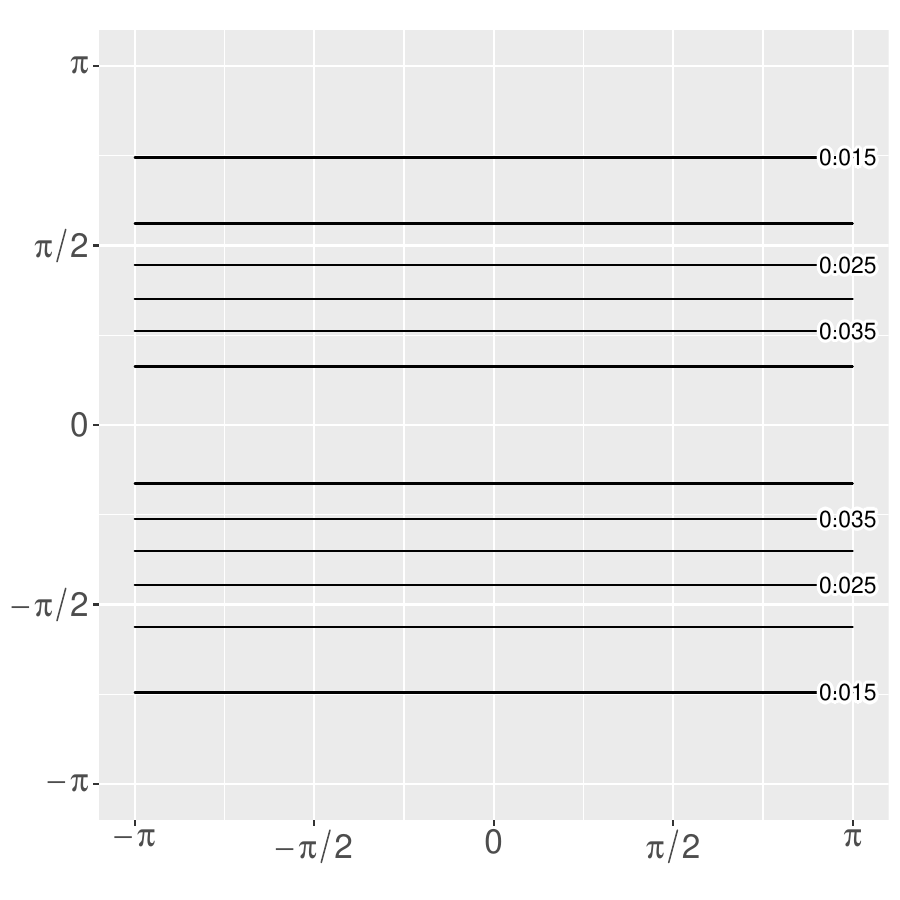}	
	\includegraphics[width=3.3cm,height=3.3cm]{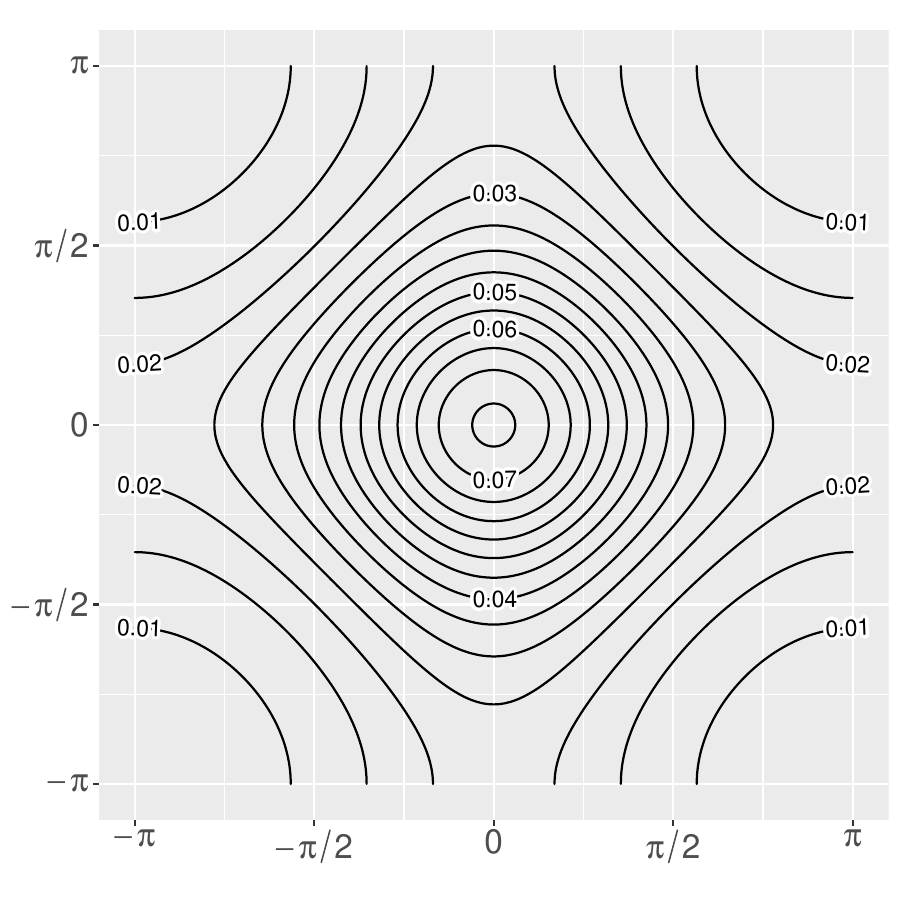}	
        \includegraphics[width=3.3cm,height=3.3cm]{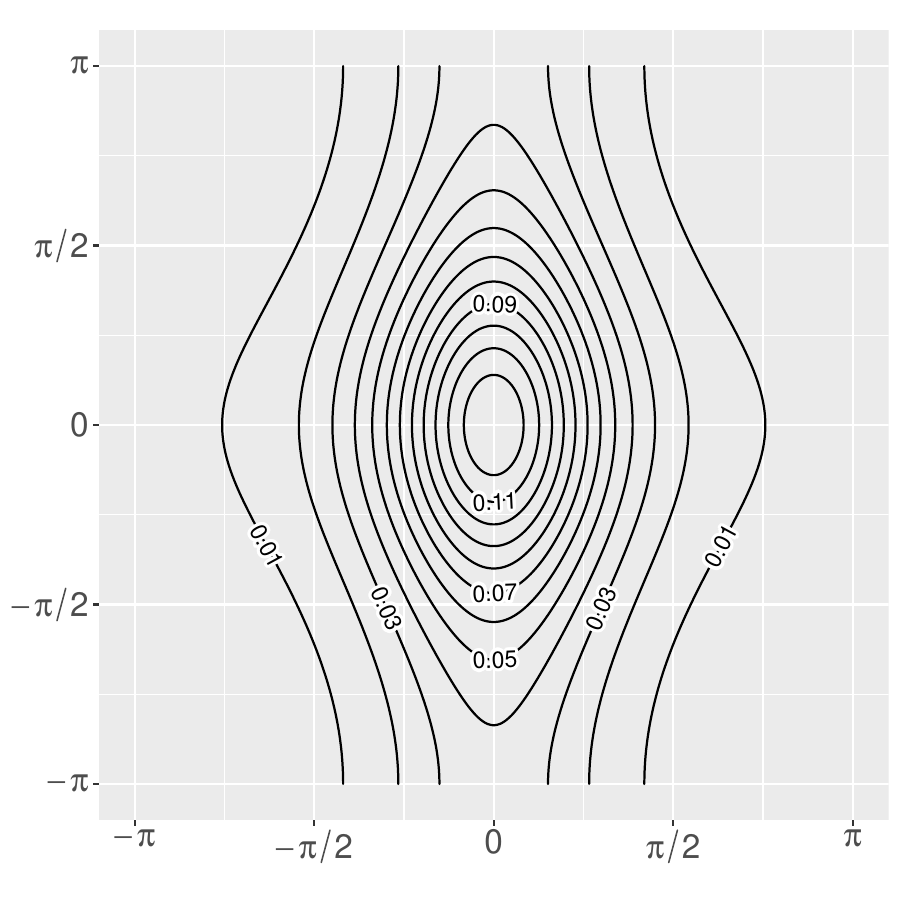}	
	\includegraphics[width=3.3cm,height=3.3cm]{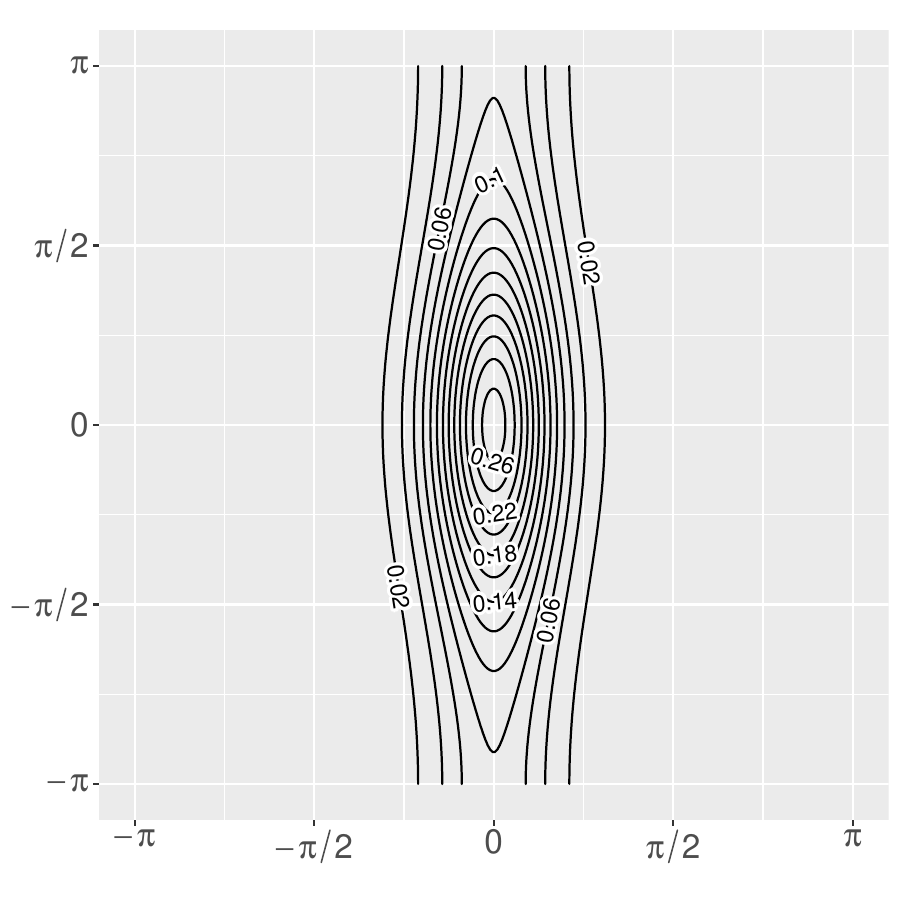}	
        \\
    
        \raisebox{1.6cm} {$\rho_{12}=0.3$} \hspace{0.12cm}
        \includegraphics[width=3.3cm,height=3.3cm]{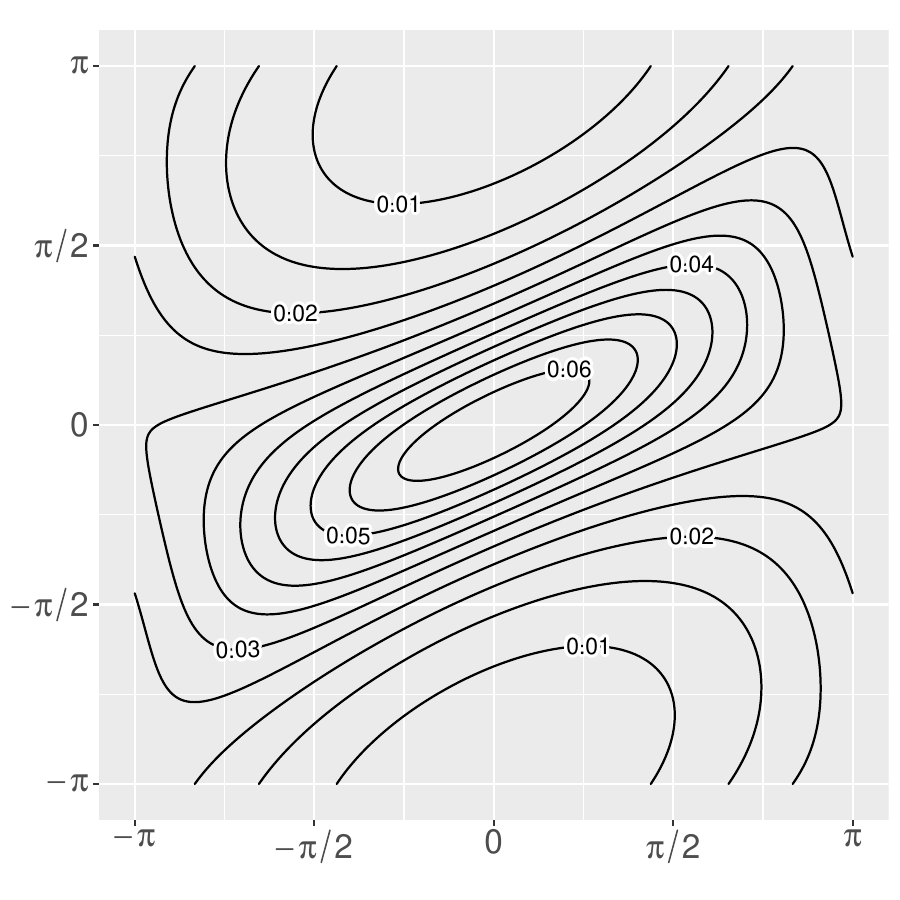}	
        \includegraphics[width=3.3cm,height=3.3cm]{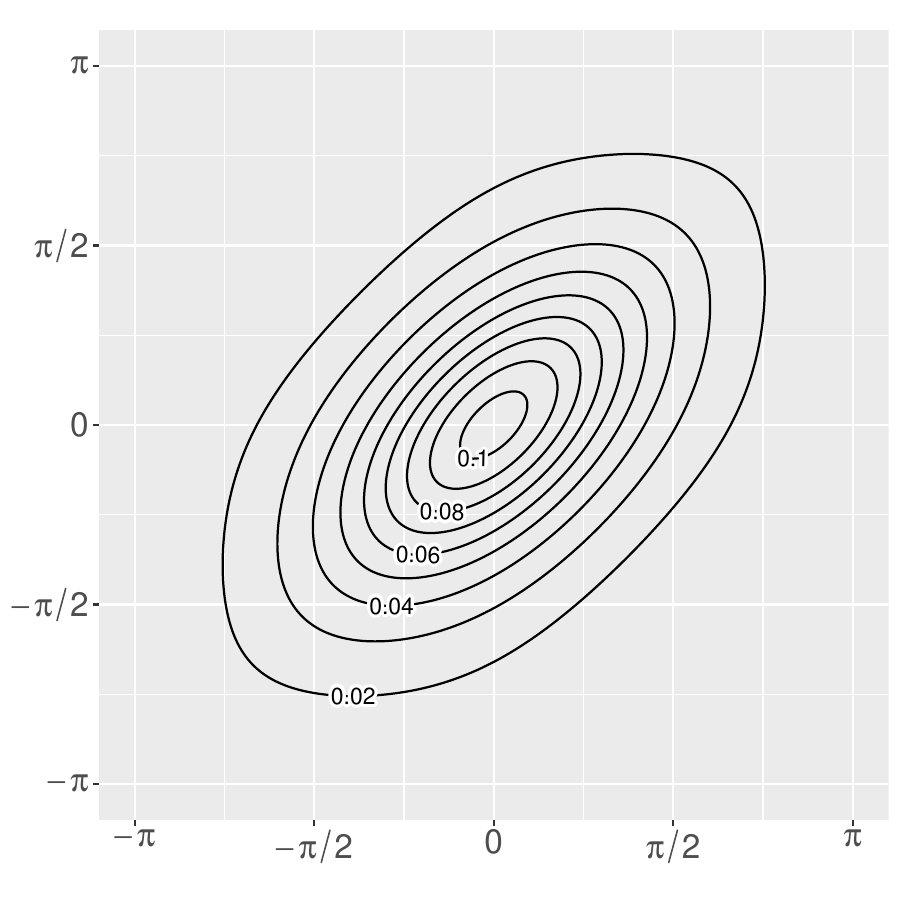}	
        \includegraphics[width=3.3cm,height=3.3cm]{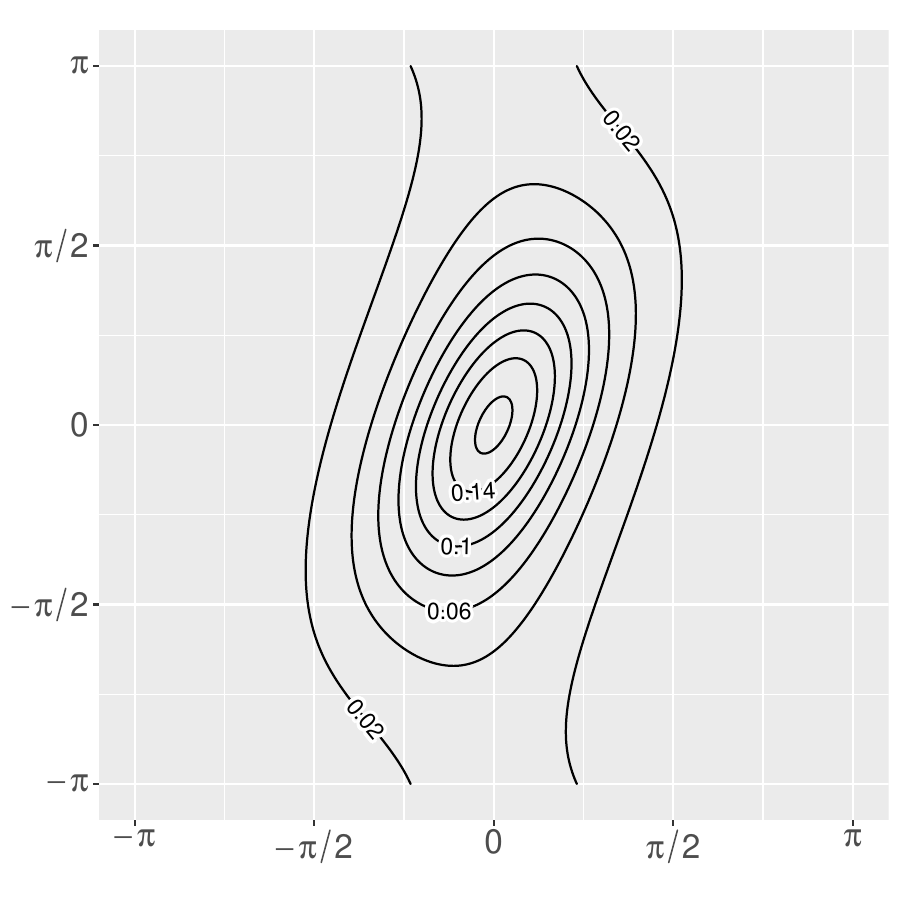}	
        \includegraphics[width=3.3cm,height=3.3cm]{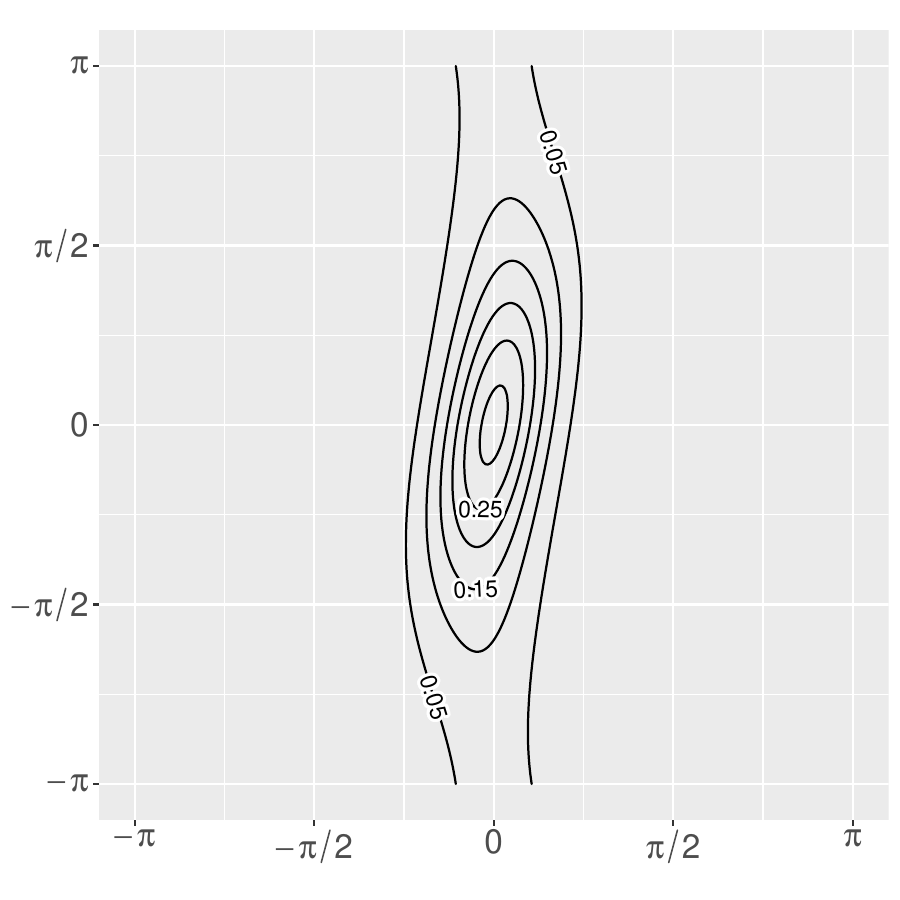}	
        \\
        
	\raisebox{1.6cm}{$\rho_{12}=0.6$} \hspace{0.12cm}
        \includegraphics[width=3.3cm,height=3.3cm]{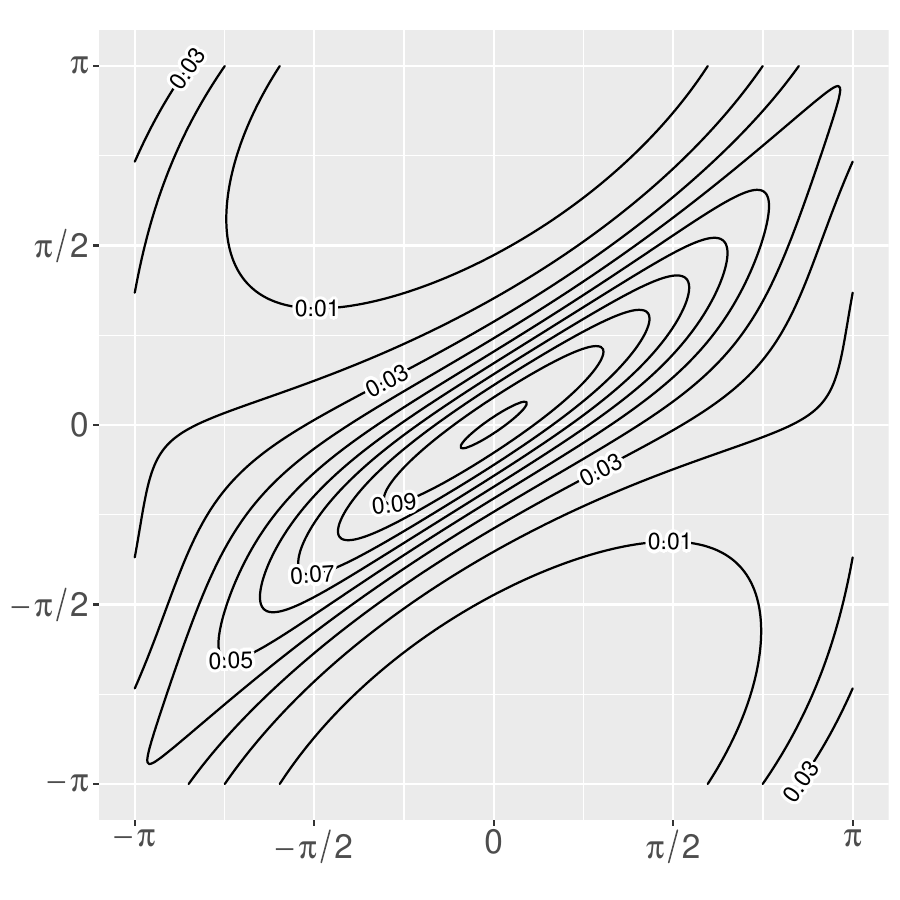}	
        \includegraphics[width=3.3cm,height=3.3cm]{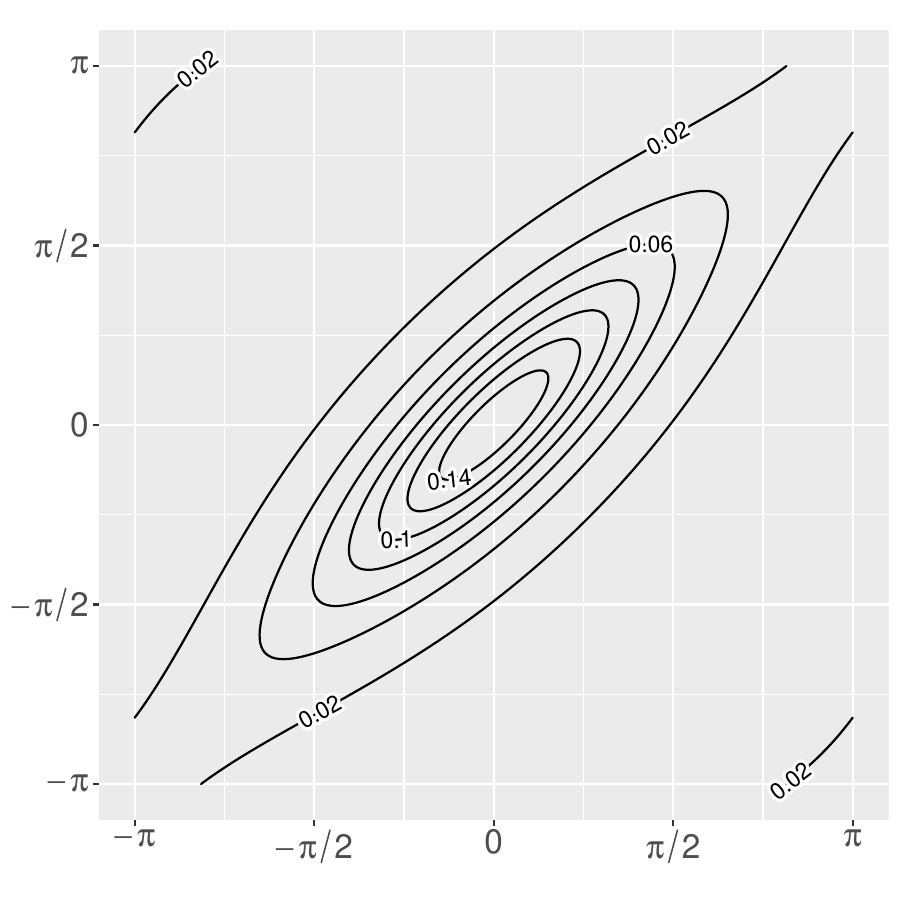}	
	\includegraphics[width=3.3cm,height=3.3cm]{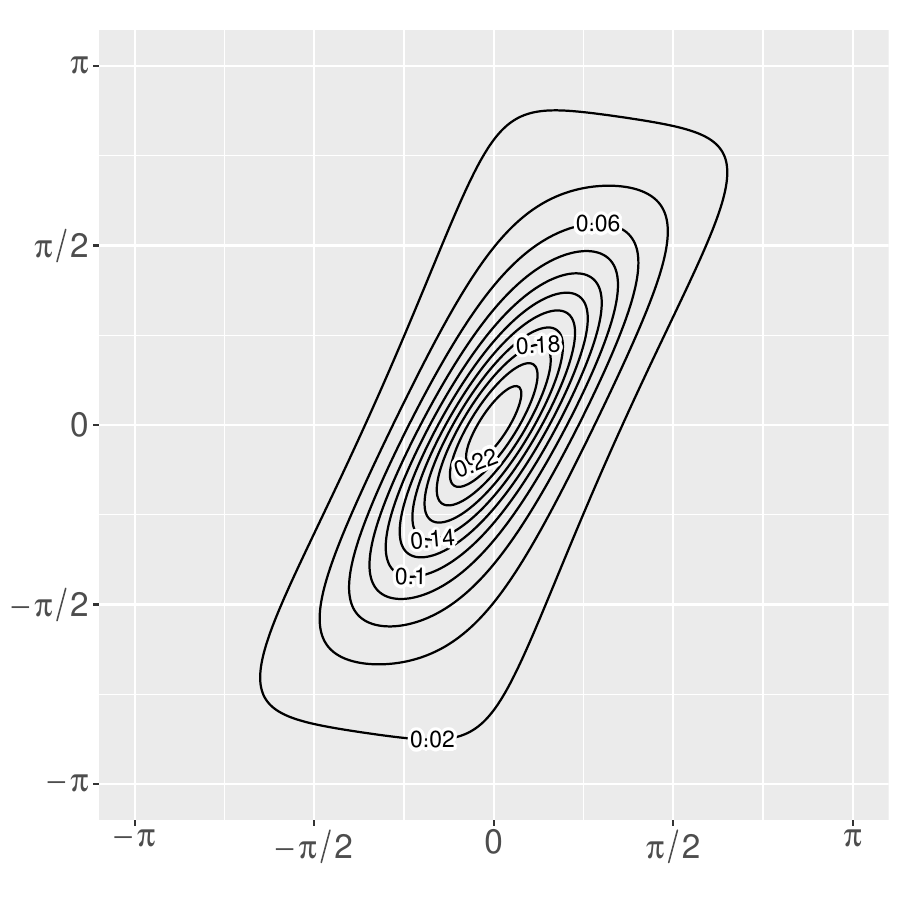}	
        \includegraphics[width=3.3cm,height=3.3cm]{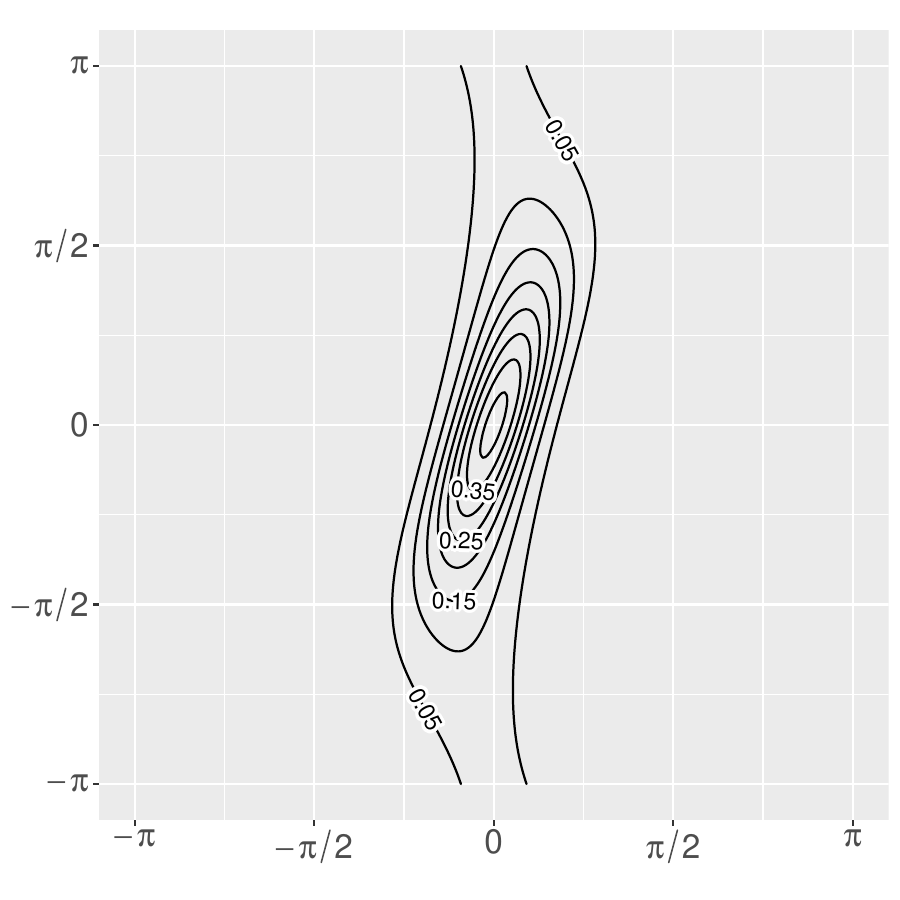}	
        \\
	\raisebox{1.6cm}{$\rho_{12}=0.9$} 
	\hspace{0.09cm}
        \includegraphics[width=3.3cm,height=3.3cm]{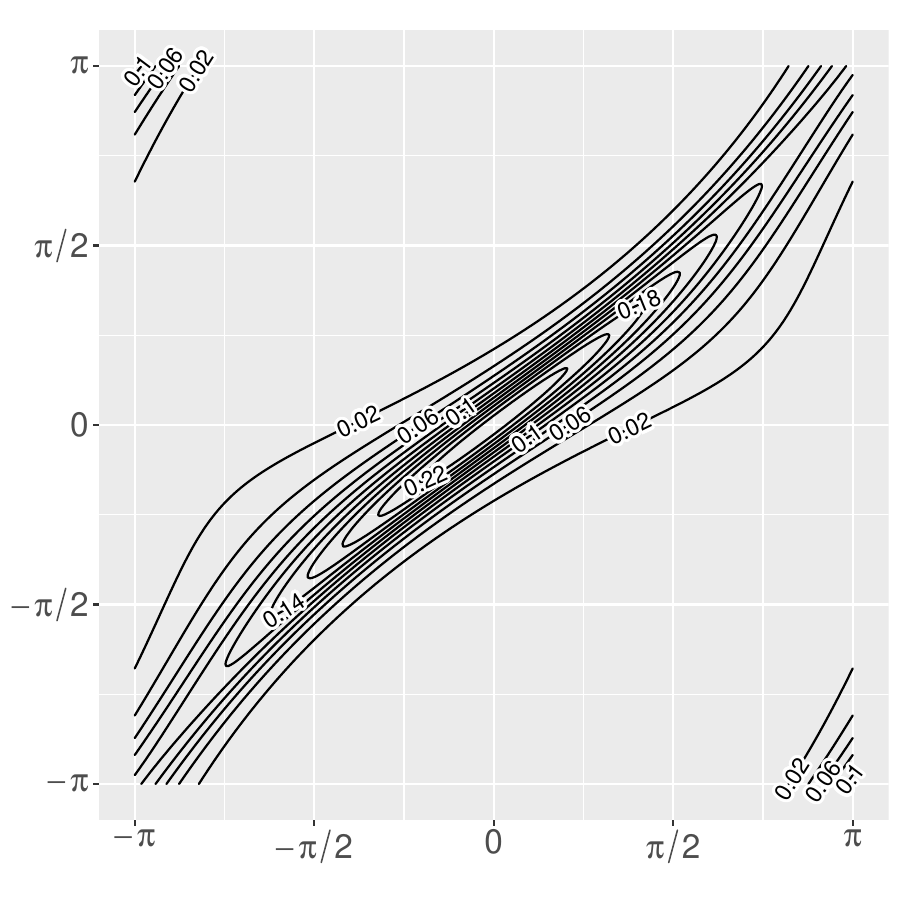}	
        \includegraphics[width=3.3cm,height=3.3cm]{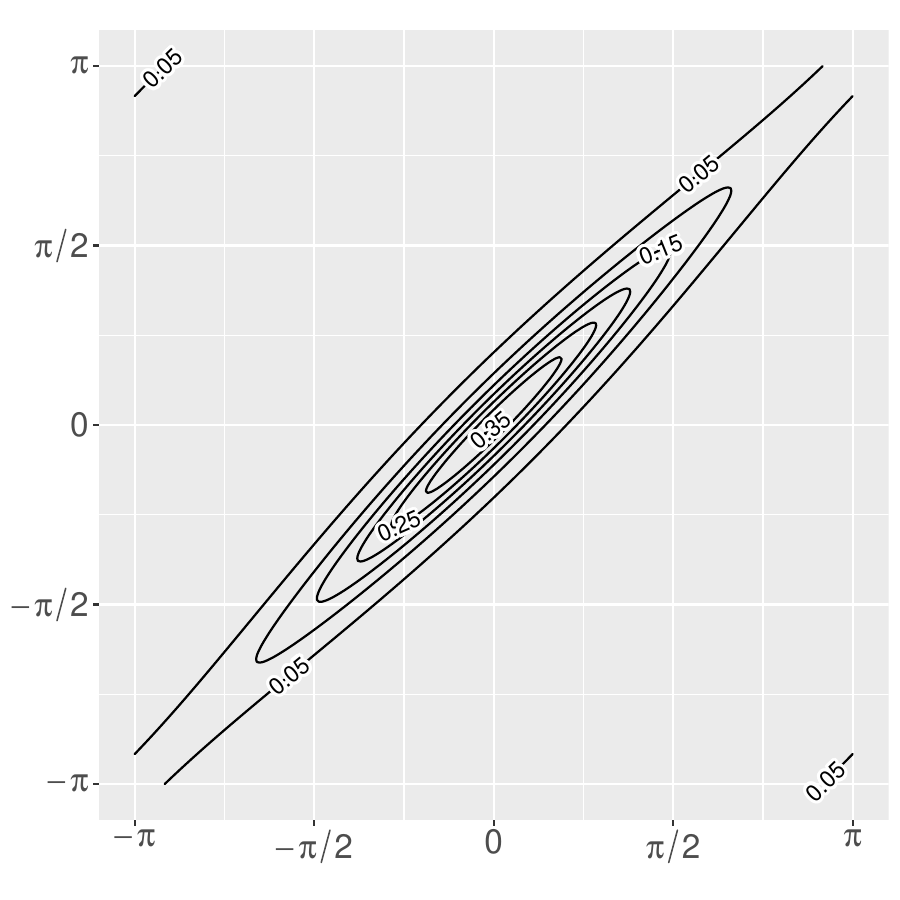}	
        \includegraphics[width=3.3cm,height=3.3cm]{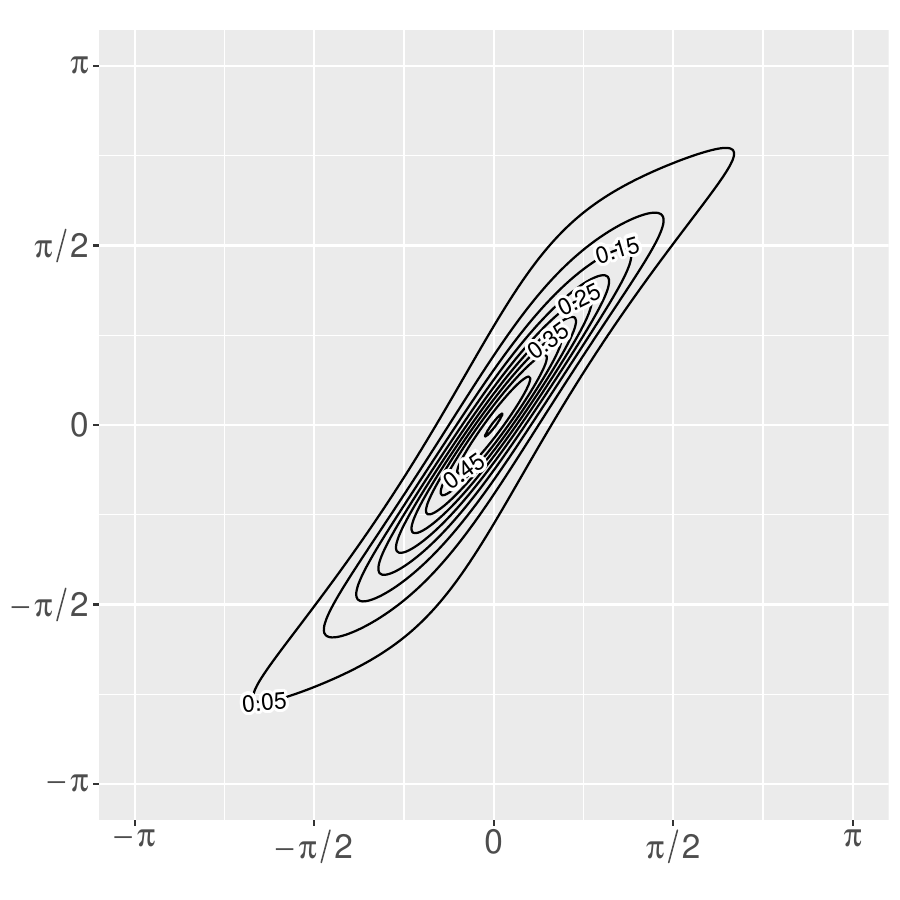}	
        \includegraphics[width=3.3cm,height=3.3cm]{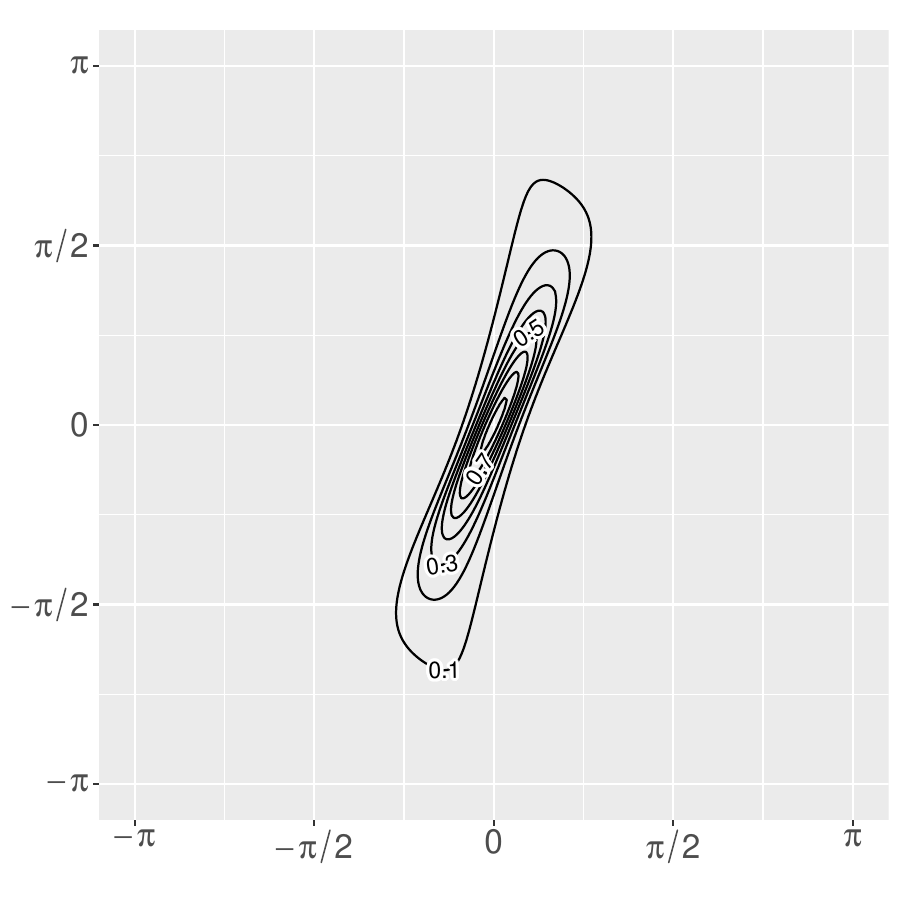}	
        \\
	\vspace{-0.1cm}
	\hspace{1.8cm}(a) \vspace{0.5cm}\\
    	
        \hspace{1.25cm} 
        \includegraphics[width=3.3cm,height=3.3cm]{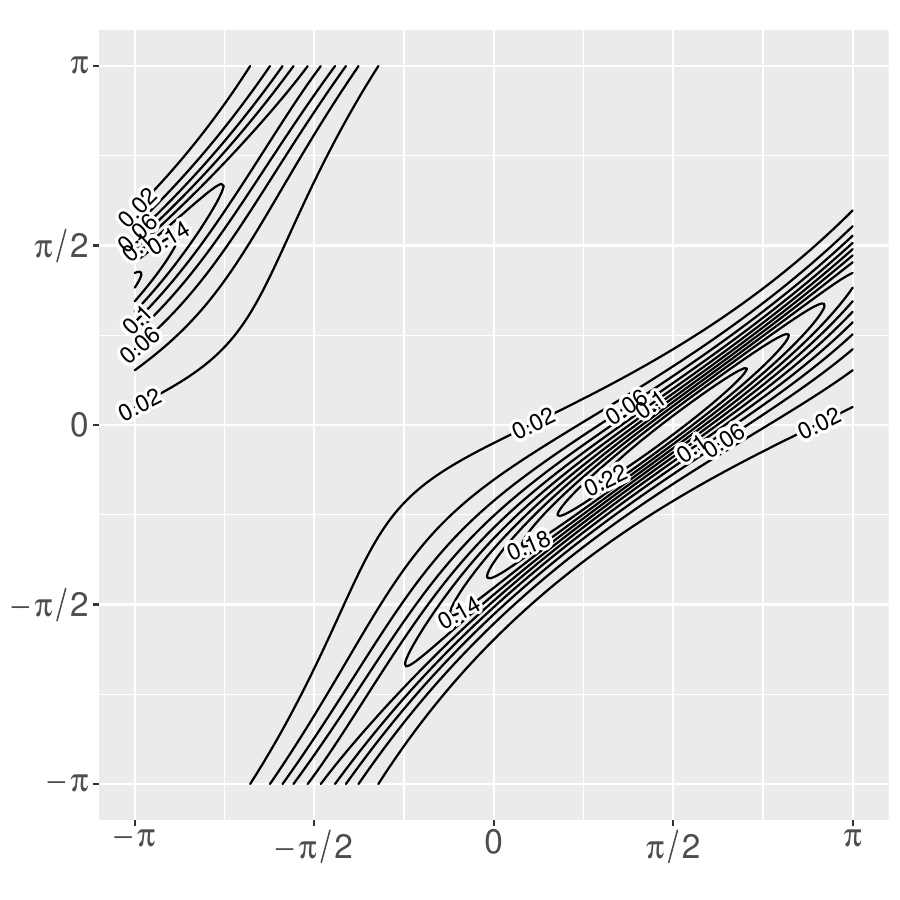}	
        \includegraphics[width=3.3cm,height=3.3cm]{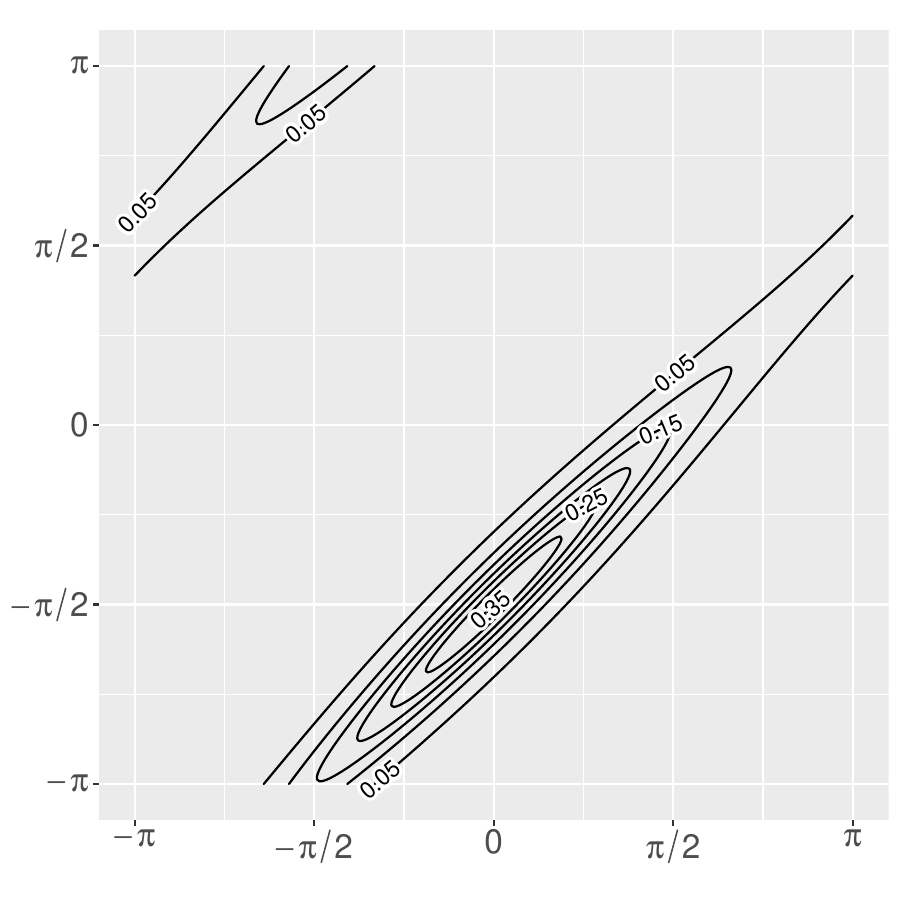}	
        \includegraphics[width=3.3cm,height=3.3cm]{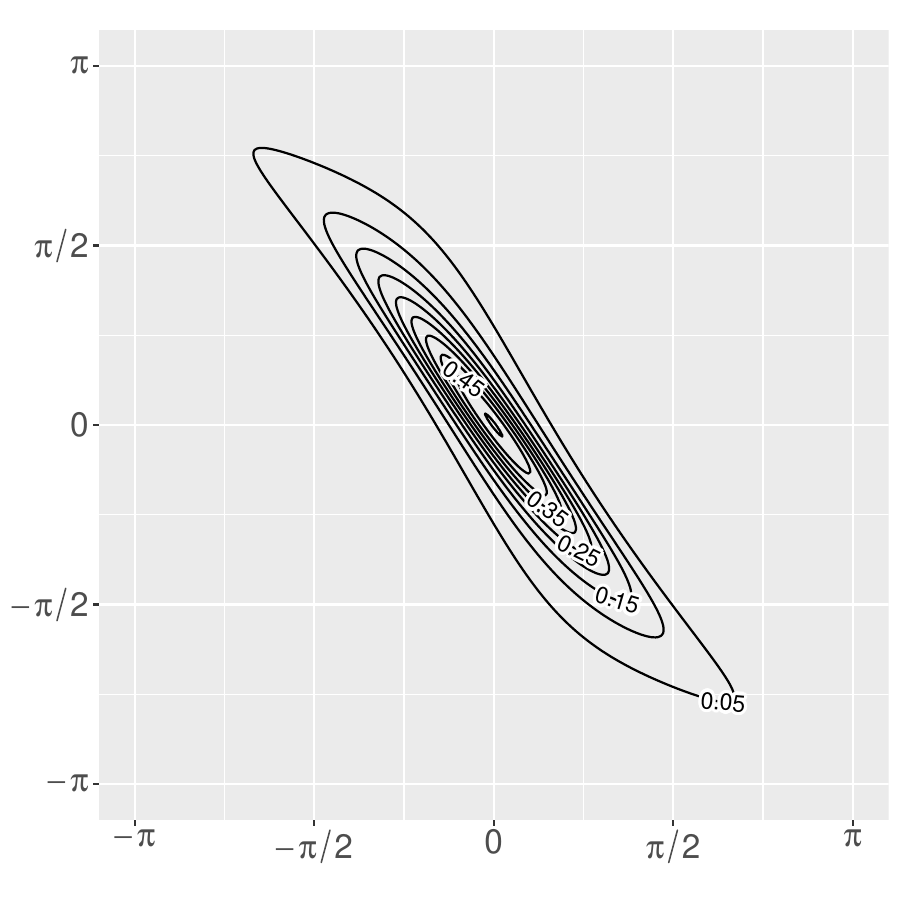}	
        \includegraphics[width=3.3cm,height=3.3cm]{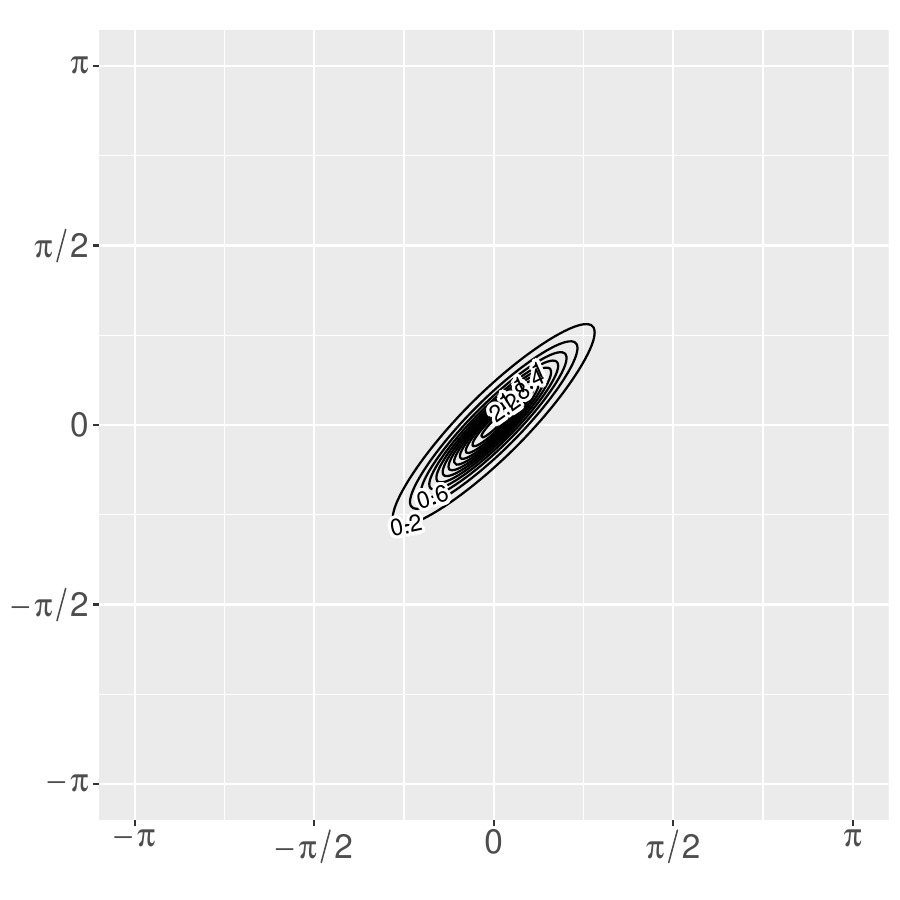}	
        \\
	\hspace{1.7cm} (i) 
        \hspace{3cm} (ii) 
        \hspace{2.8cm} (iii) 
        \hspace{2.8cm} (iv)
        \vspace{-0.2cm}\\
		\hspace{1.8cm}(b) 
		} \caption[]{Contour plots of the marginal
		density \eqref{eq:density_alternative} of $(\Theta_1,\Theta_2)$ for: (a) $\mu_1=\mu_2=0$, $\kappa_2=0.49$, with various combinations of $(\kappa_1, \rho_{12})$; and (b) the same parameters as in the upper next frame of (a), but with the following adjustments: (i) $\mu_1=0$ is replaced by $\mu_1=\pi/2$, (ii) $\mu_2=0$ is replaced by $\mu_2=-\pi/2$, (iii) $\rho_{12}=0.9$ is replaced by $\rho_{12}=-0.9$, and (iv) $\kappa_2=0.49$ is replaced by $\kappa_2=2.45$.
		\label{fig:bivariate_density} }
\end{figure}

Figure~\ref{fig:bivariate_density} displays the bivariate marginal density of $(\Theta_1,\Theta_2)$, whose functional form will be given later in \eqref{eq:density_alternative}, for selected combinations of the parameters.
In Fig.~\ref{fig:bivariate_density}(a), for fixed $\kappa_1$, it can be seen that the parameter $\rho_{12}$ controls the strength of dependence between $\Theta_1$ and $\Theta_2$.
When $\rho_{12} = 0$, $\Theta_1$ and $\Theta_2$ are independent.
As $|\rho_{12}|$ increases, the strength of dependence between $\Theta_1$ and $\Theta_2$ also increases.
The comparison between frame (b)(iii) and the frame directly above it in (a) shows that positive and negative values of $\rho_{12}$ imply positive and negative associations between the two variables, respectively.
The role of the parameter $\kappa_1$ is evident from Fig.~\ref{fig:bivariate_density}(a) for fixed $\rho_{12}$, enhancing the interpretation of $\kappa_1$ for the univariate marginal given in Fig.~\ref{fig:univariate_density}(b).
When $\kappa_1 = 0$, the TPN is highly dispersed with respect to $\theta_1$.
As $\kappa_1$ increases, the density becomes more concentrated around $\theta_1 = \mu_1$.
The parameter $\kappa_2$ plays an analogous role with respect to $\theta_2$; see frame (iv) of Fig.~\ref{fig:bivariate_density}(b).
Finally, frames (i) and (ii) of Fig.~\ref{fig:bivariate_density}(b), in comparison with the frames directly above them, suggest that the joint density achieves its maximum at $(\theta_1, \theta_2) = (\mu_1, \mu_2)$.



\subsection{Form of probability density function} \label{sec:density}

Introducing latent variables $R_1,\ldots,R_d >0$, the following relationship holds between $\bm{\Theta}$ and $(\bm{X}_c,\bm{X}_s)$ in Definition \ref{def:model}:
\begin{equation}
	\bm{X}_c =
	\begin{pmatrix}
	R_1 \cos (\Theta_1 - \mu_1 ) \\
	R_2 \cos (\Theta_2 - \mu_2 ) \\
	\vdots \\
	R_d \cos (\Theta_d - \mu_d )
	\end{pmatrix}
	\quad \mbox{and} \quad
	\bm{X}_s =
	\begin{pmatrix}
	R_1 \sin (\Theta_1 - \mu_1 ) \\
	R_2 \sin (\Theta_2 - \mu_2 ) \\
	\vdots \\
	R_d \sin (\Theta_d - \mu_d )
	\end{pmatrix}. \label{eq:xc_xs}
	\end{equation}
Let $\bm{R} = (R_1,\ldots,R_d)^\T$.
Then the joint density of $(\bm{R},\bm{\Theta})$ can be expressed as
\begin{align}
	f(\bm{r},\bm{\theta}) & = f_{\bm{X}_c} (\bm{x}_c) \, f_{\bm{X}_s} (\bm{x}_s) \prod_{j=1}^d r_j \nonumber \\
	& = \frac{1}{(2\pi)^{d} |\bm{\Sigma}_c|^{1/2} |\bm{\Sigma}|^{1/2}} \exp \left[ - \frac12 \left\{  (\bm{x}_c-\bm{\kappa})^\T \bm{\Sigma}_c^{-1} (\bm{x}_c-\bm{\kappa}) + \bm{x}_s^\T \bm{\Sigma}^{-1} \bm{x}_s \right\} \right] \prod_{j=1}^d r_j, \label{eq:pdf_gen} \\
    & \hspace{8cm} ( \bm{r} \in (0,\infty)^d , \ \bm{\theta} \in [-\pi,\pi)^d)
\end{align}
where $\bm{x}_c$ and $\bm{x}_s$ in \eqref{eq:pdf_gen} should be regarded as functions of $(\bm{r},\bm{\theta})$, as defined in \eqref{eq:xc_xs}.
It follows that the density of $\bm{\Theta}$ is 
\begin{equation}
 \frac{1}{(2\pi)^{d} |\bm{\Sigma}_c|^{1/2} |\bm{\Sigma}|^{1/2}} \int_{(\mathbb{R}^+)^d} \exp \left[ - \frac12 \left\{  (\bm{x}_c-\bm{\kappa})^\T \bm{\Sigma}_c^{-1} (\bm{x}_c-\bm{\kappa}) + \bm{x}_s^\T \bm{\Sigma}^{-1} \bm{x}_s \right\} \right] \prod_{j=1}^d r_j d \bm{r}. \label{eq:density_basic}
\end{equation}
As with existing distributions based on projections \citep{Wang2013a,Wang2014,Hernandez-Stumpfhauser2017,Mastrantonio2018}, it is also difficult to obtain a closed-form expression for \eqref{eq:pdf_gen} in the general case, apart from a special case discussed in \S \ref{sec:copula}.
However, by treating $\bm{r}$ as a latent variable, the form of our density \eqref{eq:pdf_gen} becomes useful for simplifying the MCMC algorithm; see \S \ref{sec:MCMC}.

\begin{theorem} \label{thm:density_alternative}
The density \eqref{eq:density_basic} has the alternative expression
\begin{align}
	\begin{split} \label{eq:density_alternative}
f(\bm{\theta}) = & \ \frac{1}{(2\pi)^{d} |\bm{\Sigma}_c|^{1/2} |\bm{\Sigma}|^{1/2}} \exp \left[ -\frac12 \bm{\kappa}^\T \left( \bm{I} - \bm{\Sigma}_c^{-1} \bm{T}_c \bm{V} \bm{T}_c \right) \bm{\Sigma}_c^{-1} \bm{\kappa} \right] \\
& \times \int_{(\mathbb{R}^+)^d} \exp \left[ - \frac12 (\bm{r} - \bm{\eta})^\T \bm{V}^{-1} (\bm{r}-\bm{\eta}) \right] \prod_{j=1}^d r_j \, d\bm{r},
\end{split}  
\end{align}
where 
$\bm{\eta} = \bm{V} \bm{T}_c \bm{\Sigma}_c^{-1} \bm{\kappa}$, 
$ \bm{V} = (\bm{T}_c \bm{\Sigma}_c^{-1} \bm{T}_c + \bm{T}_s \bm{\Sigma}^{-1} \bm{T}_s )^{-1} $, 
$ \bm{T}_c = \mbox{\rm diag} \{ \cos (\theta_1-\mu_1), \ldots, \cos (\theta_d-\mu_d) \}$, and
$ \bm{T}_s = \mbox{\rm diag} \{ \sin (\theta_1-\mu_1), \ldots, \sin (\theta_d-\mu_d) \}$.
\end{theorem}
Since the exponential inside the integral of \eqref{eq:density_alternative} corresponds to the kernel of a multivariate truncated normal distribution with parameters $\{\bm{\eta}, \bm{V} \}$ for the variable $\bm{r}$, the integral is equal  to 
\begin{equation} \label{eq:exv0}
E (R_1 R_2 \cdots R_d)	\int_{(\mathbb{R}^+)^d} \exp \left[ - \frac12 (\bm{r} - \bm{\eta})^\T \bm{V}^{-1} (\bm{r}-\bm{\eta}) \right]   d\bm{r},
\end{equation}
where $E(R_1 R_2 \cdots R_d)$ is the expectation with respect to the multivariate truncated normal distribution {on the positive orthant,} and the integral in \eqref{eq:exv0} is the normalizing constant of the truncated normal density.
Although $E(R_1 R_2 \cdots R_d)$  cannot be computed in closed form, numerical methods exist that allow to obtain its value; see \cite{Tallis}, \cite{LEE1983245} and \cite{Leppard}.
The integral part is proportional to the value of the cumulative distribution function of the multivariate normal distribution $ {\cal N}_d(-\bm{\eta}, \bm{V}) $ at the origin, and can be computed straightforwardly using numerical methods.

\begin{corollary} \label{cor:V}
Suppose that there exists $(\delta_1,\ldots,\delta_d)  \in \{-1,1\}^d$ such that $\bm{\Sigma}_c = \bm{D} \bm{\Sigma} \bm{D} $, where $\bm{D}=\mbox{\rm diag}(\delta_1,\ldots,\delta_d)$.
Then $\bm{V}$ is of the form
$$ 
\bm{V}= \begin{pmatrix}
s_{11} & s_{12} \cos (\delta_1 \bar{\theta}_1 - \delta_2 \bar{\theta}_2) & \cdots & s_{1d} \cos (\delta_1 \bar{\theta}_1 - \delta_d \bar{\theta}_d) \\
s_{12} \cos (\delta_1 \bar{\theta}_1 - \delta_2 \bar{\theta}_2) & s_{22} & \cdots & s_{2d} \cos (\delta_2 \bar{\theta}_2 - \delta_d \bar{\theta}_d) \\
\vdots & \vdots & \ddots & \vdots \\
s_{1d} \cos (\delta_1 \bar{\theta}_1 - \delta_d \bar{\theta}_d) & s_{2d} \cos (\delta_2 \bar{\theta}_2 - \delta_d \bar{\theta}_d) & \cdots & s_{dd}
\end{pmatrix}^{-1},
$$
where $\bar{\theta}_j=\theta_j-\mu_j$ and $s_{jk}$ is the $(j,k)$ element of $\bm{\Sigma}_c^{-1}$.
\end{corollary}

Consider the bivariate marginal distribution of $(\Theta_j,\Theta_k)$ $(j \neq k)$.
Denote the submatrix of $\bm{\Sigma}$ and $\bm{\Sigma}_c$ by
$$
\bm{\Sigma}^{(j,k)} =
\begin{pmatrix}
1 & \rho_{jk} \\
\rho_{jk} & 1
\end{pmatrix},
\quad
\bm{\Sigma}_c^{(j,k)} =
\begin{pmatrix}
1 & |\rho_{jk}| \\
|\rho_{jk}| & 1
\end{pmatrix}.
$$
Then there always exists $(\delta_j, \delta_k)$ such that $\bm{\Sigma}^{(j,k)}_c = \bm{D}\bm{\Sigma}^{(j,k)} \bm{D}$.
In fact, by assuming $\delta_j = 1$ and $\delta_k = \text{sign}(\rho_{jk})$, it is easy to see that $\bm{\Sigma}^{(j,k)}_c = \bm{D} \bm{\Sigma}^{(j,k)} \bm{D}$ holds for any $\rho_{jk}$, where $\mbox{sign}(\rho)=1$ if $\rho \geq 0$ and $\mbox{sign}(\rho)=-1$ if $\rho <0$.
Thus Theorems \ref{thm:marginals} and \ref{thm:density_alternative} imply that the bivariate marginal density of $(\Theta_j,\Theta_k)$ can be expressed in the same form as \eqref{eq:density_alternative} with $d=2$, and Corollary \ref{cor:V} suggests that the matrix $\bm{V}$ which corresponds to the bivariate case is
\begin{align}
	\bm{V} &  = \left\{ \frac{1}{1-\rho_{jk}^2} \begin{pmatrix}
	1 & \tau (\theta_j,\theta_k) \\
	\tau (\theta_j,\theta_k) & 1 
\end{pmatrix} \right\}^{-1} 
 = \frac{1-\rho_{jk}^2}{ 1 - \tau (\theta_j,\theta_k)^2 }
\begin{pmatrix}
	1 & - \tau (\theta_j,\theta_k) \\
	- \tau (\theta_j,\theta_k) & 1 
\end{pmatrix}, \label{eq:V}
\end{align}
where 
\begin{equation}
\tau(\theta_j,\theta_k) = - |\rho_{jk}| \cos (\bar{\theta}_j - \mbox{sign}(\rho_{jk}) \bar{\theta}_k) . \label{eq:tau}
\end{equation}

\section{An Extension of the Copula Case of the TPN Distribution} \label{sec:copula}


\subsection{The copula case of the TPN distribution}

Let a random vector $\bm{\Theta} =(\Theta_1,\ldots,\Theta_d)^\T$ follow $\mathrm{TPN}_d (\bm{\mu},\bm{0},\bm{\Sigma})$.
Then it follows from \eqref{eq:density_basic} and \eqref{eq:density_alternative} that the density of $\bm{\Theta}$ reduces to
\begin{align}
\begin{split} \label{eq:density_copula} 
		c(\bm{\theta}) &  = \frac{1}{(2\pi)^{d} |\bm{\Sigma}_c|^{1/2} |\bm{\Sigma}|^{1/2}} \int_{(\mathbb{R}^+)^d} \exp \left[ - \frac12 \left(  \bm{x}_c^\T \bm{\Sigma}_c^{-1} \bm{x}_c + \bm{x}_s^\T \bm{\Sigma}^{-1} \bm{x}_s \right) \right] \prod_{j=1}^d r_j d \bm{r} \\
        & = \frac{1}{(2\pi)^{d} |\bm{\Sigma}_c|^{1/2} |\bm{\Sigma}|^{1/2}} \int_{(\mathbb{R}^+)^d} \exp \left( - \frac{ \bm{r}^\T \bm{V}^{-1} \bm{r}}{2} \right) \prod_{j=1}^d r_j \, d\bm{r},
        \end{split}
\end{align}
where $\bm{V}$ is as in Theorem \ref{thm:density_alternative}.
In the second expression of the density \eqref{eq:density_copula}, the variable $\bm{\theta}$ appears only inside the matrix $\bm{V}$.
Then, under the assumption in Corollary \ref{cor:V}, the density \eqref{eq:density_copula} is a function of $(\delta_i \bar{\theta}_i - \delta_j \bar{\theta}_j )_{1 \leq i,j \leq d}$.

Theorem \ref{thm:univariate_marginal} implies that each univariate marginal of this submodel is the uniform distribution on the circle with density
\begin{equation}
c_1(\theta_k) = \frac{1}{2\pi}. \label{eq:density_copula_d1}
\end{equation}
This means that \eqref{eq:density_copula} is the density of a hypertoroidal copula model. 
We denote by $c$ the density in the copula case of our family, and by $c_j$ its $j$-variate marginal density.

{
The modes of the copula case in \eqref{eq:density_copula} become clear under certain conditions.
\begin{theorem} \label{thm:modes}
Suppose that there exists $(\delta_1,\ldots,\delta_d)  \in \{-1,1\}^d$ such that $\bm{\Sigma}_c = \bm{D} \bm{\Sigma} \bm{D} $, where $\bm{D}=\mbox{\rm diag}(\delta_1,\ldots,\delta_d)$.
Let $s_{jk} < 0$ for $1 \leq j,k \leq d, \ j \neq k$, where $s_{jk}$ is as in Corollary \ref{cor:V}.
Then the copula density \eqref{eq:density_copula} takes its maximum values if and only if $\delta_1 (\theta_1-\mu_1) =  \delta_2 (\theta_2 - \mu_2) = \cdots = \delta_d (\theta_d-\mu_k)$.
\end{theorem}
The assumption of this theorem always holds when $ d = 2 $.
Although this is not the case for $d \geq 3$ in general, our simulation results suggest that the condition of the theorem often holds for $ d = 3 $ for randomly generated $ \bm{\Sigma} $ matrices with positive off-diagonal elements.}

Interestingly, the bivariate marginal densities of the copula family have simple functional form.
\begin{theorem} \label{thm:density_copula_d2}
The bivariate marginal density of $(\Theta_j,\Theta_k)$ has the closed-form expression:
	\begin{align}
    \begin{split} \label{eq:density_copula_d2}
	c_2 (\theta_j,\theta_k) = & \ \frac{1}{(2\pi)^2} \frac{1-\rho^2}{ \{ 1-\tau(\theta_j,\theta_k)^2 \}^{3/2}} \\
    & \times \left( \left\{ 1-\tau (\theta_j,\theta_k)^2 \right\}^{1/2} - \tau (\theta_j,\theta_k) \, \text{atan}^* \left[ \frac{\left\{ 1-\tau (\theta_j,\theta_k)^2 \right\}^{1/2}}{ \tau (\theta_j,\theta_k)} \right] \right) , 
    \end{split}
	\end{align}
	where $\tau(\theta_i,\theta_j)$ and $\text{atan}^*$ are as in \eqref{eq:tau} and \S \ref{sec:derivation}, respectively.
    
\end{theorem}

This theorem implies that the distribution \eqref{eq:density_copula} can be considered a multivariate extension of the bivariate copula model of \cite{Wehrly1980} whose density  depends only on $\theta_1-\theta_2$ or $\theta_1+\theta_2$.
The model of \cite{Wehrly1980} has played an important role in directional statistics as a general model for bivariate circular data; see \cite{Jones2015} for its detailed properties.
The closed-form expression \eqref{eq:density_copula_d2} of the bivariate marginal comes from the fact that the conditions of Corollary \ref{cor:V} are always met in this case.
It is also possible to avoid the use of $\text{atan}^*$ and adopt the ordinary arctangent function which takes values on $[-\pi/2, \pi/2]$.
This can be done by using the relationship $\text{atan}^* [ \{1-\tau (\theta_j,\theta_k)^2 \}^{1/2}/ \tau (\theta_j,\theta_k) ] = \pi/2 - \arctan [ \tau (\theta_j,\theta_k) / \{ 1-\tau (\theta_j,\theta_k)^2 \}^{1/2} ] $, which gives us an alternative expression for the density \eqref{eq:density_copula_d2}:
\begin{align*}
	c_2(\theta_j,\theta_k) = & \ \frac{1}{(2\pi)^2} \frac{1-\rho^2}{ \{ 1-\tau(\theta_j,\theta_k)^2 \}^{3/2}} \\
    & \times \left( \left\{ 1-\tau (\theta_j,\theta_k)^2 \right\}^{1/2} + \tau (\theta_j,\theta_k) \left[ \arctan \left( \frac{ \tau (\theta_j,\theta_k)}{\left\{ 1-\tau (\theta_j,\theta_k)^2 \right\}^{1/2}}  \right) - \frac{\pi}{2} \right] \right). 
\end{align*}
Since the density \eqref{eq:density_copula_d2} is a function of $\theta_1 - \operatorname{sign}(\rho_{12}) \theta_2$,
the density \eqref{eq:density_copula_d2} is proportional to a univariate density on the circle, obtained by replacing $\tau(\theta_j,\theta_k)$ with $-|\rho_{jk}| \cos \theta$, where $\theta \in [-\pi,\pi)$.  


\subsection{An extension of the copula case}

The copula model \eqref{eq:density_copula} can be extended to have any univariate marginal distributions on the circle $[-\pi,\pi)$.

\begin{definition} \label{def:copula}
	
Let a random vector $(U_1,\ldots,U_d)$ follow the proposed model $\mathrm{TPN}_d(\bm{0},\bm{0},\bm{\Sigma})$.
Let $f_{j}(\theta) \ (j=1,\ldots,d)$ be densities on the circle with the mode at $\theta=0$ and $F_{j}(\theta) = \int_{0}^{\theta} f_{j}(t) dt $ be their cumulative distribution functions.
Define the circular random variables by
\begin{equation} \label{eq:coptrasform}
\Theta_1 = \mu_1 + F_{1}^{-1} \left( \frac{U_1}{2\pi} \right) , \quad  \ldots , \quad \Theta_d = \mu_d + F_{d}^{-1}\left( \frac{U_d}{2\pi} \right).	
\end{equation}
Then a new family of distributions on the hypertorus $[-\pi,\pi)^d$ is defined by the distribution of $(\Theta_1,\ldots,\Theta_d)$, which we call a copula-based extension of the (hyper-)toroidal projected normal (CTPN) distributions.
\end{definition}


It immediately follows from Sklar's theorem {\citep[e.g.,][Theorem 2.10.9]{Nelsen2006}}   that the random vector $\bm{\Theta} =(\Theta_1,\ldots,\Theta_d)^\T$ defined in Definition \ref{def:copula} has the probability density function
\begin{align}
    \lefteqn{ f(\bm{\theta}) = (2\pi)^d \overline{c} \{ 2\pi F_{1}(\theta_1 - \mu_1) , \ldots , 2 \pi F_{d}(\theta_d - \mu_d) \} \prod_{j=1}^d f_{j} (\theta_j - \mu_j) } \hspace{8cm} \label{eq:copula_extension}  \\
    & (\bm{\theta} = (\theta_1 , \ldots, \theta_d)^\T \in [-\pi,\pi)^d ), \nonumber
\end{align}
where $\overline{c}$ denotes the density \eqref{eq:density_copula} with $\mu_1= \cdots =\mu_d=0$.

The equation \eqref{eq:density_copula_d1} implies that the marginal density of $\Theta_j$ is
$$
f(\theta_j) = f_{j}(\theta_j-\mu_j).
$$
{
The assumption that the mode of $ f_{j}(\theta) $ is at $ \theta = 0 $ ensures that the mode of each marginal of the CTPN distribution is given by $ \theta_j = \mu_j $.
The lower limit of $F_{j}$ is assumed to be zero because the mode of the joint density $f(\bm{\theta})$ of the CTPN distribution is located at $\bm{\theta} = (\mu_1,\ldots,\mu_d)$ if the copula density $\overline{c}(\bm{u})$ has the mode at $\bm{u}=\bm{0}$; see \citet[\S 3.1]{Jones2015}.}
Flexible modeling is made possible through this extended model~\eqref{eq:copula_extension}. 

In addition it follows from Theorem \ref{thm:density_copula_d2} that the bivariate marginal distribution of $(\Theta_j,\Theta_k)$ has the density
\begin{equation}
f(\theta_j,\theta_k) = (2\pi)^2 \overline{c}_2 \{ 2\pi F_{j}(\theta_j) , 2\pi F_{k}(\theta_k) \} f_{j} (\theta_j) f_{k}(\theta_k),
\end{equation}
where $\overline{c}_2$ denotes the density \eqref{eq:density_copula_d2} with $\mu_j=\mu_k=0$.
This bivariate case belongs to the general family of \cite{Wehrly1980} with the circular density binding the two variables being $2\pi \overline{c}_2 (\theta_j, \theta_k )$.



\section{Coefficient of Correlation} 

As seen in Fig.~\ref{fig:bivariate_density}, the parameter $\rho_{jk}$ governs the dependence between $\Theta_j$ and $\Theta_k$.
To quantify how the value of $\rho_{jk}$ influences the dependence, we consider the circular correlation coefficient proposed by \cite{Rivest1982}.
This measure is considered here because it is a signed coefficient capable of distinguishing between positive and negative dependence.
Let 
$$
\xi_{jk}=  \frac{ \left| E \left\{ e^{i (\Theta_j - \Theta_k)} \right\} \right| - \left| E \left\{ e^{i (\Theta_j + \Theta_k)} \right\} \right| }{2},
$$
then the circular correlation coefficient between $\Theta_j$ and $\Theta_k$ proposed by \cite{Rivest1982} is defined as
$$
\rho_{jk,R} = 
\left\{ \begin{array}{ll}
\frac{\xi_{jk}}{\max \left[ E \left\{ \sin^2 (\Theta_j - \mu_j ) \right\},\ E \left\{ \sin^2 (\Theta_k - \mu_k) \right\} \right]}, & 
E \left\{ e^{i (\Theta_j - \Theta_k)} \right\} E \left\{ e^{i (\Theta_j + \Theta_k)} \right\} \neq 0, \\
2 \xi_{jk}, & E \left\{ e^{i (\Theta_j - \Theta_k)} \right\} E \left\{ e^{i (\Theta_j + \Theta_k)} \right\} = 0.
\end{array} \right. 
$$

\label{sec:corr}
\begin{figure}[t] 
	\begin{center}
		\includegraphics[width=0.40\textwidth]{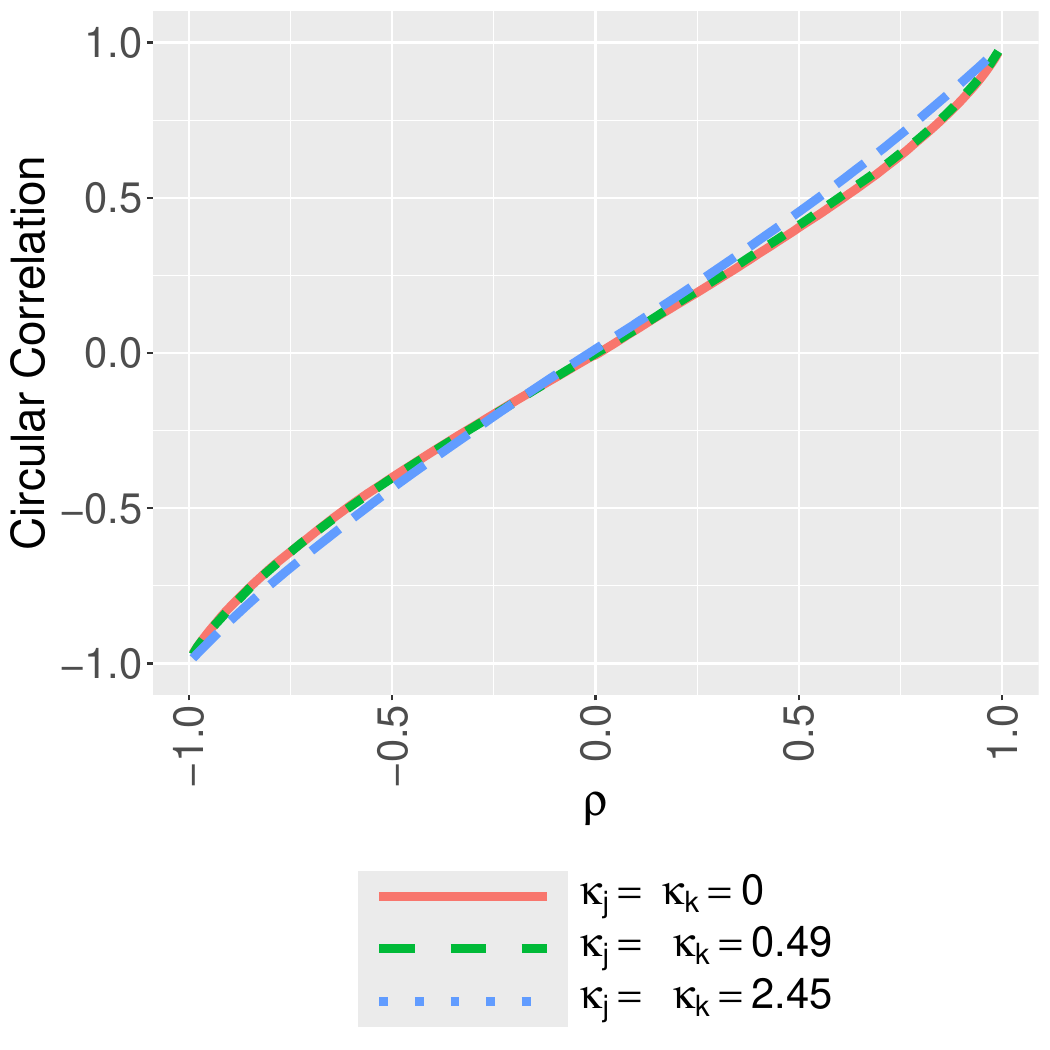} \hspace{0.5cm}
		\includegraphics[width=0.40\textwidth]{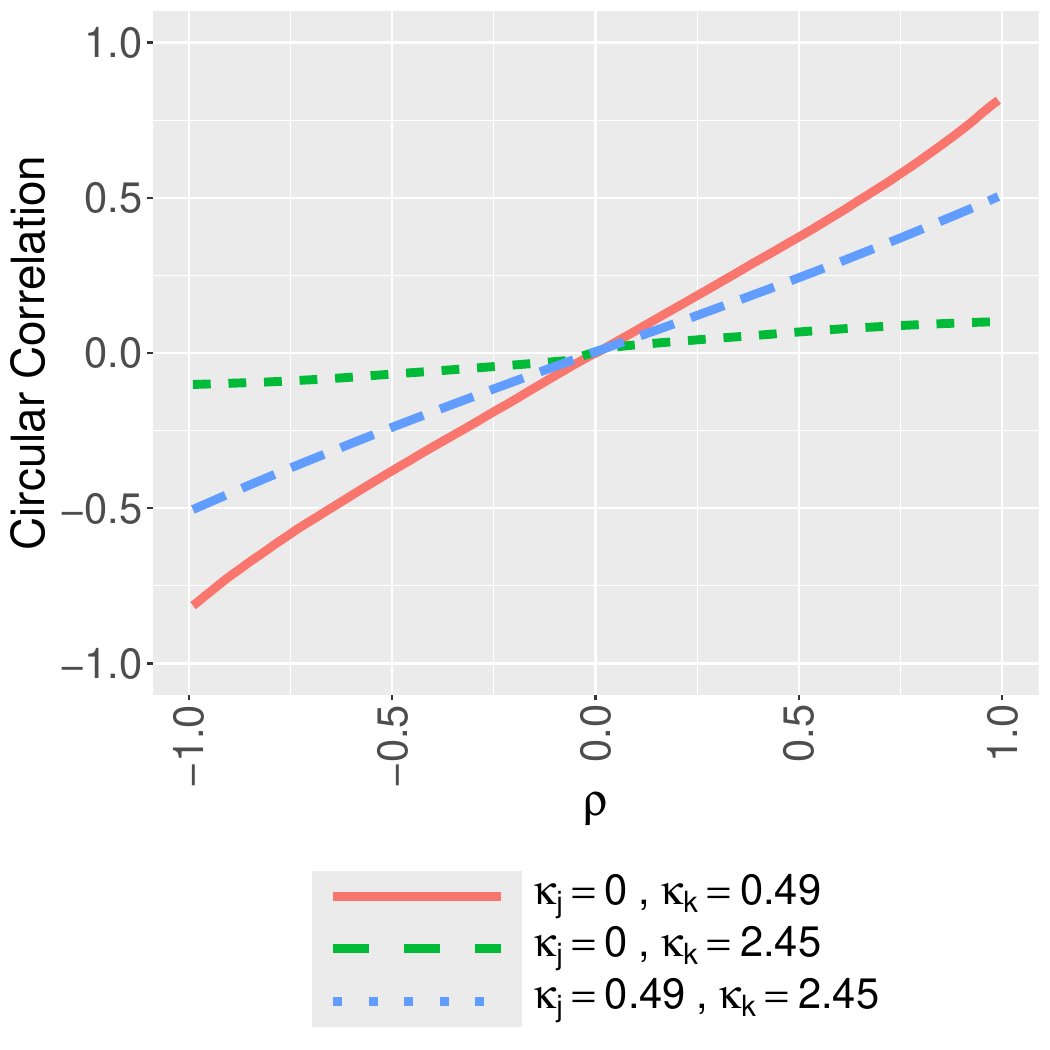}  \\ 
		{\footnotesize \hspace{0.5cm} (a) \hspace{7.1cm} (b) }
		\caption{Plot of the circular correlation coefficients of the TPN distributions, including its copula case, for different values of  $\rho_{jk}$, and values of $\kappa_j, \kappa_k \in \{0, 0.49, 2.45\}$: (a) $\kappa_j = \kappa_k$; and (b) $\kappa_j \neq \kappa_k$.}
	\label{fig:correlation}
	\end{center}
	\end{figure}
Figure \ref{fig:correlation} shows the circular correlation coefficients $ \rho_{jk,R} $ of the TPN distributions for two cases: $ \kappa_j = \kappa_k $ and $ \kappa_j \neq \kappa_k $, which include the copula case $\kappa_j=\kappa_k=0$.  
The figure suggests that the circular correlation coefficient is a strictly increasing function of $ \rho_{jk} $.  
The coefficient spans the entire range $ (-1, 1) $ when $ \kappa_j = \kappa_k $, whereas its range is limited when $ \kappa_j \neq \kappa_k $.
This can be inferred from the fact that if the range of $ \rho_{jk} $ is extended and $ \rho_{jk} = -1 $ or $ 1 $ in Definition \ref{def:model}, we have
$$
X_{cj} - \kappa_j = X_{ck} - \kappa_k, \quad X_{sj} = \operatorname{sign}(\rho_{jk}) X_{sk},
$$
and thus
$$
\Theta_k = \mu_k + \operatorname{sign}(\rho_{jk}) \, \text{atan}^* \left\{ \frac{\sin (\Theta_j - \mu_j) }{\cos (\Theta_j - \mu_j) + (\kappa_k - \kappa_j)/ R_j} \right\}.
$$
This expression implies that $ \Theta_j $ and $ \Theta_k $ have perfect linear relationship when $ \kappa_j = \kappa_k $.  
When $ \kappa_j \neq \kappa_k $, the relationship between $ \Theta_j $ and $ \Theta_k $ depends on $ R_j $, and the influence of $ R_j $ increases as $ |\kappa_k - \kappa_j| $ increases.
This suggests the advantage of the CTPN distribution, which arises from the copula case $\kappa_j=\kappa_k=0$, over the TPN distribution in terms of the attainable range of the correlation coefficient.
Nonetheless, attention must be paid to the fact that the circular correlation coefficient $ \rho_{jk,R} $ reaches $-1$ or $1$ when $ \Theta_j $ and $ \Theta_k $ have a linear relationship with probability one; see \citet[Proposition 2.3]{Rivest1982}.
Therefore, this coefficient is not designed to measure the strength of dependence between the two variables when they exhibit a nonlinear dependence, which occurs when $\kappa_j \neq \kappa_k$.

In addition, Fig.~\ref{fig:correlation} shows that the positive and negative signs of $ \rho_{jk} $ correspond to positive and negative values of the circular correlation coefficient, respectively.  
The circular correlation coefficient is zero if and only if $ \rho_{jk} = 0 $.


\section{MCMC Algorithm} \label{sec:MCMC}
\subsection{MCMC steps for the TPN}

In this section, we explain the algorithm that can be used to obtain posterior samples from the TPN distribution. 
To deal with the identification problem that the TPN inherited form the PN, we use the approach of \cite{Mastrantonio2018}, which defines a model for the unidentifiable parameters and a posterior remap is used to obtain the identifiable version of the parameters. 

While standard priors can be easily assigned to the parameters $\bm{\mu}$ and $\bm{\kappa}$, care must be taken when specifying a prior for $\bm{\Sigma}$ due to its relationship with $\bm{\Sigma}_c$. The Inverse Wishart (IW) distribution is a classical choice for modeling a covariance matrix, but not every valid covariance matrix $\bm{\Sigma}$ yields a valid $\bm{\Sigma}_c$. This makes its use problematic unless appropriate adjustments are made. The matrix $\bm{\Sigma}$ is defined over the domain $\mathcal{D}$ of all positive definite matrices, and a subset  $\mathcal{D}^*$  of this domain can be defined as the set of all positive matrices that, when taking the absolute value of their elements, remain positive definite.
We define  a new distribution, which we call Truncated IW (TIW), 
as a distribution supported on $\mathcal{D}^*$ whose density is proportional to the IW. The normalization constant of this distribution is not known, but, as we will show, it is not needed in the MCMC algorithm.
Write $\bm{\Sigma} \sim TIW(\nu,\bm{\Psi})$ if a random matrix $\bm{\Sigma}$ follows a TIW distribution with degrees of freedom $\nu$ and scale matrix $\bm{\Psi}$.

The Bayesian model is as follows:
\begin{align}
	\bm{\Theta}_i& = (\Theta_{i,1}, \ldots, \Theta_{i,d})^\T \stackrel{iid}{\sim} \mathrm{TPN}_d(\bm{\mu}, \bm{\kappa}, \bm{\Sigma}), \quad
	 \bm{\Sigma} \sim TIW(\nu, \bm{\Psi}) \quad (i= 1,\dots ,n),\\
		\mu_j &\stackrel{iid}{\sim}  CU(0, 2\pi), \quad
	\kappa_j  \stackrel{iid}{\sim} {\cal N}_1 ({m}, {v})I_{>0} \quad (j = 1 , \dots , d),
\end{align}
where $CU(0, 2\pi)$ is the uniform distribution on the circle $(0,2\pi)$ and $ {\cal N}_1({m}, {v}) I_{>0} $ denotes a normal distribution with mean $ {m} $ and variance $ {v} $ truncated to the positive real line.
As we will show below, the truncated normal ensures that the full conditional of $\kappa_j $ is available in closed form, while for $\mu_j $, regardless of the distribution, no closed-form full conditional is available.

To make the inference easier, as is often done when working with the PN, the vectors of latent variables are introduced in the model;
define $\bm{R}_i = (R_{i,1}, \dots , R_{i,d})^\T$, $\bm{X}_{c,i}=(X_{c,i,1},\ldots, X_{c,i,d})^\T$, and $\bm{X}_{s,i}=(X_{s,i,1},\ldots,X_{s,i,d})^\T$ for the $i$-th sample of $\bm{\Theta}_i$ as in \eqref{eq:xc_xs}, namely,
\begin{align}
	\bm{X}_{c,i,j} &= R_{i,j}\cos(\Theta_{i,j} - \mu_j), \quad
	\bm{X}_{s,i,j} = R_{i,j}\sin(\Theta_{i,j} - \mu_j).
\end{align}
We have that the joint density of $(\bm{R}_i, \bm{\Theta}_i)$ can be written as in \eqref{eq:pdf_gen}, as
\begin{equation} \label{eq:dens_tot}
	f(\bm{r}_i,\bm{\theta}_i) = \frac{1}{(2\pi)^{d} |\bm{\Sigma}_c|^{1/2} |\bm{\Sigma}|^{1/2}} \exp \left[ - \frac12 \left\{  (\bm{x}_{c,i}-\bm{\kappa})^\T \bm{\Sigma}_c^{-1} (\bm{x}_{c,i}-\bm{\kappa}) + \bm{x}_{s,i}^\T \bm{\Sigma}^{-1} \bm{x}_{s,i} \right\} \right] \prod_{j=1}^d r_{i,j}.
\end{equation}

For two of the steps of the MCMC algorithm, we need to use the conditional distribution of $x_{c,i, j}$ given all the other variables in the vector $\bm{x}_{c,i}$, which we indicate as $\bm{x}_{c,i, -j}$, and the conditional distribution of $x_{s,i, j}$ given $\bm{X}_{s,i, -j}=\bm{x}_{s,i, -j}$. 
Define $\bm{\Lambda}_{c} = \bm{\Sigma}_c^{-1}$, $\bm{\Lambda}_{s} = \bm{\Sigma}^{-1}$ and $\bm{\kappa}_{-j}$ as the vector $\bm{\kappa}$ without the $j$-the element.
Using standard results of a block matrix inversion and its connection with the conditional mean and variance of a multivariate normal, we know that 
\begin{align}\label{eq:cond_par}
	x_{c,i, j}|\bm{x}_{c,i, -j} &\sim {\cal N}_1(\eta_{c,i,j}, \lambda_{c, j}^{-1}), \quad
	x_{s,i, j}|\bm{x}_{s,i, -j} \sim {\cal N}_1 (\eta_{s,i,j}, \lambda_{s, j}^{-1}),   \\
	\eta_{c,i,j}& =  \kappa_j -  \lambda_{c, j}^{-1} \bm{\Lambda}_{c,j,-j}\left(\bm{x}_{c,i, -j} - \bm{\kappa}_{-j}\right), \quad 
	\eta_{s,i,j} =    -  \lambda_{s, j}^{-1} \bm{\Lambda}_{s,j,-j}\bm{x}_{s,i, -j},
\end{align}
where $\lambda_{c, j}$ and $\lambda_{s, j}$ are the $j$-th elements on the diagonal of  $\bm{\Lambda}_{c}$ and $\bm{\Lambda}_{s}$ respectively, while $\bm{\Lambda}_{c,j,-j}$ and $\bm{\Lambda}_{s,j,-j}$ are the $j$-th rows of $\bm{\Lambda}_{c}$ and $\bm{\Lambda}_{s}$ respectively, without the $j$-th element. 

\para{Latent variable $r_{i,j}$.}
We can easily see that the full conditional of $r_{i,j}$ is proportional to
\begin{equation}
	\exp \left\{ -\frac{(x_{c,i,j} - \eta_{c,i,j})^2}{2 \lambda_{c, j}^{-1}} \right\} \exp \left\{  -\frac{(x_{s,i,j} - \eta_{s,i,j})^2}{2 \lambda_{s, j}^{-1}} \right\} r_{i,j}
\end{equation}
which, by defining $\bm{\eta}_{i,j} = (\eta_{c,i,j}, \eta_{s,i,j})^\T$, $\tilde{\bm{\Sigma}}_{i,j} = \text{diag}( \lambda_{c, j}^{-1},  \lambda_{s, j}^{-1})^\T$, and $\bm{x}_{i,j}= (x_{c,i,j},x_{s,i,j})^\T $, can be written as
\begin{equation}
	\exp \left\{  -\frac{(\bm{x}_{i,j}- \bm{\eta}_{i,j})^\T \tilde{\bm{\Sigma}}_{i,j}^{-1} (\bm{x}_{i,j}- \bm{\eta}_{i,j})}{2} \right\} r_{i,j}
\end{equation} 
which shows that $\bm{\theta}_{i,j} \mid \cdots \sim \mathrm{PN}(\bm{\eta}_{i,j}, \tilde{\bm{\Sigma}}_{i,j})$, where $\bm{\theta}_{i,j} \mid \cdots$ indicates the full conditional distribution of $\bm{\theta}_{i,j}$ (i.e., the dots represent the data and all other model components with the exception of $\bm{\theta}_{i,j}$), and hence we can use the algorithm proposed by \cite{Hernandez-Stumpfhauser2017}.
First, we define
$$
\bm{\xi}_{i,j} = \{ \cos(\theta_{i,j}), \sin(\theta_{i,j})\}^\T, \quad
A_{i,j} =  \bm{\xi}_{i,j}^\T\tilde{\bm{\Sigma}}_{i,j}^{-1} \bm{\xi}_{i,j}, \quad B_{i,j} = \bm{\xi}_{i,j}^\T \tilde{\bm{\Sigma}}_{i,j}^{-1} \bm{\eta}_{i,j}.
$$
Then, we draw ${\beta}_{1,i,j} \sim U [ 0, \exp \{ - 2^{-1} A_{i,j} (r_{i,j} - B_{i,j} / A_{i,j} )^2 \} ]$ and ${\beta}_{2,i,j} \sim U(0, 1)$, and the new sample of  $r_{i,j}$ is 
$$
r_{i,j} = \left\{ (\varrho_{2,i,j}^2 - \varrho_{1,i,j}^2) {\beta}_{2,i,j} + \varrho_{1,i,j}^2 \right\}^{1/2},
$$
where $U(a,b)$ denotes the uniform distribution on $(a,b)$ and
$$
\varrho_{1,i,j} = \frac{B_{i,j}}{A_{i,j}} + \max\left\{ -\frac{B_{i,j}}{A_{i,j}}, -\left( \frac{-2 \log {\beta}_{1,i,j}}{A_{i,j}} \right)^{1/2} \right\}, \quad
\varrho_{2,i,j} = \frac{B_{i,j}}{A_{i,j}} + \left( \frac{-2 \log {\beta}_{1,i,j}}{A_{i,j}} \right)^{1/2}.
$$

\para{Parameter $\mu_j$.} Being $\mu_j$ embedded inside the sine and cosine used to compute $\bm{X}_{c,i}$ and $\bm{X}_{s,i}$, it is not generally possible to find a closed-form expression for its full conditional, which must be then sampled using a Metropolis Hastings algorithm.

\para{Parameter $\kappa_j$.} 
Since the kernel of the prior distribution of $\kappa_j$ is the same as the one of a normal distribution,  from the conditional distribution  \eqref{eq:cond_par}, we can see that 
$$
x_{c,i, j}+  \lambda_{c, j}^{-1} \bm{\Lambda}_{c,j,-j}\left(\bm{x}_{c,i, -j} - \bm{\kappa}_{-j}\right) \sim {\cal N}_1 (\kappa_j, \lambda_{c, j}^{-1}).
$$
Hence,  the full conditional of $\kappa_j$  is proportional to
\begin{equation}
	\exp \left(  -\frac{1}{2}  \left[ \kappa_j^2v^{-1}-2\kappa_jv^{-1}m + \kappa_j^2n\lambda_{c,j} - 2\kappa_j \sum_{i=1}^n \left\{ \bm{x}_{c,i} + \lambda_{c, j}^{-1} \bm{\Lambda}_{c,j,-j}\left(\bm{x}_{c,i, -j} - \bm{\kappa}_{-j}\right)\right\}   \right]   \right) 
\end{equation}
\begin{equation}
	\propto \exp \left\{ -\frac{(\kappa_j- m_{p,j})^2}{2 v_{p,j}}  \right\},
\end{equation}
where
\begin{align*}
v_{p,j}  &= (v^{-1} + n\lambda_{c,j})^{-1}, \quad
m_{p,j} = v_{p,j}\left[ \sum_{i=1}^n \left\{ \bm{x}_{c,i} +  \lambda_{c,j}^{-1} \bm{\Lambda}_{c,j,-j}(\bm{x}_{c,i, -j} - \bm{\kappa}_{-j})\right\} +v^{-1}m \right].
\end{align*}
It is easy to see that this is the kernel of ${\cal N}_1 (m_{p,j}, v_{p,j})$, and
since $\kappa_j>0$, the full conditional of $\kappa_j$ is ${\cal N}_1 (m_{p,j}, v_{p,j})\bm{I}_{\kappa_j>0}$.

	\para{Parameter $\bm{\Sigma}$.} To sample this parameter, we use a Metropolis–Hastings step. At iteration $b$, the proposal distribution is set to $IW(\tau, \bm{\Sigma}^b(\tau - (d + 1)))$, where $\bm{\Sigma}^b$ denotes the most recently accepted value of the parameter in the MCMC and $ IW(\tau, \bm{\Sigma}^b(\tau - (d + 1)))$ indicate the Inverse Wishart distribution  with degrees of freedom $\tau$ and scale matrix $\bm{\Sigma}^b \{ \tau - (d + 1) \}$.
The expected value of this proposal distribution is $\bm{\Sigma}^b$, and as the parameter $\tau$ increases, the variance of the distribution decreases, making $\tau$ a tuning parameter that controls the variability of the proposal.

After $\bm{\Sigma}^*$ is proposed, we compute $\bm{\Sigma}_c^*$ and if this is not a positive definite matrix, the proposal is rejected and we move to the next step of the algorithm. Otherwise, we use the standard Metropolis machinery to decide whether to accept or reject the sample.

\subsection{The CTPN model} \label{sec:copula_mcmc}
Let $c$ be the density of the $d$-variate TPN with uniform margins $\mathrm{TPN}_d(\bm{0},\bm{0}, \bm{\Sigma} )$.
If we wish to use the proposed density as a copula model, with the $j$-th marginal density equal to $f_{j}(\theta_{i,j})$ and the corresponding marginal cumulative distribution function $F_{j}(\theta_{i,j})$, the joint density of observation $i$ is equal to \eqref{eq:copula_extension}.
Since $c$ has the functional form of a $\mathrm{TPN}_d(\bm{0},\bm{0} ,\bm{\Sigma})$, it can be expressed as
\begin{equation}
	c\left(\theta_{i,1}^*, \dots , \theta_{i,d}^*\right) = \int_{(\mathbb{R}^+)^d}  f_{\bm{X}_{c,i}^*} (\bm{x}_{c,i}^*) f_{\bm{X}_{s,i}^*} (\bm{x}_{s,i}^*) \left( \prod_{j=1}^d r_{i,j} \right) d r_{i,1} \cdots dr_{i,d},
\end{equation}
where $f_{\bm{X}_{c,i}^*} (\bm{x}_{c,i}^*)$ and $f_{\bm{X}_{s,i}^*} (\bm{x}_{s,i}^*)$ in this case denote the densities of $d$-variate normal with mean vector of zeros and covariance matrices $\bm{\Sigma}_c$ and $\bm{\Sigma}$, respectively, and 
\begin{align}
	\theta_{i,j}^*& = 2\pi F_{j}(\theta_{i,j} - \mu_j), \\
	\bm{x}_{c,i}^*& = \left(  r_{i,1}\cos( \theta_{i,1}^*), \dots , r_{i,d}\cos( \theta_{i,d}^*) \right), \quad
	\bm{x}_{s,i}^* = \left(  r_{i,1}\sin( \theta_{i,1}^*), \dots , r_{i,d}\sin( \theta_{i,d}^*) \right).
\end{align}
This means that if we introduce the latent variables $\bm{r}_i=(r_{i,1}, \dots , r_{i,d})^\T$, we can work with the joint density
\begin{equation}\label{eq:copulamodel}
	f(\bm{\theta}_i, \bm{r}_i) =  \left(f_{\bm{X}_{c,i}^*} (\bm{x}_{c,i}^*) f_{\bm{X}_{s,i}^*} (\bm{x}_{s,i}^*) \prod_{j=1}^d r_{i,j}\right) (2 \pi)^d \prod_{j=1}^df_{j}(\theta_{i,j}- \mu_j).
\end{equation}
{Hence, the parameter $\bm{\Sigma}$ and the latent variables $\bm{r}_i$ can be sampled as in the previous section, whereas the variables $\bm{\theta}_i^*$, $\bm{x}_{c,i}^*$, and $\bm{x}_{s,i}^*$ must be used in place of $\bm{\theta}_i$, $\bm{x}_{c,i}$, and $\bm{x}_{s,i}$. The marginal density can depend on some parameters,
that  must be sampled using the Metropolis-Hastings algorithm, if closed-form full conditionals are not available.
}

\section{Real Data Example} \label{sec:example}

\begin{figure}[t]
	\centering
\includegraphics[width=0.70\textwidth]{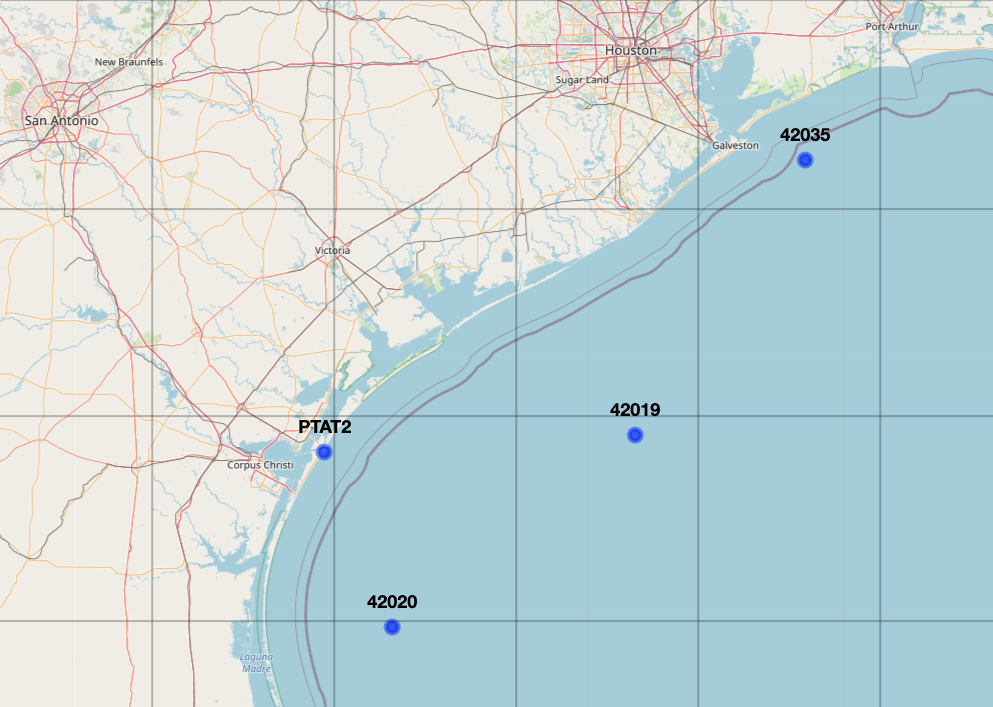}
\caption{Map showing the locations of four buoys in the Gulf of Mexico. A  grid with 1-degree intervals of latitude and longitude is overlaid on the map to provide geographic reference.} \label{fig:map}
\end{figure}

\begin{figure}
	\centering
\includegraphics[width=0.9\textwidth]{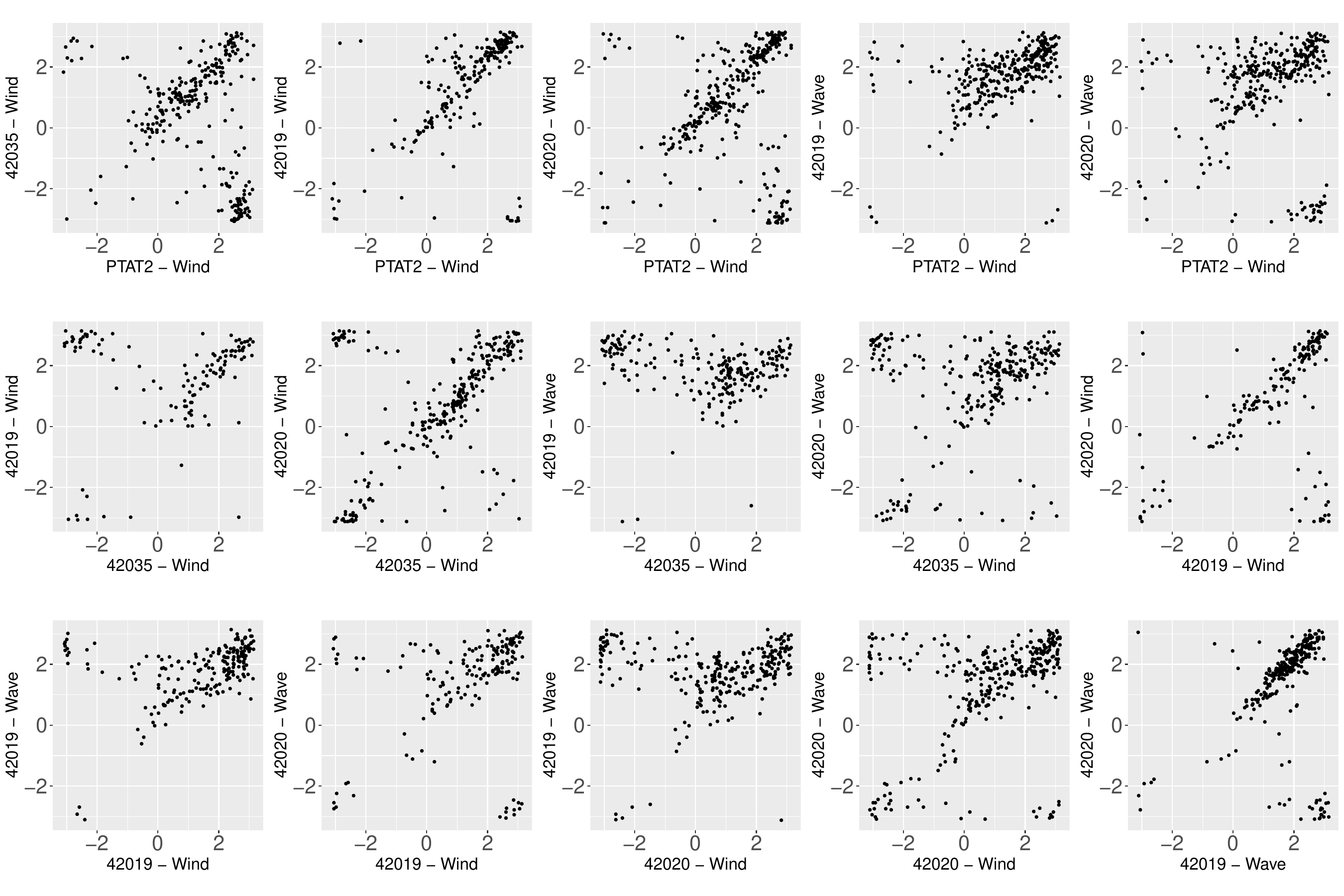}
\caption{Scatterplot of all possible combinations of observed data.
}  \label{fig:data}
\end{figure}





To demonstrate the practical applicability of the proposed distribution, in this section, we apply it to a real dataset obtained from the National Data Buoy Center website (http://www.ndbc.noaa.gov/).
We use data from four buoys located in the Gulf of Mexico, an ocean basin and a marginal sea of the Atlantic Ocean, off the Texas coast. Two buoys (Station \texttt{PTAT2}: 27.826° N, 97.051° W, and Station \texttt{42035}: 29.235° N, 94.410° W) provide data only on wind direction, while the other two (Station 42019: 27.908° N, 95.343° W, and Station 42020: 26.970° N, 96.679° W) include both wind and wave directions: the spatial location of the buoys is shown in Fig.~\ref{fig:map}. The data we use are daily measurements taken at 6 PM, from January 1st, 2020 to December 31st, 2020, yielding 366 recordings of a 6-dimensional vector of circular random variables.  In the selected time window,
station \texttt{PTAT2} has 1 missing value. Station \texttt{42035} has 92 missing values, station \texttt{42019} has 157 missing values for wind direction and 57 for wave direction, while station \texttt{42020} has 46 missing values in wave direction and 51 in wind direction. The missing values are treated as additional random variables, and posterior values are simulated during the MCMC algorithm.
A scatterplot for each pair of data is shown in Fig.~\ref{fig:data}.

{
By using the proposed MCMC, we estimate the parameters of a 6 dimensional TPN as well as a CTPN with wrapped Cauchy marginals. The wrapped Cauchy density is defined as
\begin{equation}
	f_j (\theta_{i,j}| \mu_j^*, \lambda_j) = \frac{1}{2\pi} \cdot \frac{1 - \lambda_j^2}{1 + \lambda_j^2 - 2 \lambda_j \cos(\theta_{i,j} -  \mu_j^*)} \quad (-\pi \leq \theta_{i,j} < \pi),
\end{equation}
where it depends on the location parameter $\mu_j^*$, which is also the mode of the distribution and its center of symmetry, and a concentration parameter $\lambda_j \in [0,1]$. Its cumulative distribution function can also be written in closed form as
\begin{equation}\label{eq:Fwc}
	F_j (\theta_{i,j} \leq a|\mu_j^*, \lambda_j) = 
\begin{cases} 
\frac{1}{2\pi} \arccos \left\{ \frac{(1 + \lambda_j^2) \cos(a - \mu_j^*) - 2 \lambda_j}{1 + \lambda_j^2 - 2 \lambda_j \cos(a - \mu_j^*)} \right\}, & \text{if } \sin(a - \mu_j^*) \geq 0 \\ 
1 - \frac{1}{2\pi} \arccos \left\{ \frac{(1 + \lambda_j^2) \cos(a - \mu_j^*) - 2 \lambda_j}{1 + \lambda_j^2 - 2 \lambda_j \cos(a - \mu_j)} \right\}, & \text{if } \sin(a - \mu_j^*) < 0 
\end{cases}
\end{equation}
where $a \in [0, 2\pi)$.
By construction, the cumulative distribution function at $\mu_j^*$ is equal to zero. Interestingly, it is possible to find in closed form also the quantile function of the wrapped Cauchy which is 
\begin{equation} \label{eq:F-1wc}
F_j^{-1}(u|\mu_j^*, \lambda_j)  = 
\mu_j^* + \operatorname{sign}\left(0.5 - u\right) \cdot \arccos \left\{ 
\frac{2\lambda_j + (1 + \lambda_j^2) \cos(2\pi u)}{1 + \lambda_j^2 + 2\lambda_j \cos(2\pi u)}
\right\}.
\end{equation}
Equations \eqref{eq:Fwc} and \eqref{eq:F-1wc}, under the assumption $\mu_j^* = 0$, can be used in \eqref{eq:coptrasform}.
Since the wrapped Cauchy is symmetrical, $\mu_j$  is the mean direction of the $j$-th marginal distribution.
}

We assume circular uniform prior distributions for the mean directions of both models, we assume $\kappa_j  \stackrel{iid}{\sim} {\cal N}_1(0, 100000)I_{>0}$, $\lambda_j \sim U(0,1)$ and $\bm{\Sigma} \sim TIW(8, \bm{I}_6)$.
The MCMC is implemented with 30,000 iterations, with a burn-in of 10,000 and a thinning factor of 10.
To select the best model between the two,   we use the Continuous Ranked Probability Score (CRPS) \citep{Gneiting2007}. The CRPS is a metric that compares the true value of a random variable with its predictive distribution and is straightforward to compute in a Bayesian setting. In this real data example, it is computed using 10\% of the data, which is set aside and not used in the model fitting. The CRPS for the TPN is 0.133 while for the wrapped Cauchy is 0.117: since  the wrapped Cauchy is the one with lower CRPS index, results are only presented for this model.

\begin{figure}[t]
	\centering 
\includegraphics[width=0.70\textwidth]{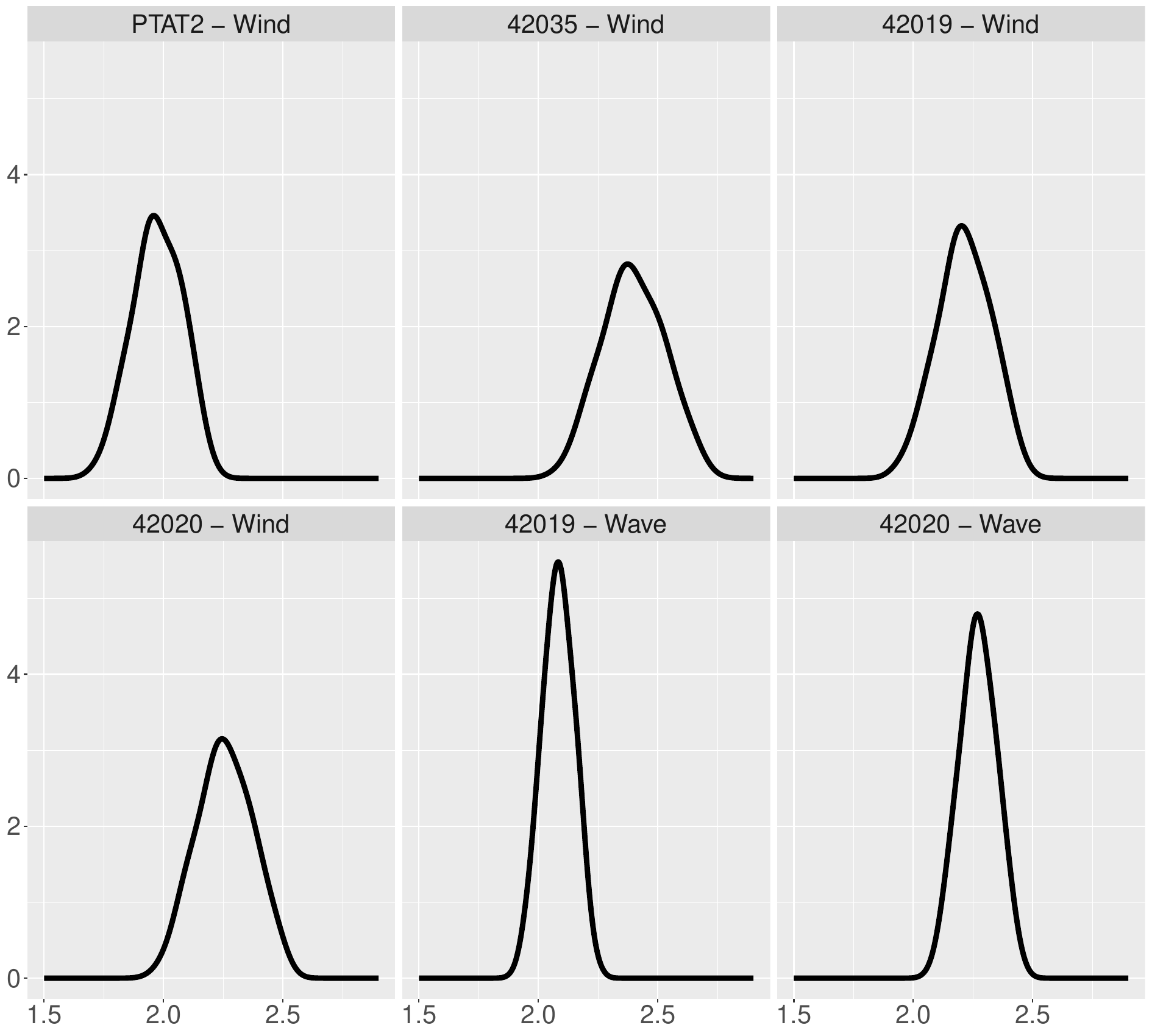}
\caption{Posterior density of parameter $\mu_{k,d}$ for the 6 circular variables. The y-axis is represented in logarithmic scale to better visualize the densities.
} \label{fig:posterior_mu}
\end{figure} 
 
\begin{figure}[t]
	\centering 
\includegraphics[width=0.70\textwidth]{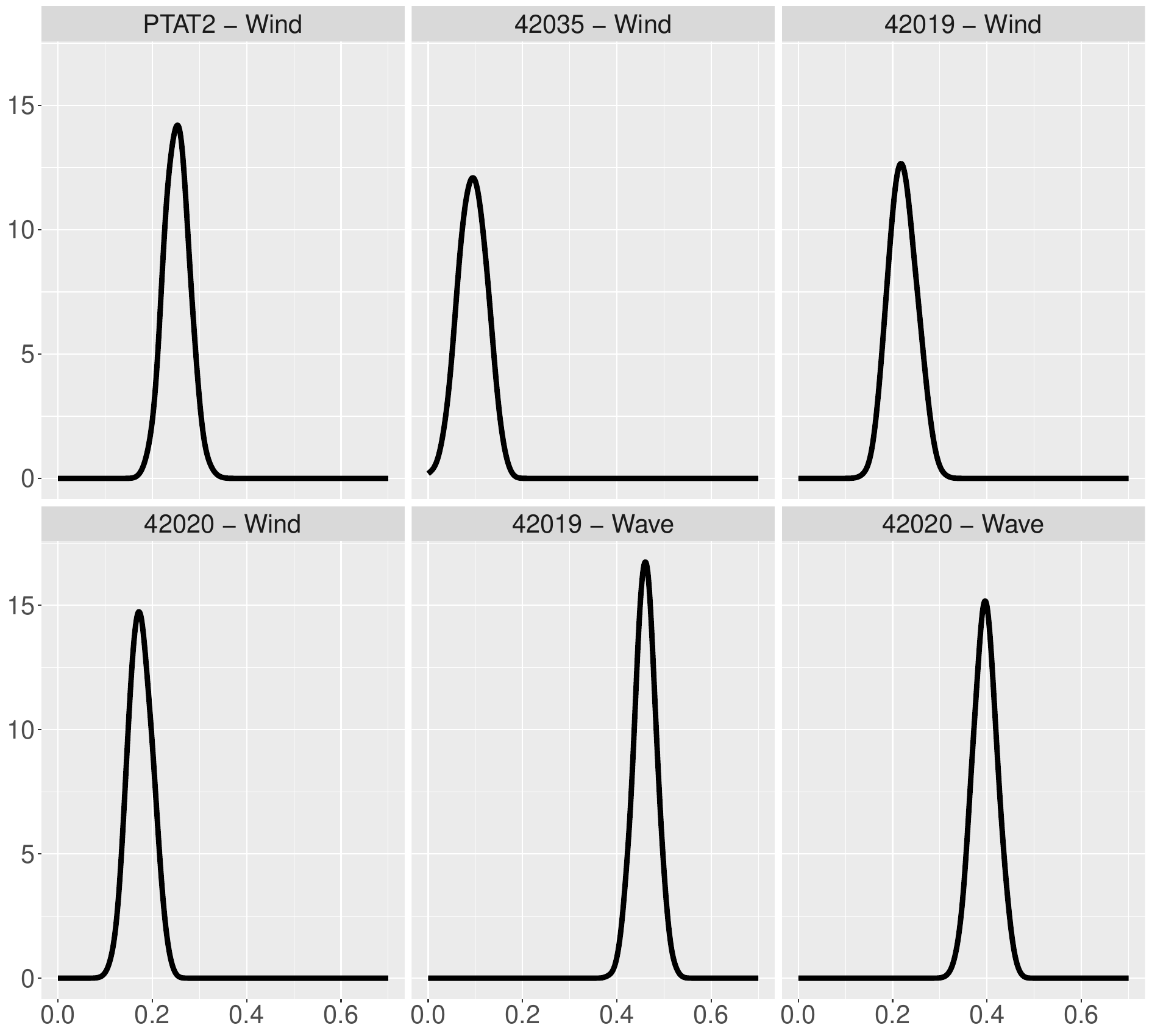}
\caption{Posterior density of parameter $\lambda_{k,d}$  for the 6 circular variables. 
} \label{fig:posterior_kappa}
\end{figure} 

\begin{figure}[t]
	\centering
\includegraphics[width=0.60\textwidth]{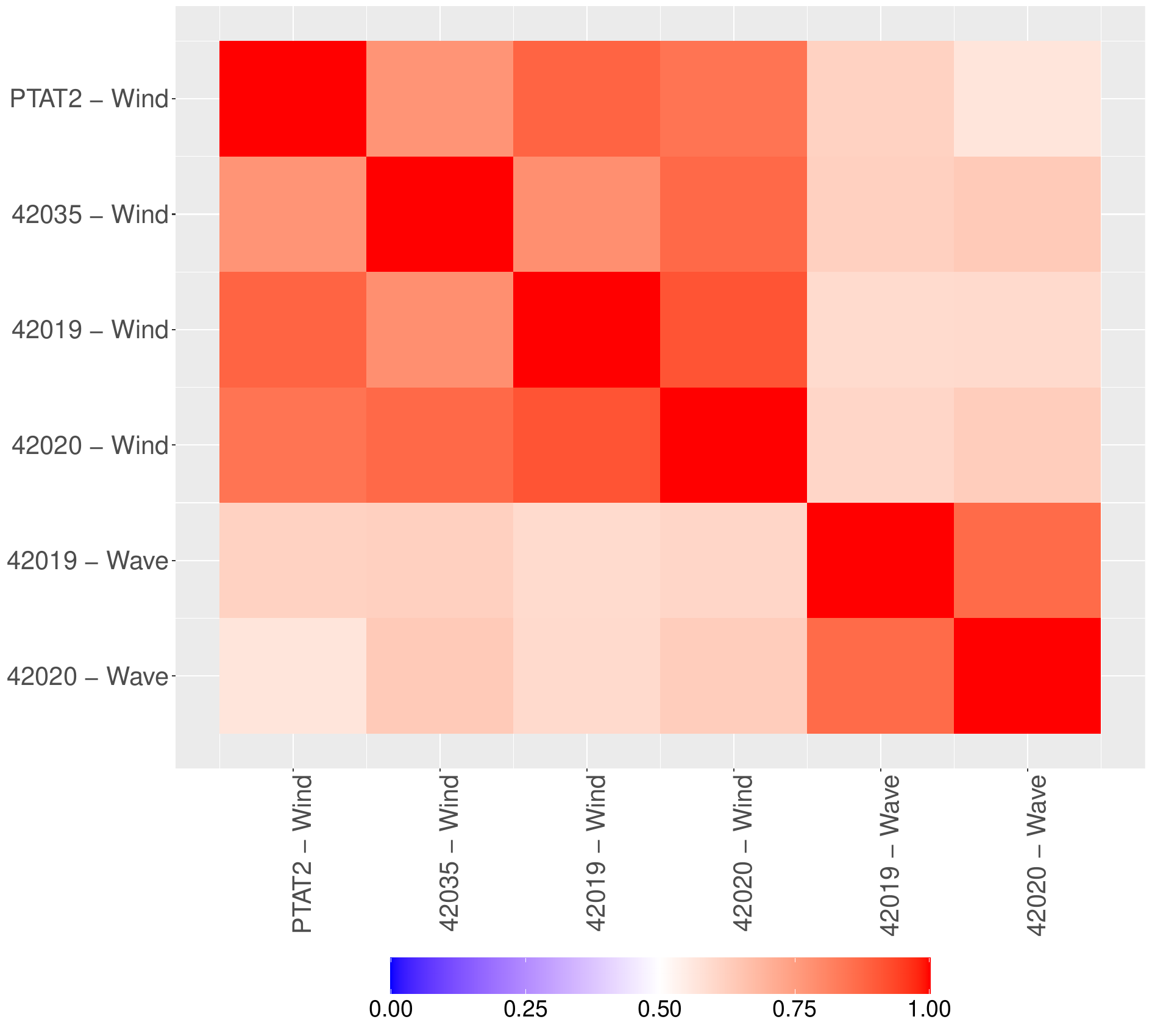}
\caption{Heatmap of the posterior means of matrix $\bm{\Sigma}$.} \label{fig:posterior_Sigma}
\end{figure} 
 
In terms of both the mean direction $\mu_{k,d}$ (Fig.~\ref{fig:posterior_mu}) and the concentration parameter $\lambda_{k,d}$ (Fig.~\ref{fig:posterior_kappa}), the distributions of the two wave variables appear quite similar, which is not the case for the wind directions.
Figure \ref{fig:posterior_Sigma} highlights a strong correlation between variables of the same type, while the correlation between wave and wind directions is weaker, though still around 0.5 for all combinations, even when recorded at the same buoy.
Overall, the three summaries in Figs.~\ref{fig:posterior_mu}--\ref{fig:posterior_Sigma} convey a coherent picture. For wave directions, the posterior means $\mu_{k,d}$ and concentrations $\lambda_{k,d}$ are closely aligned across buoys, suggesting that the underlying generating mechanism is broadly shared across sites. By contrast, wind directions exhibit larger between-buoy variability in both location and concentration, which is consistent with local effects (e.g., orographic shielding, fetch) and differing measurement conditions.
The dependence structure captured by $\bm{\Sigma}$ complements the marginal behavior: strong within-type correlations point to pronounced co-movement within the wave block and within the wind block, whereas cross-type correlations around 0.5 indicate a moderate but systematic coupling between waves and winds—even for co-located instruments. This pattern aligns with a physical interpretation in which wave fields integrate wind forcing over space and time, leading to smoother (and therefore more strongly correlated) dynamics for waves than for winds.

As a point of reference, more flexible hypertoroidal models can recover similar qualitative patterns but at the cost of reduced parameter interpretability. In our application, the proposed parametrization provides direct roles for $\mu_{k,d}$ (mean direction), $\lambda_{k,d}$ (concentration), and $\bm{\Sigma}$ (latent linear dependence), enabling clear, domain-relevant interpretations while capturing the main features of the data. A formal head-to-head comparison is beyond the scope of this section, but the qualitative agreement of locations, concentrations, and the sign/magnitude of cross-type dependence supports the practical interpretability of the proposed model.

\section*{Acknowledgement}
Shogo Kato and Gianluca Mastrantonio are grateful to Mitsui Sumitomo Insurance Co., Ltd. for financial support.
The work of Kato is supported by JSPS KAKENHI Grant Number 25K15037.

\section*{Supplementary material}
\label{SM}
The Supplementary Material includes the proofs of the theoretical claims from the main article and simulation examples of the Bayesian estimation of the TPN and CTPN distributions.

\bibliographystyle{apalike}

\bibliography{DirectionalStats}


\end{document}


\title{\textbf{\Large Supplementary material for ``An interpretable family of projected normal distributions and a related copula model for Bayesian analysis of hypertoroidal data''}}
\author{\scshape{Shogo\ Kato\,$^{a}$, \quad Gianluca\ Mastrantonio\,$^{b}$,} \vspace{0.1cm}\\
\scshape{ \quad and \quad Masayuki\ Ishikawa\,$^{c}$} \vspace{0.5cm}\\
\textit{\(^{a}\) Institute of Statistical Mathematics, Japan} \vspace{0.1cm}\\
\textit{\(^{b}\) Department of Mathematics, Polytechnic of Turin, Italy} \vspace{0.1cm}\\
\textit{\(^{c}\) Mitsui Sumitomo Insurance Co., Ltd., Japan} \vspace{0.5cm}\\ 
}
\date{\today}
\maketitle

\begin{abstract}
The Supplementary Material presents the proofs of the theoretical claims from the main article in \S \ref{sec:proofs} and simulation examples of the Bayesian estimation of the TPN and CTPN distributions in \S \ref{sec:simulations}.
\end{abstract}

\section{Proofs} \label{sec:proofs}

\subsection{Proof of Theorem 1}

\begin{proof}
	Let $\bm{X}_c$ and $\bm{X}_s$ be defined as in Definition 1, and let $R_j$ be as in \eqref{eq:xc_xs}.
	Then
	$$
	\begin{pmatrix}
		R_1 \cos (\Theta_1 - \mu_1 ) \\
		\vdots \\
		R_d \cos (\Theta_d - \mu_d ) \\
		R_1 \sin (\Theta_1 - \mu_1 ) \\
		\vdots \\
		R_d \sin (\Theta_d - \mu_d )
	\end{pmatrix} 
	=
	\begin{pmatrix}
		\bm{X}_c \\
		\bm{X}_s 
	\end{pmatrix}
	\stackrel{\rm d}{=} 
	\begin{pmatrix}
		\bm{X}_c \\
		-\bm{X}_s 
	\end{pmatrix} 
	=
	\begin{pmatrix}
		R_1 \cos (\Theta_1 - \mu_1 ) \\
		\vdots \\
		R_d \cos (\Theta_d - \mu_d ) \\
		-R_1 \sin (\Theta_1 - \mu_1 ) \\
		\vdots \\
		-R_d \sin (\Theta_d - \mu_d )
	\end{pmatrix} 
	= 
	\begin{pmatrix}
		R_1 \cos (\mu_1 - \Theta_1 ) \\
		\vdots \\
		R_d \cos (\mu_d - \Theta_d ) \\
		R_1 \sin (\mu_1 - \Theta_1 ) \\
		\vdots \\
		R_d \sin (\mu_d - \Theta_d )
	\end{pmatrix} .
	$$
	Here the equality in distribution holds because the multivariate normal distribution ${\cal N}_d (\bm{0},\bm{\Sigma})$ is symmetric about $\bm{0}$.
	Thus we have $ \bm{\Theta} - \bm{\mu} \stackrel{\rm d}{=} \bm{\mu} - \bm{\Theta} $.
\end{proof}

\subsection{Proof of Theorem 2}

\begin{proof}
	Define $\bm{X}_c= (X_{c,1},\ldots,X_{c,d})^\T$ and $\bm{X}_s = (X_{s,1},\ldots,X_{s,d})^\T$ as in Definition 1.
	Let 
	\begin{align*}
		\bm{X}_{c1} & =(X_{c,1},\ldots,X_{c,d_1})^\T, \quad \bm{X}_{c2}=(X_{c,d_1+1},\ldots,X_{c,d})^\T, \\
		\quad \bm{X}_{s1} & =(X_{s,1},\ldots,X_{s,d_1})^\T, \quad \bm{X}_{s2}=(X_{s,d_1+1},\ldots,X_{s,d})^\T.
	\end{align*}
	Then it follows from the theory of the multivariate normal distribution that
	$$
	\bm{X}_{c,1} \sim {\cal N}_d (\bm{\kappa}_1,\bm{\Sigma}_{c,11}) , \quad \bm{X}_{c,2} \sim {\cal N}_d (\bm{\kappa}_2,\bm{\Sigma}_{c,22}), \quad 
	\bm{X}_{s,1} \sim {\cal N}_d (\bm{0},\Sigma_{11}) , \quad \bm{X}_{s,2} \sim N_d (\bm{0},\Sigma_{22}),
	$$
	where $\bm{\Sigma}_{c,\ell\ell}$ is a $d_\ell \times d_\ell$ matrix, $ [\bm{\Sigma}_{c,\ell\ell}]_{j,k} =|\rho_{\ell\ell,j,k}| $, and $\rho_{\ell\ell,j,k}$ is the $(j,k)$ element of $\bm{\Sigma}_{\ell\ell}$ $(\ell=1,2)$.
	Clearly, $\bm{\Sigma}_{11}$ and $\bm{\Sigma}_{22}$ are positive definite matrices whose diagonal elements are equal to 1.
	Also $\bm{X}_{c,\ell}$ and $\bm{X}_{s,\ell}$ are independent for $\ell=1,2$.
	Then it follows from the derivation of the proposed model given in Definition 1 that
	$
	\bm{\Theta}_1 \sim \textsc{TPN}_{d_1}(\bm{\mu}_1, \bm{\kappa}_1, \bm{\Sigma}_{11})
	$ and 
	$\bm{\Theta}_2 \sim \textsc{TPN}_{d_2}(\bm{\mu}_2, \bm{\kappa}_2, \bm{\Sigma}_{22})
	$ hold.
\end{proof}

\subsection{Proof of Theorem 4}

\begin{proof} 
	The part of the argument of the exponential function in equation (4) can be expressed as
	\begin{align*}
		\lefteqn{ (\bm{x}_c-\bm{\kappa})^\T \bm{\Sigma}_c^{-1} (\bm{x}_c-\bm{\kappa}) + \bm{x}_s^\T \bm{\Sigma}^{-1} \bm{x}_s } \hspace{2cm} \\
		& = (\bm{T}_c \bm{r} -\bm{\kappa})^\T \bm{\Sigma}_c^{-1} (\bm{T}_c \bm{r} -\bm{\kappa}) + (\bm{T}_s \bm{r})^\T \bm{\Sigma}^{-1} (\bm{T}_s \bm{r}) \\
		& = \bm{r}^\T \bm{T}_c \bm{\Sigma}_c^{-1} \bm{T}_c \bm{r} - 2 \bm{\kappa}^\T \bm{\Sigma}_c^{-1} \bm{T}_c \bm{r} + \bm{\kappa}^\T \bm{\Sigma}_c^{-1} \bm{\kappa} + \bm{r}^\T \bm{T}_s \bm{\Sigma}^{-1} \bm{T}_s \bm{r}  \\
		& = \bm{r}^\T \bm{V}^{-1} \bm{r} - 2 \bm{\kappa}^\T \bm{\Sigma}_c^{-1} \bm{T}_c \bm{r} + \bm{\kappa}^\T \bm{\Sigma}_c^{-1} \bm{\kappa} \\
		& = (\bm{r}-\bm{\eta})^\T \bm{V}^{-1} (\bm{r}-\bm{\eta}) + \bm{\kappa}^\T \bm{\Sigma}_c^{-1} \bm{\kappa}  - \bm{\eta}^\T \bm{V}^{-1} \bm{\eta} \\
		& = (\bm{r}-\bm{\eta})^\T \bm{V}^{-1} (\bm{r}-\bm{\eta}) + \bm{\kappa}^\T \left( I - \bm{\Sigma}_c^{-1} \bm{T}_c \bm{V} \bm{T}_c \right) \bm{\Sigma}_c^{-1} \bm{\kappa}.
	\end{align*}
	Substituting this result into equation (4), we obtain the alternative expression (5).
\end{proof}

\subsection{Proof of Corollary 1}

\begin{proof}
	It is a straightforward exercise to show that
	\begin{align*}
		\bm{V}^{-1} = \ & \bm{T}_c \bm{\Sigma}_c^{-1} \bm{T}_c +\bm{T}_s \bm{\Sigma}^{-1} \bm{T}_s 
		= \bm{T}_c \bm{\Sigma}_c^{-1} \bm{T}_c + \bm{T}_s \bm{D} \bm{\Sigma}_c^{-1} \bm{D} \bm{T}_s \\ 
		= \ & \begin{pmatrix}
			s_{11} \cos^2 \bar{\theta}_1 & s_{12} \cos \bar{\theta}_1 \cos \bar{\theta}_2 & \cdots & s_{1d} \cos \bar{\theta}_1 \cos \bar{\theta}_d \\
			s_{12} \cos \bar{\theta}_1 \cos \bar{\theta}_2 & s_{22} \cos^2 \bar{\theta}_2 & \cdots & s_{2d} \cos \bar{\theta}_2 \cos \bar{\theta}_d \\
			\vdots & \vdots & \ddots & \vdots \\
			s_{1d} \cos \bar{\theta}_1 \cos \bar{\theta}_d & s_{2d} \cos \bar{\theta}_2 \cos \bar{\theta}_d & \cdots & s_{dd} \cos^2 \bar{\theta}_d
		\end{pmatrix} \\
		& + \begin{pmatrix}
			s_{11} \sin^2 \bar{\theta}_1 & s_{12} \sin (\delta_1 \bar{\theta}_1) \sin (\delta_2 \bar{\theta}_2) & \cdots & s_{1d} \sin (\delta_1 \bar{\theta}_1) \sin (\delta_d \bar{\theta}_d) \\
			s_{12} \sin (\delta_1 \bar{\theta}_1) \sin (\delta_2 \bar{\theta}_2)  & s_{22} \sin^2 \bar{\theta}_2 & \cdots & s_{2d} \sin (\delta_2 \bar{\theta}_2) \sin (\delta_d \bar{\theta}_d) \\
			\vdots & \vdots & \ddots & \vdots \\
			s_{1d} \sin (\delta_1 \bar{\theta}_1) \sin (\delta_d \bar{\theta}_d) & s_{2d} \sin (\delta_2 \bar{\theta}_2) \sin (\delta_d \bar{\theta}_d) & \cdots & s_{dd} \sin^2 \bar{\theta}_d
		\end{pmatrix} \\
		= \ & \begin{pmatrix}
			s_{11} & s_{12} \cos (\delta_1 \bar{\theta}_1 - \delta_2 \bar{\theta}_2) & \cdots & s_{1d} \cos (\delta_1 \bar{\theta}_1 - \delta_d \bar{\theta}_d) \\
			s_{12} \cos (\delta_1 \bar{\theta}_1 - \delta_2 \bar{\theta}_2) & s_{22} & \cdots & s_{2d} \cos (\delta_2 \bar{\theta}_2 - \delta_d \bar{\theta}_d) \\
			\vdots & \vdots & \ddots & \vdots \\
			s_{1d} \cos (\delta_1 \bar{\theta}_1 - \delta_d \bar{\theta}_d) & s_{2d} \cos (\delta_2 \bar{\theta}_2 - \delta_d \bar{\theta}_d) & \cdots & s_{dd}
		\end{pmatrix}.
	\end{align*}
	Thus we obtain the expression for $\bm{V}$ in the corollary as required.
\end{proof}

\subsection{Proof of Theorem 5}

\begin{proof}
Let $\Theta^=  = \{ (\theta_1,\ldots,\theta_d)^\T \in [-\pi,\pi)^d \, |\, \delta_1(\theta_1-\mu_1) = \cdots = \delta_d (\theta_d-\mu_d) \}$.
Denote $\bm{x}_c^=$ and $\bm{x}_s^=$ by $\bm{x}_c = \bm{x}_c (\bm{\theta})$ and $\bm{x}_s = \bm{x}_s (\bm{\theta})$, respectively, when $\bm{\theta} = \bm{\theta}^= \in \Theta^=$.
In order to prove the theorem, it suffices to show that the integrand of the first expression of the copula density (8) satisfies
\begin{equation} \label{eq:mode_inequality}
\exp \left\{ - \frac12 \left(  \bm{x}_c^\T \bm{\Sigma}_c^{-1} \bm{x}_c 
+ \bm{x}_s^\T \bm{\Sigma}^{-1} \bm{x}_s \right) \right\} \prod_{j=1}^d r_j
\;\leq\;
\exp \left\{ - \frac12 \left(  {\bm{x}_c^=}^\T \bm{\Sigma}_c^{-1} \bm{x}_c^=
+ {\bm{x}_s^=}^\T \bm{\Sigma}^{-1} {\bm{x}_s^=} \right) \right\} \prod_{j=1}^d r_j,
\end{equation}
for any $(r_1,\ldots,r_d) \in (\mathbb{R}^+)^d$, with equality if and only if 
$\bm{\theta} \in \Theta^=$.
The argument of the exponential function on the left-hand side of \eqref{eq:mode_inequality} can be evaluated as
\begin{align}
\begin{split} \label{eq:mode_inequality2}
\lefteqn{ \bm{x}_c^\T \bm{\Sigma}_c^{-1} \bm{x}_c 
+ \bm{x}_s^\T \bm{\Sigma}^{-1} \bm{x}_s } \hspace{0.2cm} \\
& = \bm{x}_c^\T \bm{\Sigma}_c^{-1} \bm{x}_c 
+ \bm{x}_s^\T (\bm{D} \Sigma_c \bm{D})^{-1} \bm{x}_s = \bm{x}_c^\T \bm{\Sigma}_c^{-1} \bm{x}_c 
+ (\bm{D} \bm{x}_s)^\T \Sigma_c^{-1} (\bm{D} \bm{x}_s) \\
& = \sum_{j,k=1}^d \left\{ s_{jk} \cdot r_j \cos (\theta_j-\mu_j) \cdot r_k \cos (\theta_k-\mu_k) + s_{jk} \cdot \delta_j r_j  \sin (\theta_j-\mu_j) \cdot \delta_k r_k \sin (\theta_k-\mu_k) \right\} \\
& = \sum_{j,k=1}^d s_{jk} r_j r_k \cos \{ \delta_j (\theta_j-\mu_j) - \delta_k (\theta_k-\mu_k) \} \\
& = \sum_{j=1}^d s_{jj} r_j^2 + \sum_{j \neq k} s_{jk} r_j r_k \cos \{ \delta_j (\theta_j-\mu_j) - \delta_k (\theta_k-\mu_k) \} \\
& \geq \sum_{j=1}^d s_{jj} r_j^2 + \sum_{j \neq k}  s_{jk} r_j r_k = \sum_{j,k=1}^d s_{jk} r_j r_k = {\bm{x}_c^{=}}^\T \bm{\Sigma}_c^{-1} {\bm{x}_c^{=}} 
+ (\bm{D} \bm{x}_s^=)^\T \Sigma_c^{-1} (\bm{D} \bm{x}_s^= ) \\
 & = {\bm{x}_c^=}^\T \bm{\Sigma}_c^{-1} \bm{x}_c^= 
+ {\bm{x}_s^=}^\T \bm{\Sigma}^{-1} \bm{x}_s^=,
\end{split}
\end{align}
where the inequality holds because of the assumption $s_{jk} < 0$.
Equality in \eqref{eq:mode_inequality2} holds if and only if $\bm{\theta} \in \Theta^=$, 
since $\cos \{ \delta_j (\theta_j-\mu_j) - \delta_k (\theta_k-\mu_k) \} < 1$ 
whenever $\delta_j (\theta_j-\mu_j) \neq \delta_k (\theta_k-\mu_k)$.
The result \eqref{eq:mode_inequality2} implies the inequality \eqref{eq:mode_inequality}.
Thus, for any $\bm{\theta} \in [-\pi,\pi)^d$, we have
\begin{align}
c (\bm{\theta}) & = \frac{1}{(2\pi)^{d} |\bm{\Sigma}_c|^{1/2} |\bm{\Sigma}|^{1/2}} \int_{(\mathbb{R}^+)^d} \exp \left\{ - \frac12 \left(  \bm{x}_c^\T \bm{\Sigma}_c^{-1} \bm{x}_c + \bm{x}_s^\T \bm{\Sigma}^{-1} \bm{x}_s \right) \right\} \prod_{j=1}^d r_j d \bm{r} \\
 & \leq \frac{1}{(2\pi)^{d} |\bm{\Sigma}_c|^{1/2} |\bm{\Sigma}|^{1/2}} \int_{(\mathbb{R}^+)^d} \exp \left\{ - \frac12 \left(  {\bm{x}_c^=}^\T \bm{\Sigma}_c^{-1} \bm{x}_c^= + {\bm{x}_s^=}^\T \bm{\Sigma}^{-1} \bm{x}_s^= \right) \right\} \prod_{j=1}^d r_j d \bm{r} \\
 & = c (\bm{\theta}^=),
\end{align}
with equality if and only if $\bm{\theta} \in \Theta^=$, since the inequality in \eqref{eq:mode_inequality2} is strict whenever $\bm{\theta} \notin \Theta^= $ for any $(r_1,\ldots,r_d) \in (\mathbb{R}^+)^d$.
Therefore the copula density (8) attains its maximum if and only if $\delta_1 (\theta_1-\mu_1) =  \delta_2 (\theta_2 - \mu_2) = \cdots = \delta_d (\theta_d-\mu_k)$.
\end{proof}

\subsection{Proof of Theorem 6}
 
\begin{proof}
	First we simplify the form of the integrand in (8) by putting $\tilde{\bm{r}}=\bm{V}^{-1/2} \bm{r}$.
	Then $c(\bm{\theta})$ can be expressed as
	\begin{align}
		c(\bm{\theta}) & = C \int_{\bm{V}^{-1/2}(\mathbb{R}^+)^2} \exp \left( - \frac{\|\tilde{\bm{r}}\|^2}{2} \right) \left( \bm{v}_1^\T \tilde{\bm{r}} \right) \left( \bm{v}_2^\T \tilde{\bm{r}} \right) |\bm{V}|^{1/2} d\tilde{\bm{r}} \nonumber \\
		& = C \cdot |\bm{V}|^{1/2} \, \bm{v}_1^\T \left\{ \int_{\bm{V}^{-1/2}(\mathbb{R}^+)^2} \exp \left( - \frac{\|\tilde{\bm{r}}\|^2}{2} \right)
		\begin{pmatrix}
			\tilde{r}_1^2 & \tilde{r}_1 \tilde{r}_2 \\
			\tilde{r}_1 \tilde{r}_2 & \tilde{r}_2^2
		\end{pmatrix}
		d\tilde{\bm{r}} \right\}  \bm{v}_2, \label{eq:integral_matrix}
	\end{align}
	where $C= \{ (2\pi)^2 (1-\rho^2) \}^{-1}$, $\bm{v}_j^\T$ denotes the $j$-th row of $\bm{V}^{1/2}$, and $\tilde{r}_j$ is the $j$-th component of $\tilde{\bm{r}}$ $(j=1,2)$.
	Here we calculate the integrals for each entry of the matrix inside the closing curly bracket in \eqref{eq:integral_matrix}.
	We first consider the integral 
	$$  
	\int_{\bm{V}^{-1/2}(\mathbb{R}^+)^2} \exp \left( - \frac{\|\tilde{\bm{r}}\|^2}{2} \right)
	\tilde{r}_1^2 d\tilde{\bm{r}} \equiv R_{11}.
	$$
	Transforming the variable $\tilde{\bm{r}}$ into polar-coordinate form $r_1=s \cos t$ and $r_2 = s \sin t$, it holds that
	$$
	R_{11} = \int_D \exp \left( - \frac{s^2}{2} \right) s^2 \cos^2 t \cdot s \, ds dt.
	$$
	In order to express the region of the integration $D$ in polar-coordinate form, we note that the eigenvalue decomposition of $\bm{V}^{-1}$ implies
	\begin{align}
	\begin{split} \label{eq:V_half_inv}
		\bm{V}^{-1/2} & = \begin{pmatrix}
			1/\surd{2} & -1/\surd{2} \\
			1/\surd{2} & 1/\surd{2}
		\end{pmatrix} 
		\begin{pmatrix}
			\lambda_+ & 0 \\
			0 & \lambda_-
		\end{pmatrix} 
		\begin{pmatrix}
			1/\surd{2} & 1/\surd{2} \\
			-1/\surd{2} & 1/\surd{2}
		\end{pmatrix} \\
		& = \begin{pmatrix}
			\cos (-\pi/4) & - \sin (-\pi/4) \\
			\sin (-\pi/4) & \cos (-\pi/4)
		\end{pmatrix}^\T
		\begin{pmatrix}
			\lambda_+ & 0 \\
			0 & \lambda_-
		\end{pmatrix} 
		\begin{pmatrix}
			\cos (-\pi/4) & - \sin (-\pi/4) \\
			\sin (-\pi/4) & \cos (-\pi/4)
		\end{pmatrix},
		\end{split}
	\end{align}
	where $\lambda_+ =[ \{1+\tau(\bm{\theta}) \}/ (1-\rho^2) ]^{1/2}$ and $\lambda_- = [ \{1-\tau(\bm{\theta}) \}/ (1-\rho^2) ]^{1/2}$ are squared roots of the eigenvalues of $\bm{V}^{-1}$.
	This expression implies that $D$ can be expressed as
	$$
	D = \{ (s,t) \, | \, s \geq 0 , \ \alpha_1 \leq t \leq \alpha_2  \},
	$$
	where $\alpha_1=\mbox{atan}^* \{ (\lambda_+-\lambda_-)/(\lambda_+ + \lambda_-) \}$ and $\alpha_2=\mbox{atan}^* \{ (\lambda_+ +\lambda_-)/(\lambda_+ - \lambda_-) \}$. 
	Then the integration in $R_{11}$ simplifies to
	\begin{align*}
		R_{11} & = \left\{ \int_0^{\infty} \exp \left( - \frac{s^2}{2} \right) s^3 ds \right\} \times \left( \int_{\alpha_1}^{\alpha_2} \cos^2 t dt \right) = \alpha_2 - \alpha_1 ,
	\end{align*}
	where the last equality holds by showing $\int_0^{\infty} \exp \left( - s^2/2 \right) s^3 ds = 2$ via integration by parts and using the relationship $\alpha_1=\pi/2-\alpha_2$.
	Using the formula $\arctan (x) + \arctan(y) = \arctan[ (x+y)/(1+xy) ]$, $R_{11}$ can be expressed in terms of $\tau (\bm{\theta})$ as 
	$$
	R_{11} = \mbox{atan}^* \left[ \frac{\{ 1-\tau (\bm{\theta})^2 \}^{1/2}}{ \tau (\bm{\theta})} \right]. 
	$$
	Similarly, it is possible to calculate the integrals for the other entries of the matrix inside the closing curly bracket in \eqref{eq:integral_matrix} as
	$$
	R_{22} \equiv \int_{\bm{V}^{-1/2}(\mathbb{R}^+)^2} \exp \left( - \frac{\|\tilde{\bm{r}}\|^2}{2} \right)
	\tilde{r}_2^2 d\tilde{\bm{r}} =  \mbox{atan}^* \left[ \frac{\{ 1-\tau (\bm{\theta})^2 \}^{1/2}}{ \tau (\bm{\theta})} \right] ,
	$$
	$$
	R_{12} \equiv \int_{\bm{V}^{-1/2}(\mathbb{R}^+)^2} \exp \left( - \frac{\|\tilde{\bm{r}}\|^2}{2} \right)
	\tilde{r}_1 \tilde{r}_2 d\tilde{\bm{r}} = - \cos (2\alpha_2).
	$$
	Here $R_{12}$ can be expressed in terms of $\tau(\bm{\theta})$ as
	\begin{align*}
		R_{12} & = - (2 \cos^2 \alpha_2 -1) = - \left( 2 \left[ \frac{\lambda_+ - \lambda_-}{\{ (\lambda_+-\lambda_-)^2 + (\lambda_+ + \lambda_-)^2 \}^{1/2} } \right]^2 - 1 \right) \\
		& = \left\{ 1-\tau (\bm{\theta})^2 \right\}^{1/2}.
	\end{align*}
	Then the quadratic form in \eqref{eq:integral_matrix} can be expressed as
	\begin{align*}
		\lefteqn{ \bm{v}_1^\T \left\{ \int_{\bm{V}^{-1/2}(\mathbb{R}^+)^2} \exp \left( - \frac{\|\tilde{\bm{r}}\|^2}{2} \right)
			\begin{pmatrix}
				\tilde{r}_1^2 & \tilde{r}_1 \tilde{r}_2 \\
				\tilde{r}_1 \tilde{r}_2 & \tilde{r}_2^2
			\end{pmatrix}
			d\tilde{\bm{r}} \right\}  \bm{v}_2 } \hspace{2cm} \\
		&= \frac{1}{4\lambda_+^2 \lambda_-^2} \left\{ 2 (\lambda_+^2 + \lambda_-^2) R_{12} + (\lambda_+^2-\lambda_-^2) (R_{11}+R_{22}) \right\} \\
		& = \frac{1-\rho^2}{1-\tau (\bm{\theta})} \left( \left\{ 1-\tau (\bm{\theta})^2 \right\}^{1/2} - \tau (\bm{\theta}) \, \mbox{atan}^* \left[ \frac{\{ 1-\tau (\bm{\theta})^2 \}^{1/2} }{ \tau (\bm{\theta})} \right] \right),
	\end{align*}
	where $\bm{v}_1$ and $\bm{v}_2$ can be calculated from the expression of $\bm{V}$ with $d=2$, given in \S 2.4 of the main article, as
	$$ 
	\begin{pmatrix}
		\bm{v}_1^\T \\ \bm{v}_2^\T
	\end{pmatrix} 
	= \bm{V}^{1/2} = \frac{1}{2 \lambda_+ \lambda_-}
	\begin{pmatrix}
		\lambda_+ + \lambda_- & \lambda_- - \lambda_+ \\
		\lambda_- - \lambda_+ & \lambda_+ + \lambda_-
	\end{pmatrix}.
	$$
	It is also clear from \eqref{eq:V_half_inv} that $|\bm{V}|^{1/2} = (\lambda_+ \lambda_-)^{-1} = (1-\rho^2)/\{1-\tau (\bm{\theta})^2\}^{1/2}$.
	Substituting these results into \eqref{eq:integral_matrix}, we obtain the expression (10) of the density.
\end{proof}

\section{Simulated Examples} \label{sec:simulations}

\begin{figure}
	\centering 
    \begin{subfigure}[b]{0.750\textwidth}
        \centering
        \includegraphics[width=\textwidth]{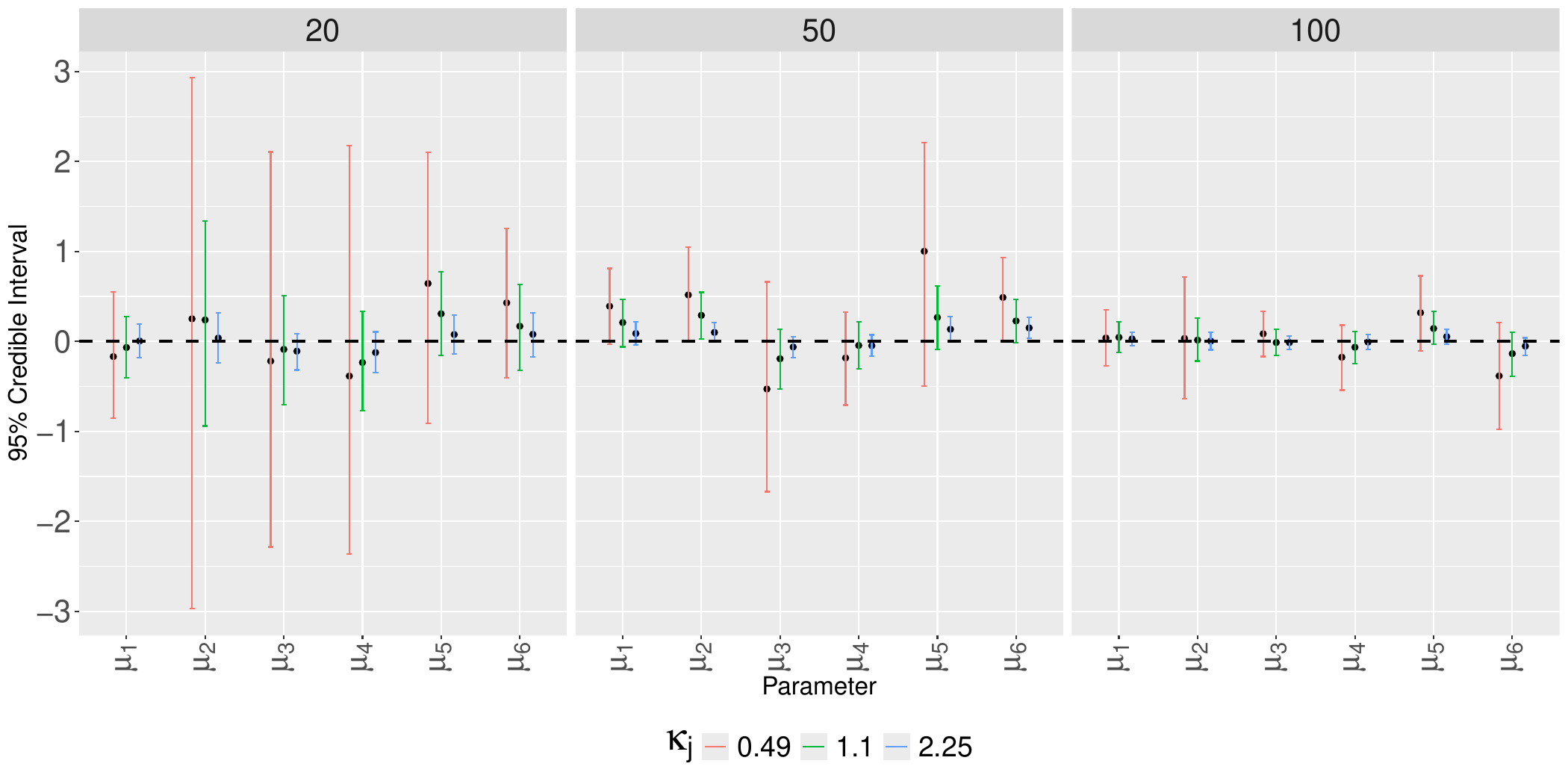}
        \caption{Diagonal $\bm{\Sigma}$}
    \end{subfigure}
    \hfill
    \begin{subfigure}[b]{0.750\textwidth}
        \centering
        \includegraphics[width=\textwidth]{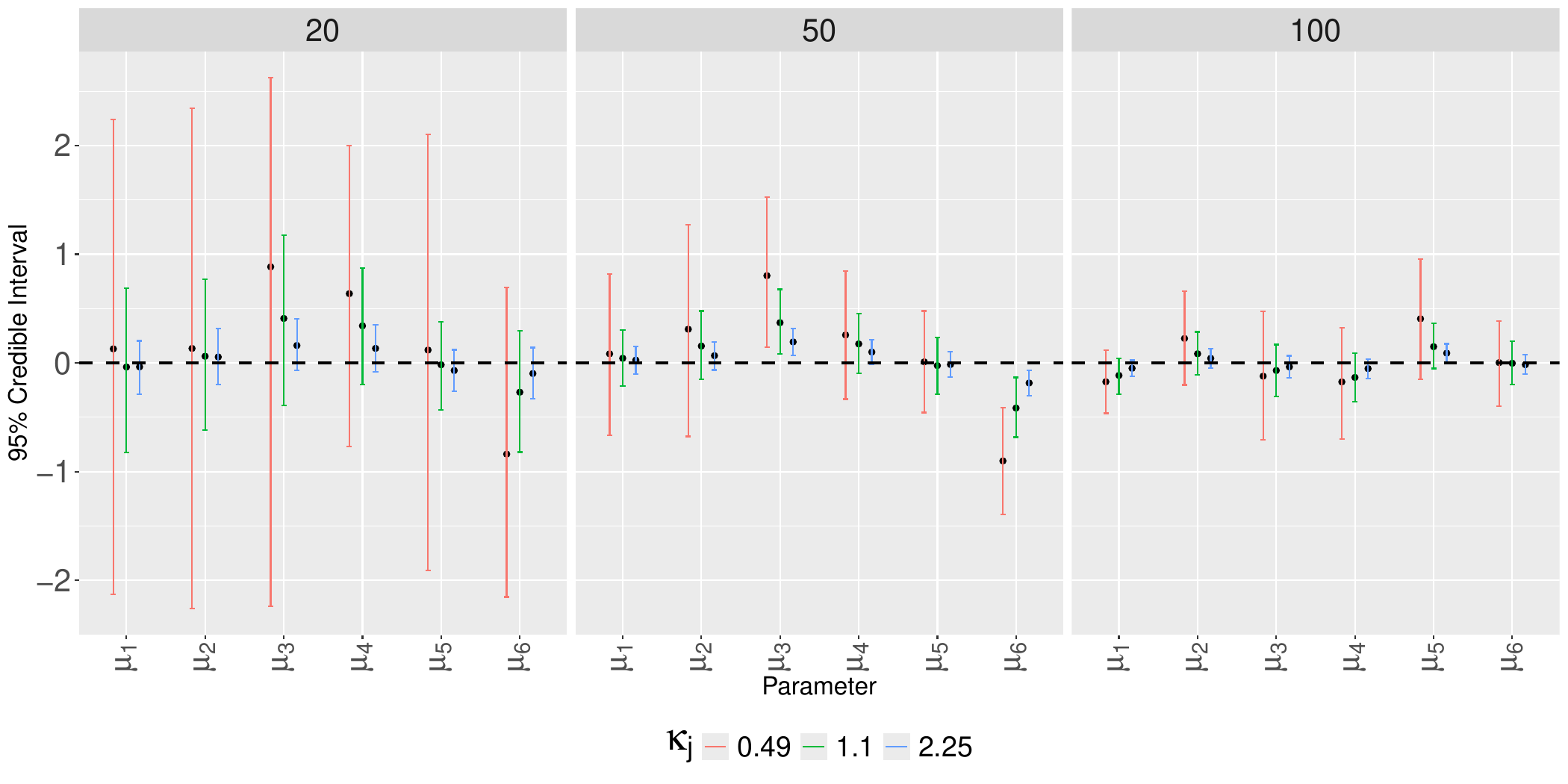}
        \caption{Full $\bm{\Sigma}$}
    \end{subfigure}
    \caption{TPN results - 95\% confidence intervals of the posterior distribution for the circular difference between the parameter $\mu_j$ and its true value (lines). The dot represents the posterior means. The first row presents results with a diagonal $\bm{\Sigma}$, while the second row shows results with a full $\bm{\Sigma}$. The left panel displays results for $n = 20$, the center panel for $n = 50$, and the right panel for $n = 100$.}
    \label{fig:sim_res_1}
\end{figure}
\begin{figure}
	\centering
    \begin{subfigure}[b]{0.750\textwidth}
        \centering
        \includegraphics[width=\textwidth]{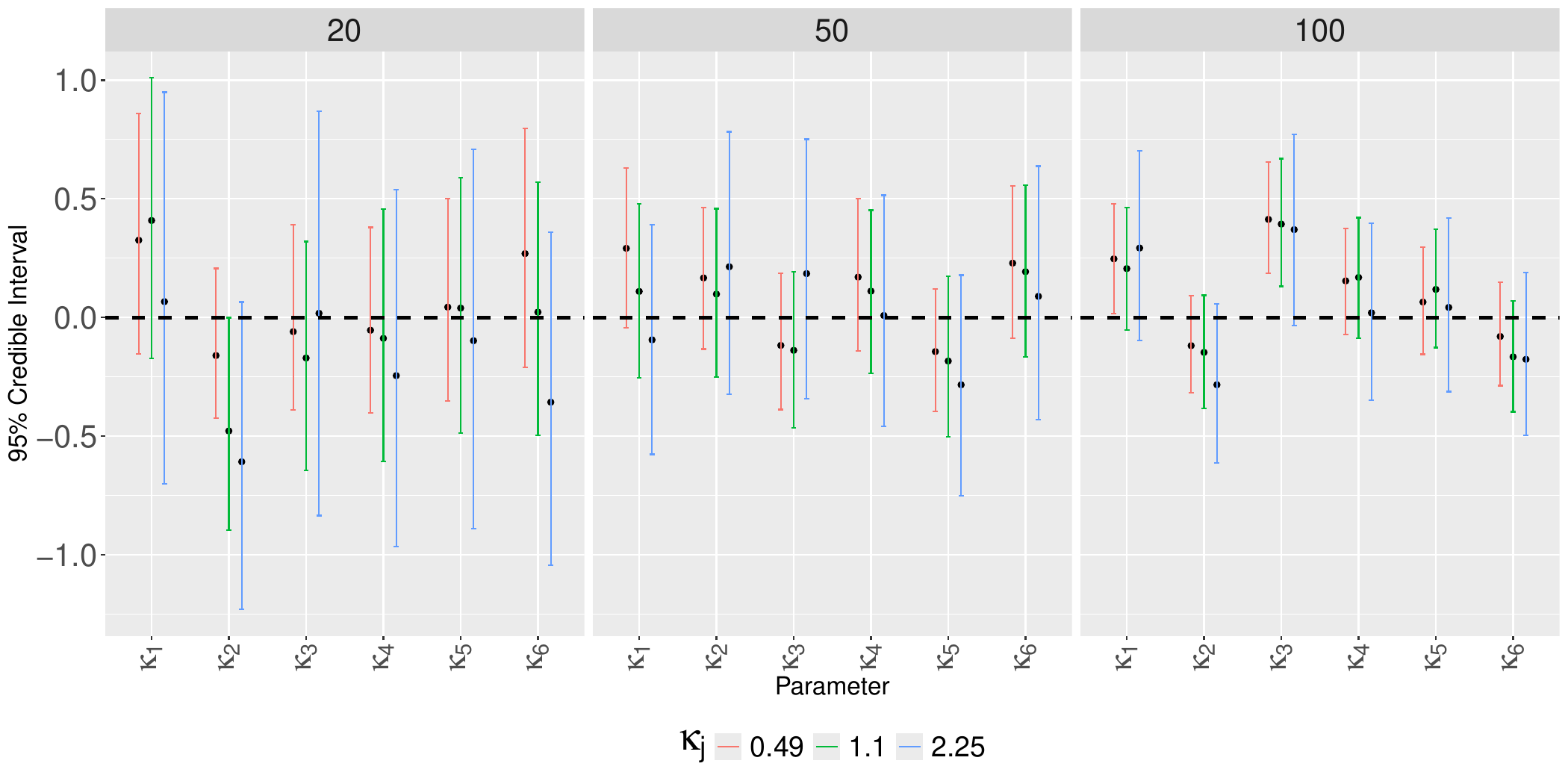}
        \caption{Diagonal $\bm{\Sigma}$}
    \end{subfigure}
    \hfill
    \begin{subfigure}[b]{0.750\textwidth}
        \centering
        \includegraphics[width=\textwidth]{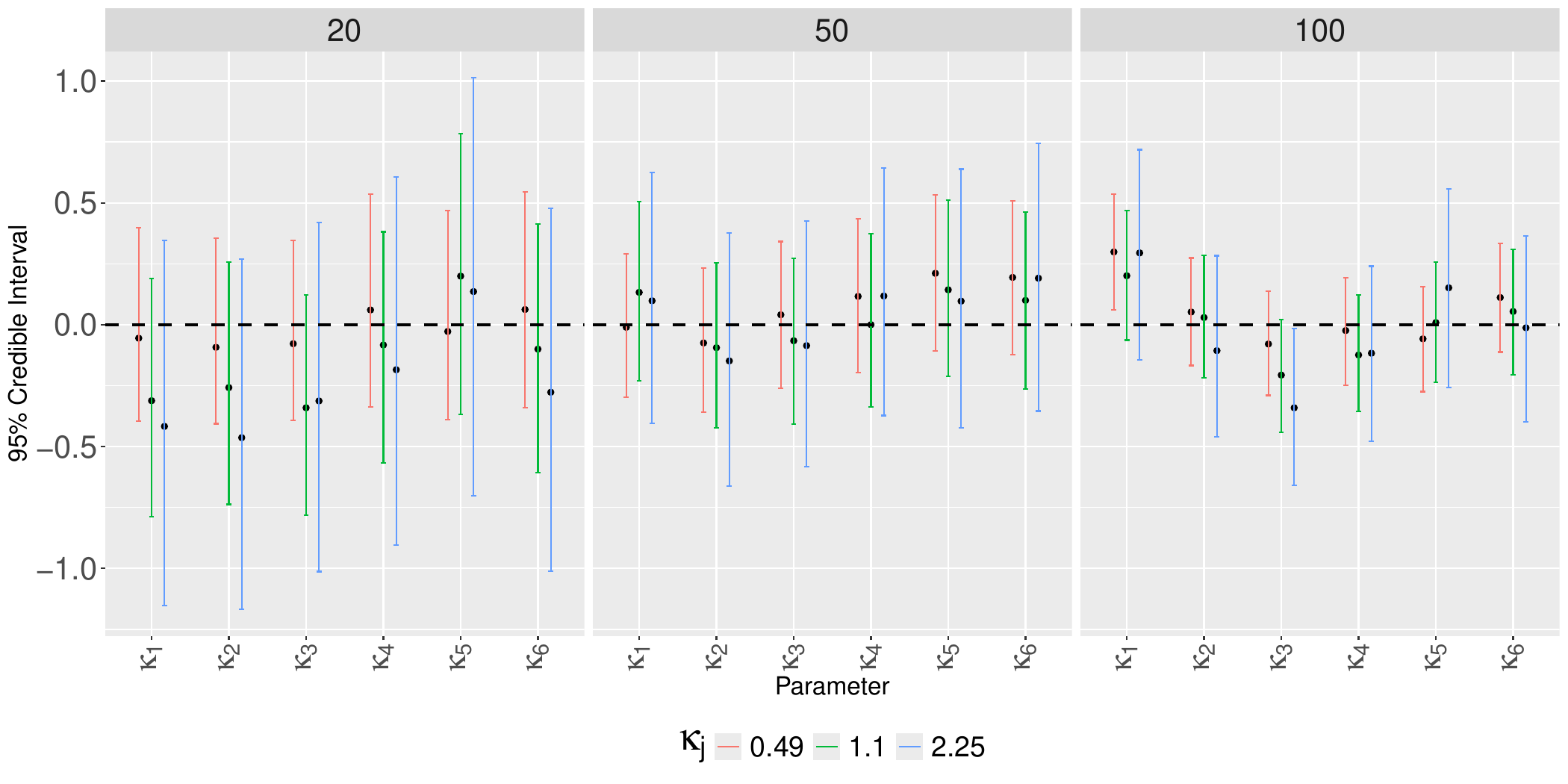}
        \caption{Full $\bm{\Sigma}$}
    \end{subfigure}
    \caption{TPN results - 95\% confidence intervals of the posterior distribution for the difference between the parameter $\kappa_j$ and its true value (lines). The dot represents the posterior means. The first row presents results with a diagonal $\bm{\Sigma}$, while the second row shows results with a full $\bm{\Sigma}$. The left panel displays results for $n = 20$, the center panel for $n = 50$, and the right panel for $n = 100$.}
    \label{fig:sim_res_2}
\end{figure}

\begin{figure}[t]
	\centering
    \begin{subfigure}[b]{0.750\textwidth}
        \centering
        \includegraphics[width=\textwidth]{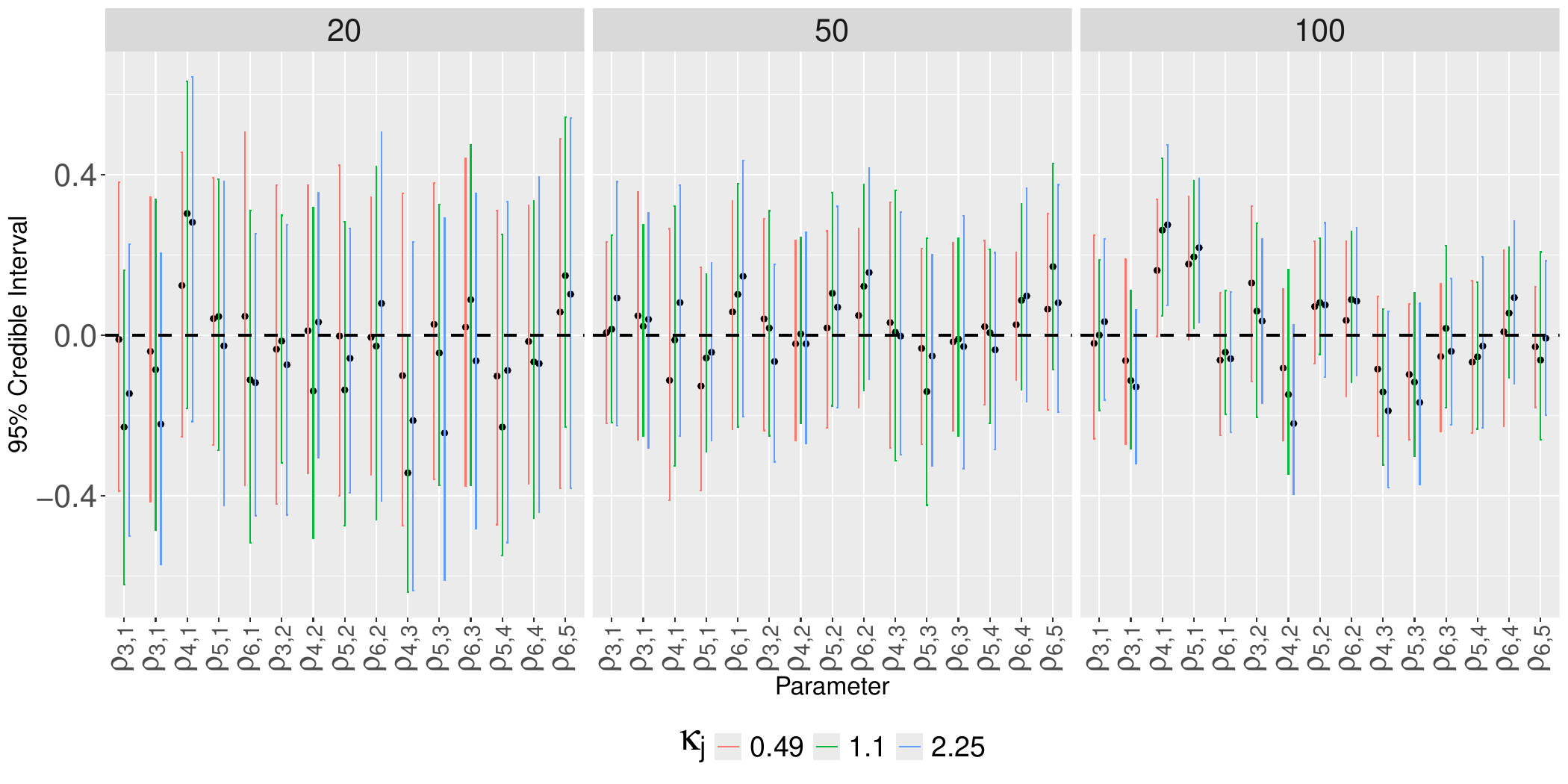}
        \caption{Diagonal $\bm{\Sigma}$}
    \end{subfigure}
    \hfill
    \begin{subfigure}[b]{0.750\textwidth}
        \centering
        \includegraphics[width=\textwidth]{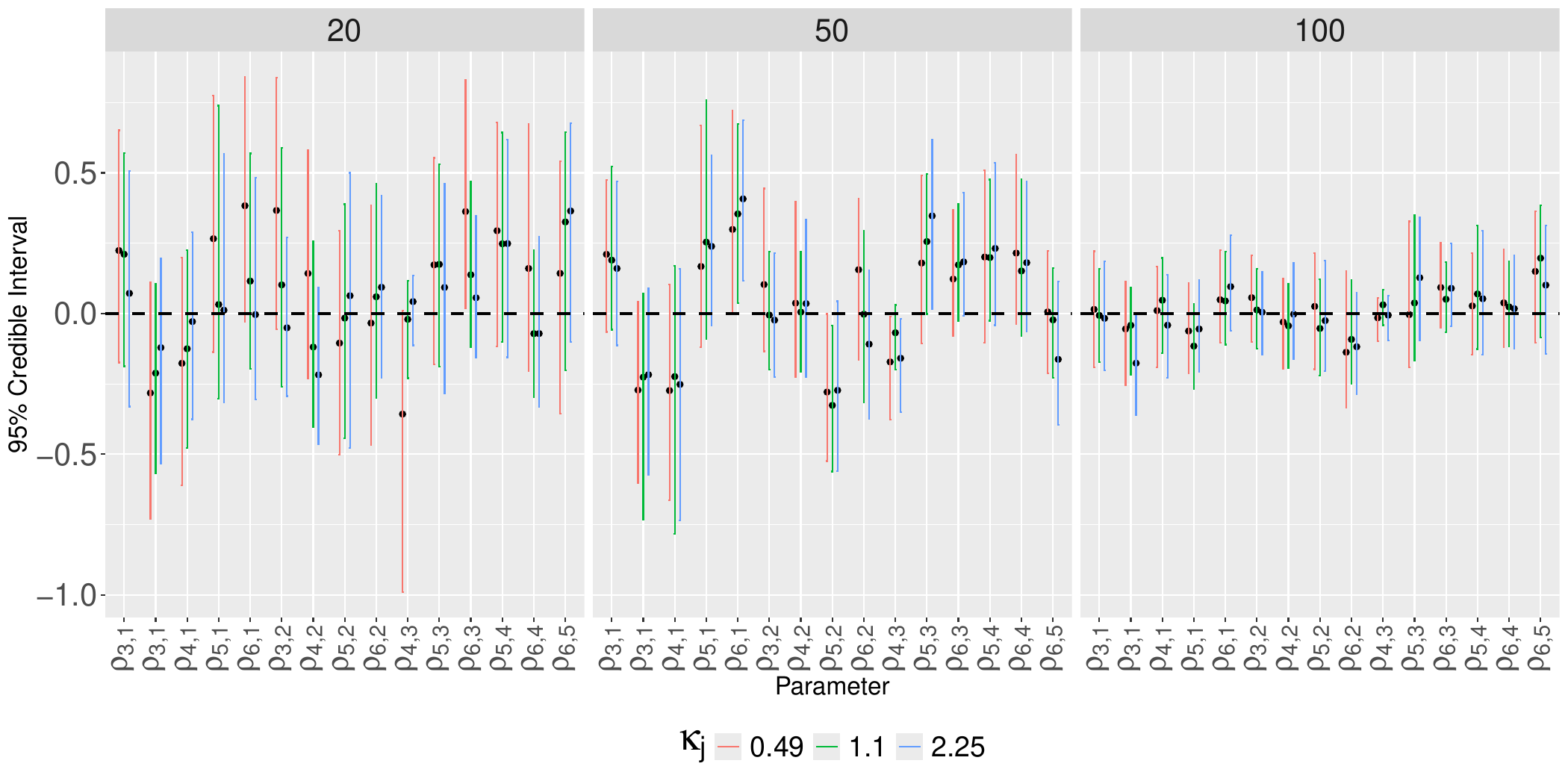}
        \caption{Full $\bm{\Sigma}$}
    \end{subfigure}
    \caption{TPN results - 95\% confidence intervals of the posterior distribution for the difference between the parameter $\rho_{j,k}$ and its true value (lines). The dot represents the posterior means. The first row presents results with a diagonal $\bm{\Sigma}$, while the second row shows results with a full $\bm{\Sigma}$. The left panel displays results for $n = 20$, the center panel for $n = 50$, and the right panel for $n = 100$.}
    \label{fig:sim_res_3}
\end{figure}

\begin{figure}[t]
	\centering
    \begin{subfigure}[b]{0.750\textwidth}
        \centering
        \includegraphics[width=\textwidth]{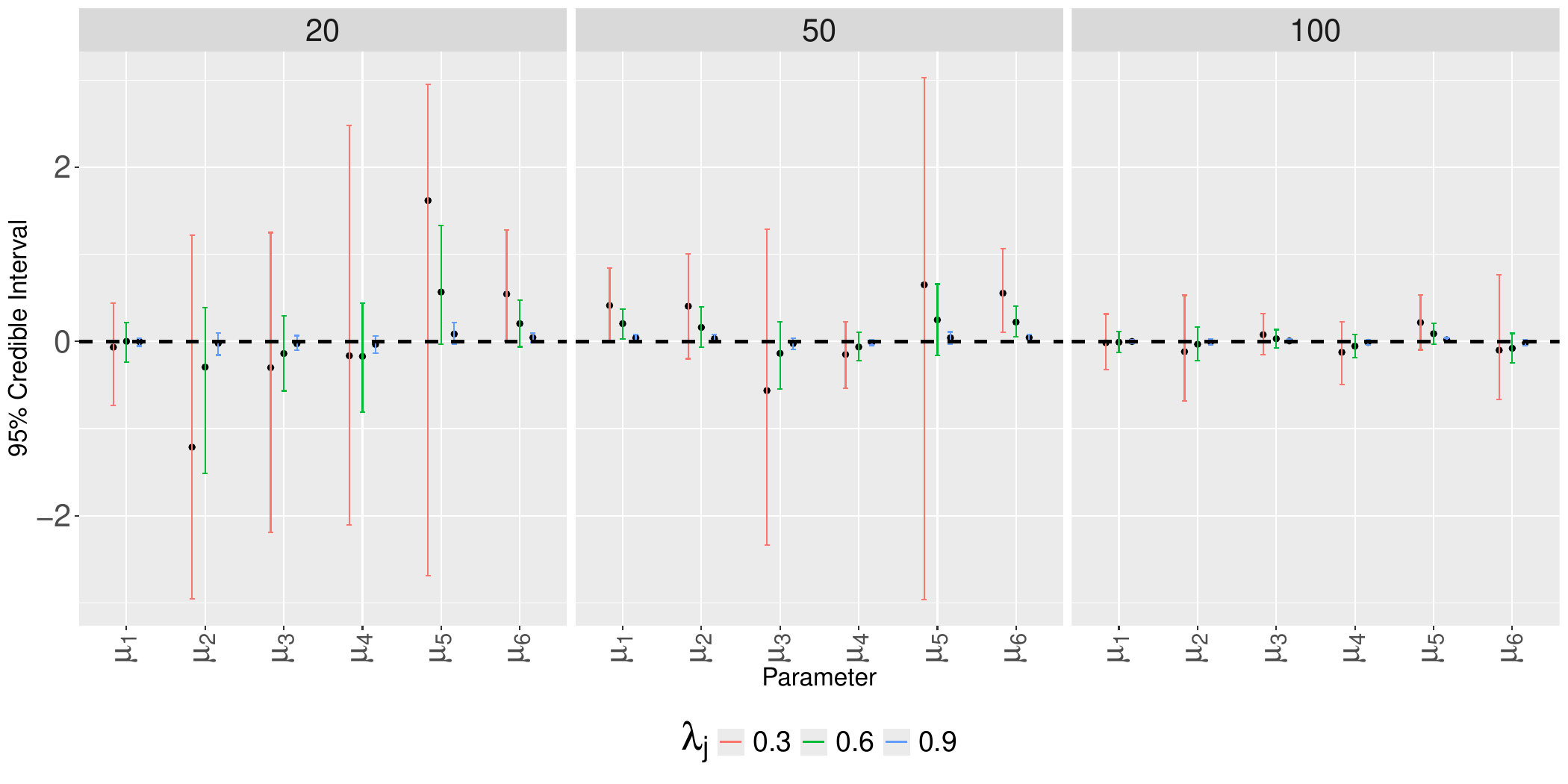}
        \caption{Diagonal $\bm{\Sigma}$}
    \end{subfigure}
    \hfill
    \begin{subfigure}[b]{0.750\textwidth}
        \centering
        \includegraphics[width=\textwidth]{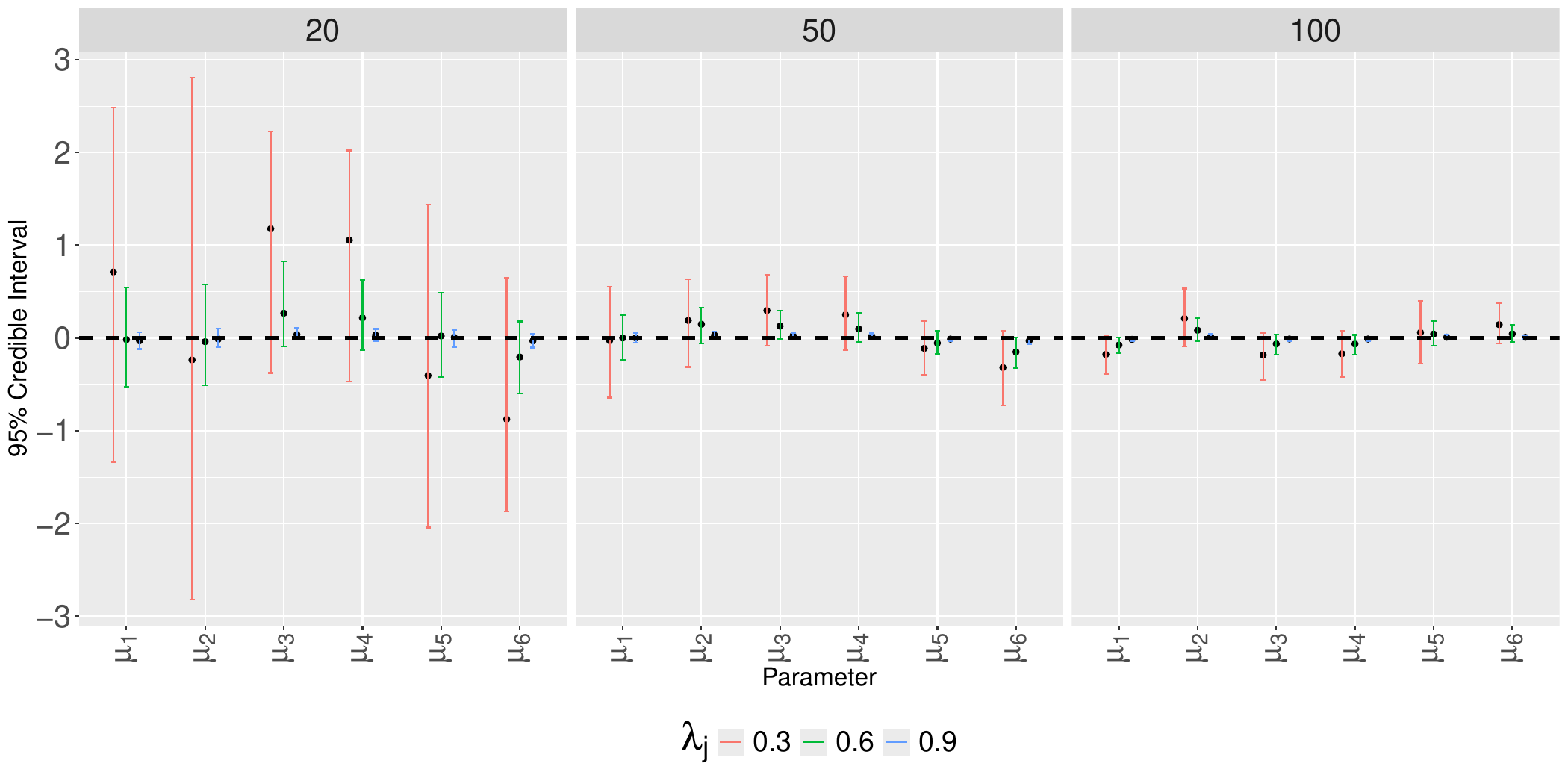}
        \caption{Full $\bm{\Sigma}$}
    \end{subfigure}
    \caption{CTPN results - 95\% confidence intervals of the posterior distribution for the circular difference between the parameter $\mu_j$ and its true value  (lines). The dot represents the posterior means. The first row presents results with a diagonal $\bm{\Sigma}$, while the second row shows results with a full $\bm{\Sigma}$. The left panel displays results for $n = 20$, the center panel for $n = 50$, and the right panel for $n = 100$.}
    \label{fig:sim_res_4}
\end{figure}
\begin{figure}[t]
	\centering
    \begin{subfigure}[b]{0.750\textwidth}
        \centering
        \includegraphics[width=\textwidth]{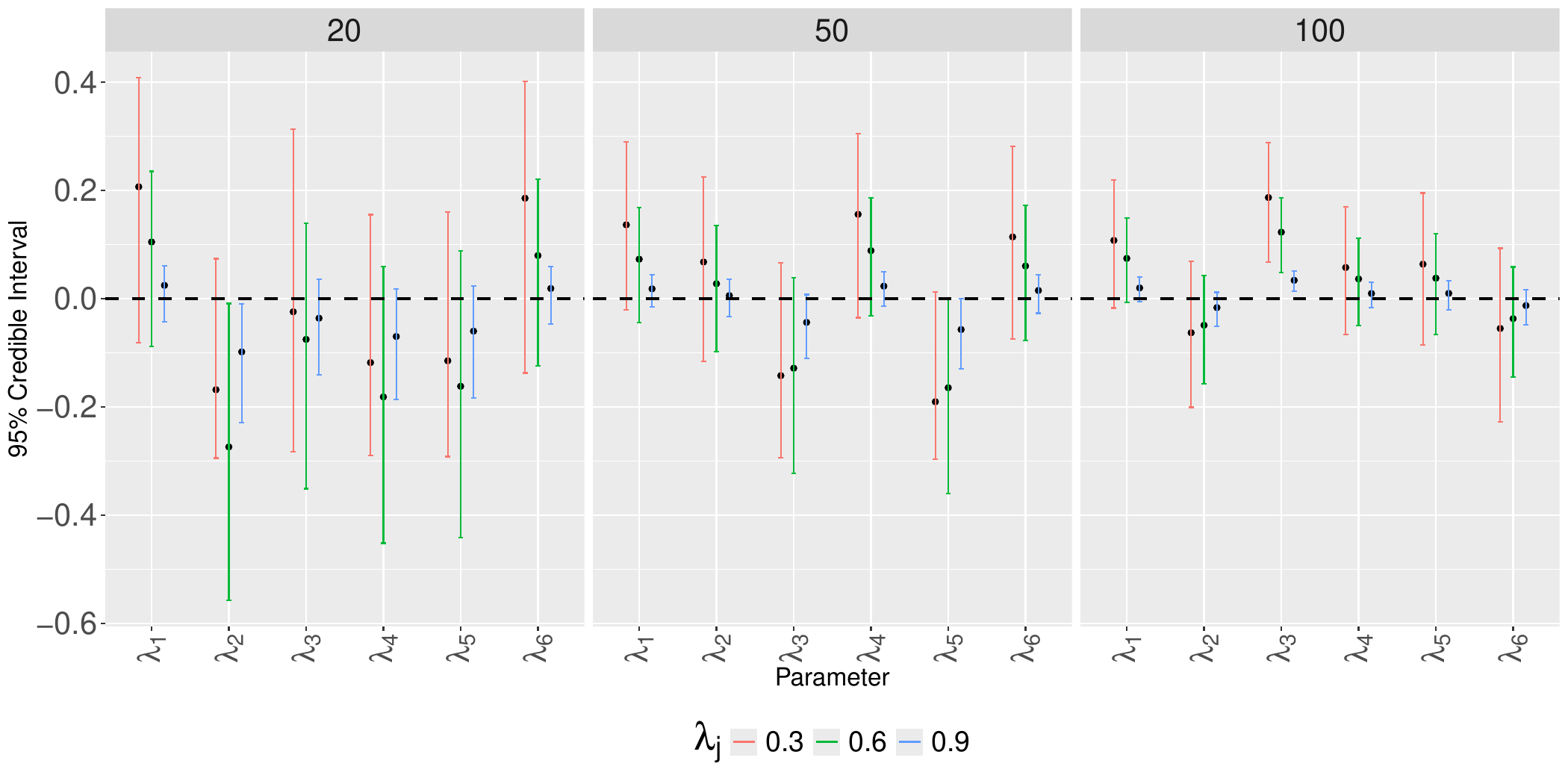}
        \caption{Diagonal $\bm{\Sigma}$}
    \end{subfigure}
    \hfill
    \begin{subfigure}[b]{0.750\textwidth}
        \centering
        \includegraphics[width=\textwidth]{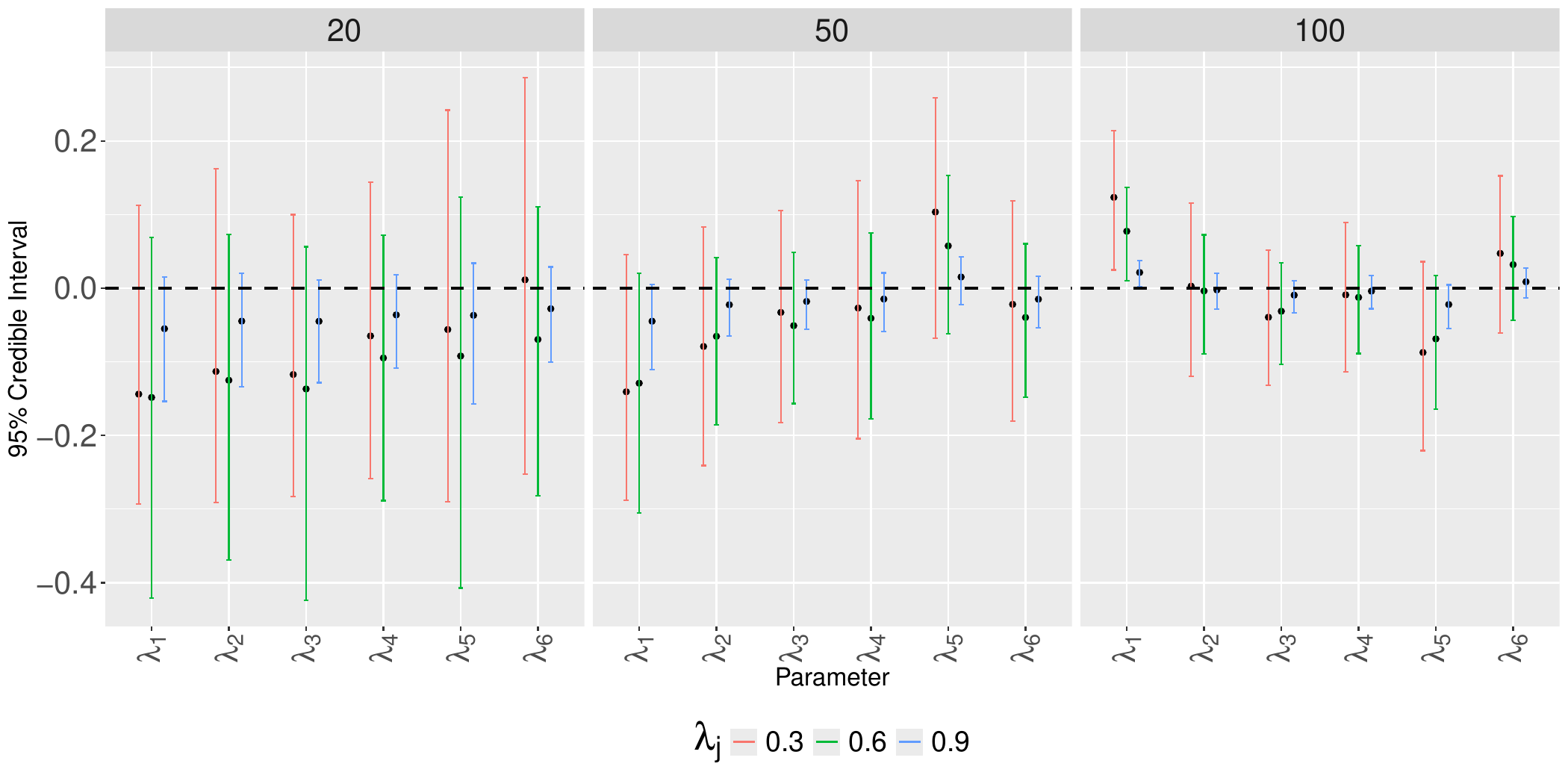}
        \caption{Full $\bm{\Sigma}$}
    \end{subfigure}
    \caption{CTPN results - 95\% confidence intervals of the posterior distribution for the difference between the parameter $\lambda_{j}$ and its true value (lines). The dot represents the posterior means. The first row presents results with a diagonal $\bm{\Sigma}$, while the second row shows results with a full $\bm{\Sigma}$. The left panel displays results for $n = 20$, the center panel for $n = 50$, and the right panel for $n = 100$.}
    \label{fig:sim_res_5}
\end{figure}

\begin{figure}[t]
	\centering
    \begin{subfigure}[b]{0.750\textwidth}
        \centering
        \includegraphics[width=\textwidth]{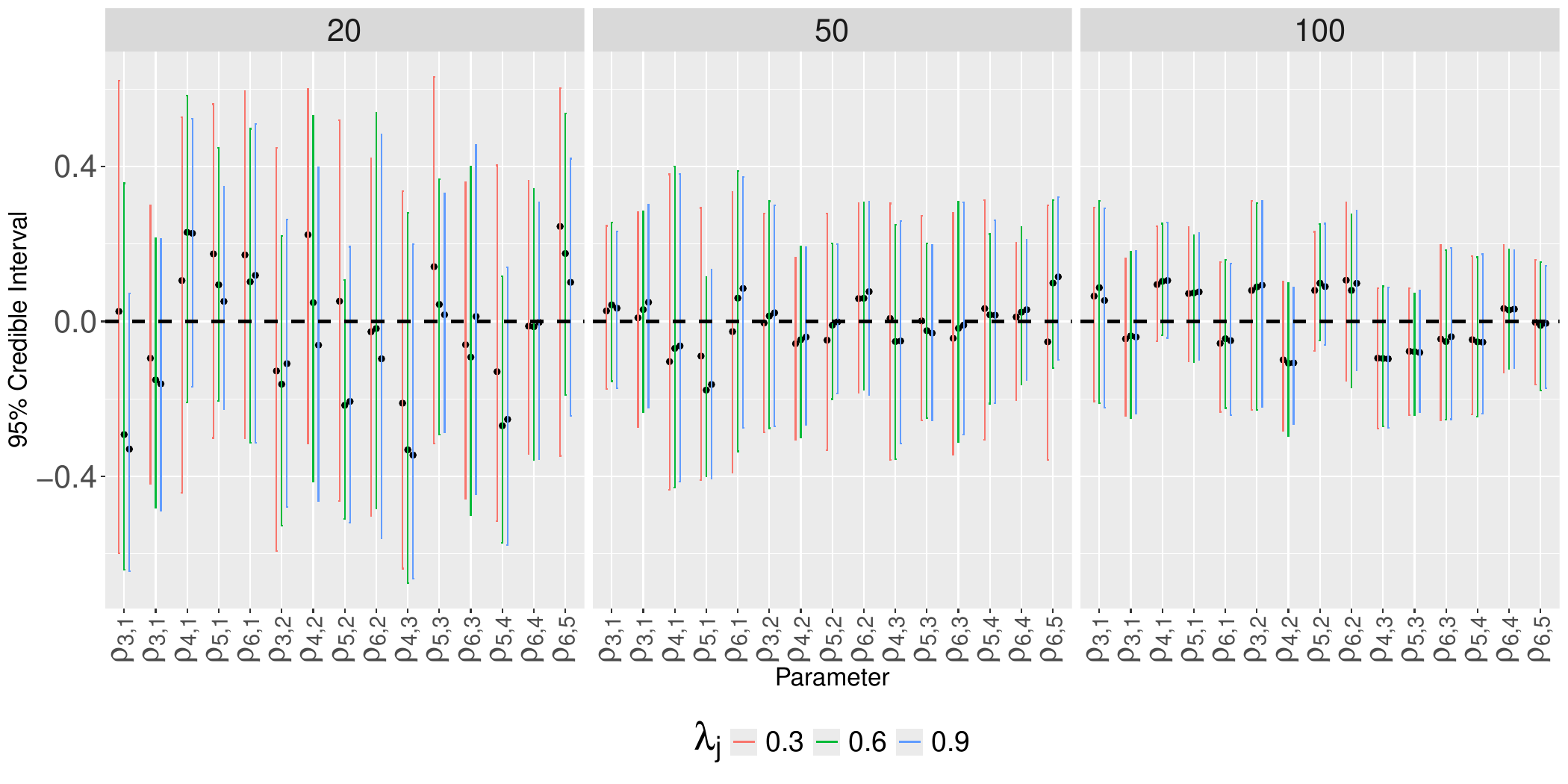}
        \caption{Diagonal $\bm{\Sigma}$}
    \end{subfigure}
    \hfill
    \begin{subfigure}[b]{0.750\textwidth}
        \centering
        \includegraphics[width=\textwidth]{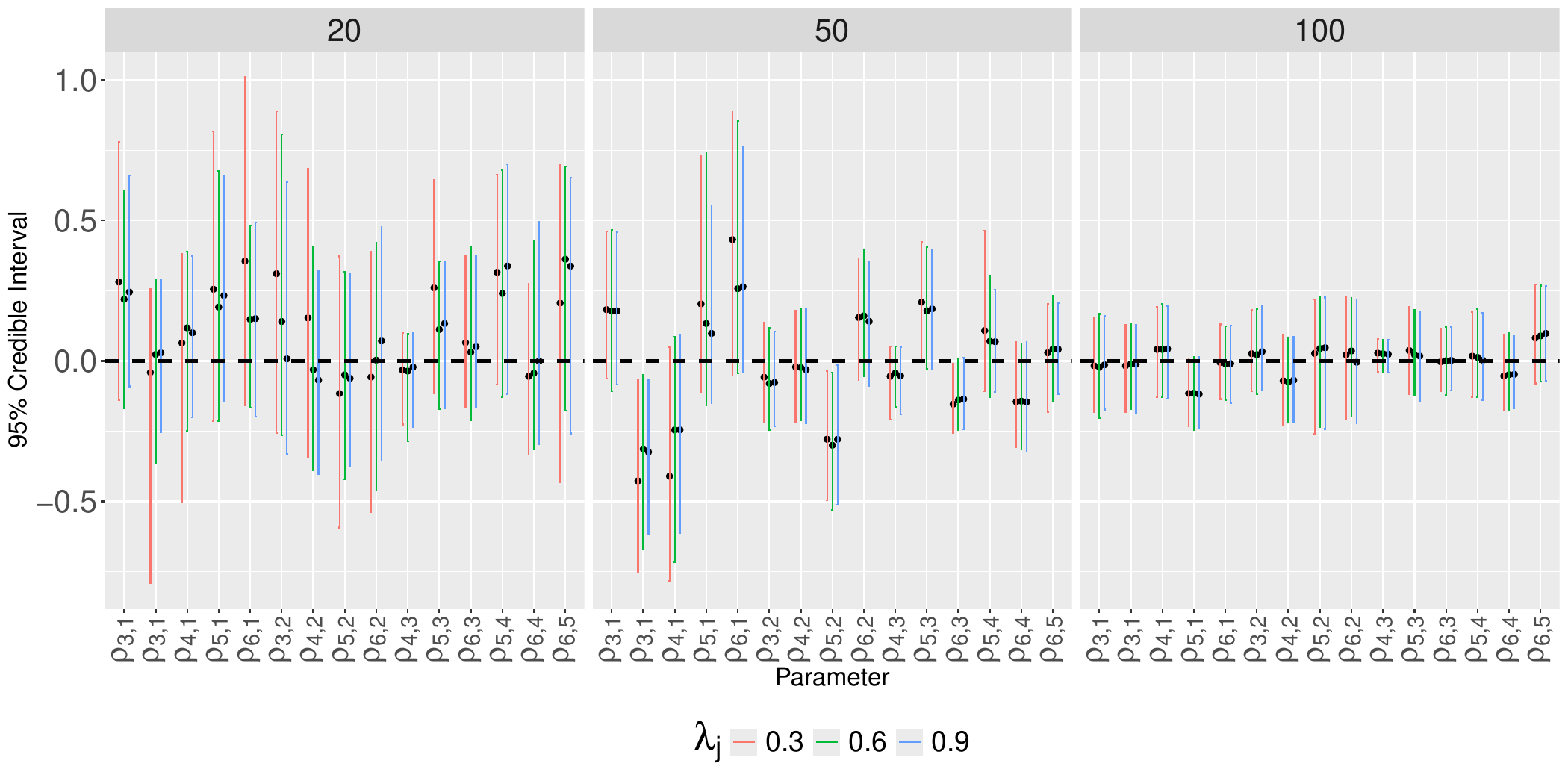}
        \caption{Full $\bm{\Sigma}$}
    \end{subfigure}
    \caption{CTPN results -  95\% confidence intervals of the posterior distribution for the difference between the parameter $\rho_{j,k}$ and its true value (lines). The dot represents the posterior means. The first row presents results with a diagonal $\bm{\Sigma}$, while the second row shows results with a full $\bm{\Sigma}$. The left panel displays results for $n = 20$, the center panel for $n = 50$, and the right panel for $n = 100$.}
    \label{fig:sim_res_6}
\end{figure}

With these simulations, we aim to evaluate whether the parameters of the TPN and CTPN can be estimated.

\para{TPN model:} In this setting, we assume we have $n$ vectors of circular variables $\bm{\theta}_i$ which are distributed as a $\textsc{TPN}_d(\bm{\mu}, \bm{\kappa}, \bm{\Sigma})$. In all simulations, we assume $d=6$ and 
\begin{equation} \label{eq:mu}
	\bm{\mu} = \left(0, \frac{\pi}{6} , \frac{2\pi}{6}, \frac{3\pi}{6}, \frac{4\pi}{6}, \frac{5\pi}{6}\right)^\T. 
\end{equation} 
For the parameter $\bm{\kappa}$, we consider three different settings: $\bm{\kappa} = 0.49\bm{1}_6$, $\bm{\kappa} = 1.1\bm{1}_6$, and $\bm{\kappa} = 2.45 \bm{1}_6$, which correspond to circular concentrations equal to 0.3, 0.6 and 0.9 respectively,  and are used to test the model's performance under high, medium, and low variability. For $ \bm{\Sigma}$, we consider a case where the random variables are independent, i.e., $ \bm{\Sigma} = \bm{I}_6$, and the following $\bm{\Sigma}$ which has been generated from the TIW
\begin{equation}
	\left(
\begin{array}{cccccc}
	 1.0000& -0.2491&  0.4159 & 0.2955 &-0.3639& -0.4327\\
	-0.2491&  1.0000 & -0.4133 &-0.2469 & 0.1064&  0.0461\\
		0.4159& -0.4133&  1.0000 & 0.8052 &-0.2054& -0.5939\\
	  0.2955& -0.2469&  0.8052 & 1.0000 &-0.3317 &-0.4140\\
	 -0.3639&  0.1064& -0.2054& -0.3317 & 1.0000 &-0.0489\\
	 -0.4327&  0.0461& -0.5939& -0.4140& -0.0489 & 1.0000
\end{array}	 
\right).
\end{equation}
The datasets are then generated by all possible combinations of the three vectors for $\bm{\kappa}$, the two different $\bm{\Sigma}$, and $n \in \{20,50,100\}$.

\para{CTPN model:} To use the CTPN, we need to specify the density for the marginals. In this case, since the main interest is in parameter recovery, we select the wrapped Cauchy marginals, which have closed-form expressions for  the density the cumulative distribution function, and the quantile function, necessary for the model: these closed-form expressions are in \S 6 of the main article. For the  copula component, we only need to specify the value of $\bm{\Sigma}$, which remains the same as in the previous section. 
For the vector of mean directions of the marginals, we use the same values as in \eqref{eq:mu}. 
Similarly to the TPN case, we test vectors of marginal concentrations $\bm{\lambda}$, which are set to $\bm{\lambda} = 0.3\bm{1}_6$, $\bm{\lambda} = 0.6\bm{1}_6$, and $\bm{\lambda} = 0.9 \bm{1}_6$, representing small, medium, and large concentration, respectively. Then, we test the model for all possible combinations of the three vectors for $\bm{\lambda}$, the two different $\bm{\Sigma}$, and $n \in \{20,50,100\}$.
The prior distributions are the same of the one used in \S 7 of the main article, while the iterations are 30,000, with a burn-in of 10,000 and a thinning factor of 10.

The results are shown in Figs.~\ref{fig:sim_res_1}--\ref{fig:sim_res_6}, depicting the posterior distribution of the parameters relative to the true values. Although, for some parameters, the 95\% CI does not contain zero, it is evident that, in most cases, regardless of the parameter value or whether the TPN or CTPN is used, the MCMC method successfully recovers the true parameters used to simulate the data. As expected, this recovery improves with larger sample sizes.

